\documentclass[12pt]{article}\usepackage[]{graphicx}\usepackage[]{color}
\makeatletter
\def\maxwidth{ %
  \ifdim\Gin@nat@width>\linewidth
    \linewidth
  \else
    \Gin@nat@width
  \fi
}
\makeatother

\definecolor{fgcolor}{rgb}{0.345, 0.345, 0.345}
\newcommand{\hlkwb}[1]{\textcolor[rgb]{0.69,0.353,0.396}{#1}}%

\usepackage{framed}
\makeatletter
 {\par\unskip\endMakeFramed%
 \at@end@of@kframe}
\makeatother

\definecolor{shadecolor}{rgb}{.97, .97, .97}
\definecolor{messagecolor}{rgb}{0, 0, 0}
\definecolor{warningcolor}{rgb}{1, 0, 1}
\definecolor{errorcolor}{rgb}{1, 0, 0}
\newenvironment{knitrout}{}{} 

\usepackage{alltt}

\newcommand\PartOfBoundOffDiagonal{3\efourB+4(K + \dmax)\,\efive}
\newcommand\boundOffDiagonal{\efourA + \PartOfBoundOffDiagonal}
\newcommand\K{K^\prime}
\newcommand\ThmOneVarBound{Q^{4b} \, \Unit\Time  \big(\altb + \etwo \, (\Unit\Time-\altb) \big)}

\newcommand\BvConstant{\mathcal{C}_0}
\newcommand\dmax{d_{\max}}

\newcommand\boundLemmaUniform{\efourB+(K+\dmax)\efive}

\newcommand\altB{C}
\newcommand\el{\epsilon}   
\newcommand\eone{\epsilon^{}_{\!\mathrm{A1}}}
\newcommand\etwo{\epsilon^{}_{\!\mathrm{A4}}}
\newcommand\altb{c}
\newcommand\Ai{
\begin{assumption}\label{A1}
There is an $\eone>0$, independent of $\Unit$ and $\Time$, and a collection of neighborhoods $\{B_{\unit\comma\time}\subset A_{\unit\comma\time}, \unit\in 1\mycolon\Unit,\time\in 1\mycolon\Time\}$ such that, for all  $\unit$ and $\time$, any  bounded real-valued function $|h(x)|\le 1$, and any value of $x_{B^c_{\unit\comma\time}}$, 
\begin{eqnarray}
\nonumber
&& \hspace{-12mm} \Bigg| \int h(x_{\unit\comma\time}) f_{X_{\unit\comma\time}|Y_{B_{\unit\comma\time}},X_{B^c_{\unit\comma\time}}}(x_{\unit\comma\time}\given  \data{y}_{B_{\unit\comma\time}}, x_{B^c_{\unit\comma\time}}) \, dx_{\unit\comma\time}
\\
\siOnly{\label{eq:weak_coupling2}}
\msOnly{\nonumber}
&& \hspace{-3mm} - \int h(x_{\unit\comma\time})f_{X_{\unit\comma\time}|Y_{B_{\unit\comma\time}}}(x_{\unit\comma\time}\given \data{y}_{B_{\unit\comma\time}})
\, dx_{\unit\comma\time}
\Bigg| \hspace{1mm} < \hspace{1mm} \eone.
\end{eqnarray}
\end{assumption}
}

\newcommand\Aii{
\begin{assumption}\label{A1b}
For the collection of neighborhoods in Assumption~\ref{A1}, with $B^+_{\unit\comma\time}=B^{}_{\unit\comma\time}\cup (\unit,\time)$, there is a constant $\Bsize$, depending on $\eone$ but not on $\Unit$ and $\Time$, such that
\begin{equation}
\nonumber
\sup_{\unit\in 1:\Unit, \, \time\in 1:\Time} \big| B^+_{\unit\comma\time}\big| \le \Bsize.
\end{equation}
\end{assumption}
}

\newcommand\Aiii{
\begin{assumption}\label{A2}
There is a constant $Q$, independent of $\Unit$ and $\Time$, such that, for all $\unit$ and $\time$,
\begin{equation}
\nonumber
Q^{-1} <f_{Y_{\unit\comma\time}|X_{\unit\comma\time}}(\data{y}_{\unit\comma\time}\given x_{\unit\comma\time}) < Q
\end{equation}
\end{assumption}
}

\newcommand\Aiv{
  \begin{assumption}\label{A:unconditional:mix}
    There exists $\etwo > 0$, independent of $\Unit$ and $\Time$, such that the following holds.
    For each $\unit, \time$, a set $\altB^{}_{\unit\comma\time} \subset (1\mycolon U)\times(0\mycolon N)$ exists such that $(\altUnit, \altTime) \notin \altB_{\unit\comma\time}$ implies $B_{\unit\comma\time}^+ \cap B_{\altUnit\comma\altTime}^+ = \emptyset$ and
      \begin{equation} 
\siOnly{\label{eq:A:unconditional:mix}}
\msOnly{\nonumber}
        \big| f_{X_{B^{+}_{\altUnit\comma\altTime}}|X^{}_{B^{+}_{\unit\comma\time}}} - f_{X_{B^{+}_{\altUnit\comma\altTime}}} \big|
        < \etwo \, f_{X_{B^{+}_{\altUnit\comma\altTime}}}
      \end{equation}
      Further, there is a uniform bound $|\altB_{\unit\comma\time}| \le \altb$.
\end{assumption}
}

\newcommand\TheoremI{
\begin{theorem}  \label{thm:tif}
Let $\MC{\loglik}$ denote the Monte Carlo likelihood approximation constructed by {\UBF}.
Consider a limit with a growing number of bootstrap replicates, $\Rep\to\infty$, and suppose assumptions~\ref{A1},~\ref{A1b} and~\ref{A2}.
There are quantities $\el(\Unit,\Time)$ and $V(\Unit,\Time)$, with bounds $|\el|<\eone Q^{2}$ and $V < Q^{4\Bsize} \, U^2 N^2$, such that
\begin{equation}
\label{th1:lik:bound}
\Rep^{1/2}\big[ \MC{\loglik}-\loglik-\el\Unit\Time \big]  \xrightarrow[\Rep \rightarrow \infty]{d} \normal\big[0,V\big],
\end{equation}
where $\xrightarrow[\Rep \rightarrow \infty]{d}$ denotes convergence in distribution and $\normal[\mu,\Sigma]$ is the normal distribution with mean $\mu$ and variance $\Sigma$.
If additionally Assumption~\ref{A:unconditional:mix} holds, we obtain an improved variance bound
\begin{equation}
\label{th1:lik:bound2}
V < \ThmOneVarBound.
\end{equation}
\end{theorem}
}

\newcommand\ethree{\epsilon^{}_{\mathrm{B1}}}
\newcommand\efourA{\epsilon^{}_{\mathrm{B4}}}
\newcommand\efourB{\epsilon^{}_{\mathrm{B5}}}
\newcommand\efive{\epsilon_{\mathrm{B6}}}

\newcommand\Bi{

\begin{assumptionB}\label{B1}
  There is an $\ethree>0$, independent of $\Unit$ and $\Time$, and a collection of neighborhoods $\{B_{\unit\comma\time}\subset A_{\unit\comma\time}, \unit\in 1\mycolon\Unit,\time\in 1\mycolon\Time\}$ such that the following holds for all $\unit$ and $\time$, and any bounded real-valued function $|h(x)|\le 1$.
  Setting $A=A_{u,n}$, $B=B_{u,n}$, $f_{A}(x_A) = f_{Y_{A}|X_{A}}(\data y_{A}|x_A)$, and $f_{B}(x_B) = f_{Y_{B}|X_{B}}(\data y_{B}|x_B)$, so that we have the identity
    \[
    f_{X_{u,n}|Y_A}(x | \data{y}_A) = 
    \frac{
      \E_{g} \! \big[ 
      f_{A}\big(X_{A}^P\big) f_{X_{u,n}|X_{A^{[n]}},\myvec X_{n-1}} \big(x | X_{A^{[n]}}^P, \myvec X_{n-1}\big) 
      \big]
    } {
      \E_{g} \! \big[f_A\big(X_A^P\big)\big]
    },
  \]
we require that
\begin{eqnarray}
\nonumber
&&\hspace{-30mm}
\left| 
\int \hspace{-1mm} h(x)
\left\{
\frac{
  \E_{g} \! \big[ 
    f_{A}\big(X_{A}^P\big) f_{X_{u,n}|X_{A^{[n]}},\myvec X_{n-1}} \big(x | X_{A^{[n]}}^P, \myvec X_{n-1}\big) 
  \big]
} {
  \E_{g} \! \big[f_A\big(X_A^P\big)\big]
} - 
\right.
\right.
\\
\nonumber
&& 
\siOnly{\hspace{20mm}}
\hspace{-6.5mm} 
\siOnly{\hspace{2mm}}
\left.
\left.
\frac{
  \E_{g} \! \big[
    f_{B}\big(X_{B}^P\big) f_{X_{u,n}|X_{B^{[n]}},\myvec X_{n-1}}\big(x | X_{B^{[n]}}^P, \myvec X_{n-1}\big) \big] 
} {
  \E_{g} \! \big[f_B\big(X_B^P\big)\big]
} 
\!
\right\}
\!
dx
\right| \hspace{1mm} < \hspace{1mm} \ethree.
\end{eqnarray}
\end{assumptionB}
}

\newcommand\Bii{
\begin{assumptionB}\label{B1b}
The bound $\sup_{\unit\in 1:\Unit,\time\in 1:\Time} \big| B^+_{\unit\comma\time}\big| \le \Bsize$ in Assumption~\ref{A1b} applies for the neighborhoods defined in Assumption~\ref{B1}. 
This also implies there is a finite maximum temporal depth for the collection of neighborhoods, defined as
\begin{equation}
\nonumber
\dmax=\sup_{(\unit,\time)}\hspace{1mm} \sup_{(\altUnit,\altTime)\in B_{\unit,\time}}|\time-\altTime|.
\end{equation}
\end{assumptionB}
}

\newcommand\Biii{
\begin{assumptionB}\label{B2} Identically to Assumption~\ref{A2}, 
$Q^{-1} <f_{Y_{\unit\comma\time}|X_{\unit\comma\time}}(\data{y}_{\unit\comma\time}\given x_{\unit\comma\time}) < Q$.
\end{assumptionB}
}

\newcommand\BivA{
\begin{assumptionB}\label{B:unconditional:mix}
  We use subscripts of $g$ to denote marginal and conditional densities derived from \myeqref{eq:g}.
Suppose there is an $\efourA$, independent of $\Unit$ and $\Time$, such that the following holds.
For each $\unit$ and $\time$, a set $\altB^{}_{\unit\comma\time} \subset (1\mycolon U)\times(0\mycolon N)$ exists such that $(\altUnit, \altTime) \notin \altB_{\unit\comma\time}$ implies $B_{\unit\comma\time}^+ \cap B_{\altUnit\comma\altTime}^+ = \emptyset$ and
\begin{equation}
\nonumber
\big| 
   g^{}_{{X}^{P}_{B^{}_{\altUnit\comma\altTime} \cup B^{}_{\unit\comma\time}}} 
   - g^{}_{{X}^{P}_{B^{}_{\altUnit\comma\altTime}}} \, g^{}_{{X}^{P}_{B^{}_{\unit\comma\time}}}
\big|
<
(1/2) \, \efourA \, \,  g^{}_{{X}^{P}_{{B^{}_{\altUnit\comma\altTime}} \cup {B^{}_{\unit\comma\time}}}}
\end{equation}
\begin{eqnarray}
\nonumber
\big|
g^{}_{X^P_{B^{}_{\altUnit\comma\altTime}}|\myvec{X}_{0:\Time}} 
\, g^{}_{X^P_{B^{}_{\unit\comma\time}}|\myvec{X}_{0:\Time}} 
-
g^{}_{X^P_{B^{}_{\altUnit\comma\altTime} \cup B^{}_{\unit\comma\time}}|\myvec{X}_{0:\Time}}
\big|
&&
\\
\nonumber
&&\hspace{-30mm} < (1/2) \, \efourA \, \, 
 g^{}_{X^P_{B^{}_{\altUnit\comma\altTime} \cup B^{}_{\unit\comma\time}}|\myvec{X}_{0:\Time}}
      \end{eqnarray}
      Further, there is a uniform bound $|\altB_{\unit\comma\time}| \le \altb$.
\end{assumptionB}
}
\newcommand\BivB{
\begin{assumptionB}\label{B:temporal:mix}
There is a constant $K$, independent of $\Unit$ and $\Time$, such that, for any $0\leq d \leq \dmax$, any $\time\ge K+d$, and any set $D\subset (\UnitSet)\times(n{\mycolon}n-d)$, 
\begin{eqnarray}
\nonumber
&&\hspace{-6mm} \big|
g^{}_{{X}_{D}|\myvec{X}_{\time-d-K}}(x^{}_{D}\given\myvec{x}^{(1)}_{\time-d-K})
-
g^{}_{{X}_{D}|\myvec{X}_{\time-d-K}}(x^{}_{D}\given\myvec{x}^{(2)}_{\time-d-K})
\big|
\\
\nonumber
&& \siOnly{\hspace{10mm}} \hspace{4mm}
< 
\efourB  \,  \,
g^{}_{{X}_{D}|\myvec{X}_{\time-d-K}}(x^{}_{D}\given\myvec{x}^{(1)}_{\time-d-K})
\end{eqnarray}
holds for all $\myvec{x}^{(1)}_{\time-d-K}$, $\myvec{x}^{(2)}_{\time-d-K}$, and $x^{}_{D}$.
\end{assumptionB}
}

\newcommand\Bv{
\begin{assumptionB}\label{B:girf}
Let $h$ be a bounded function with $|h(x)|\le 1$. 
Let $\myvec{X}^{\GR}_{\time,\Ninter,\np,\rep}$ be the Monte Carlo quantity constructed in {\ABFIR}, conditional on $\myvec{X}^{\IF}_{\time-1,\Ninter,\rep}=\myvec{x}^{\IF}_{\time-1,\Ninter,\rep}$.
There is a constant $\BvConstant(\Unit,\Time,\Ninter)$ such that, for all $\efive^{}>0$ and $\myvec{x}^{\IF}_{\time-1,\Ninter,\rep}$, whenever the number of particles satisfies $J > \BvConstant(\Unit,\Time,\Ninter)/\efive^{3}$,
\begin{equation}
\nonumber
\hspace{-0.5mm}
\left|  
\, 
  \EMC\Big[ 
    \frac{1}{\Np} \sum_{\np=1}^{\Np} h(\myvec{X}^{\GR}_{\time,\Ninter,\np,\rep}) 
   \Big] \!
-  \E_g \! \big[h(\myvec{X}_\time)\given \myvec{X}_{\time-1}=\myvec{x}^{\IF}_{\time-1,\Ninter,\rep}\big] 
\right| \!
< \efive^{}.
\end{equation}
\end{assumptionB}
}

\newcommand\Bvi{
\begin{assumptionB}\label{B:XG_XA_ind}
For $1\leq \time \leq N$, the Monte Carlo random variable $X^A_{\time,i}$ is independent of $w^M_{\unit,\time,i,j}$ conditional on $X^A_{\time-1,i}$.
\end{assumptionB}
}

\newcommand\TheoremII{
\begin{theorem}  \label{thm:abf}
Let $\MC{\loglik}$ denote the Monte Carlo likelihood approximation constructed by {\ABFIR}, or by {\ABF} since this is the special case of {\ABFIR} with $\Ninter=1$.
Consider a limit with a growing number of bootstrap replicates, $\Rep\to\infty$, and suppose assumptions~\ref{B1},~\ref{B1b}, \ref{B2}, \ref{B:temporal:mix}, \ref{B:girf} and~\ref{B:XG_XA_ind}. 
Suppose the number of particles $\Np$ exceeds the requirement for~\ref{B:girf}.
There are quantities $\el(\Unit,\Time)$ and $V(\Unit,\Time)$ with $|\el|<
Q^2\ethree + 2 Q^{2\Bsize}\big(\boundLemmaUniform\big)$
and $V<Q^{4\Bsize}\Unit^2\Time^2$ 
such that
\begin{equation}
\siOnly{\label{th:abf:lik:bound}}
\msOnly{\nonumber}
\Rep^{1/2}
  \big[ 
    \MC{\loglik}-\loglik-\el\Unit\Time 
  \big]  
  \xrightarrow[\Rep \rightarrow \infty]{d} 
  \normal\big[ 0,V \big].
\end{equation}
If additionally Assumption~\ref{B:unconditional:mix} holds, we obtain an improved rate of
\begin{equation}
\siOnly{\label{th:abf:lik:bound2}}
\msOnly{\nonumber}
V < Q^{4\Bsize}\Time\Unit
\big\{ c + \big(\boundOffDiagonal \big) \big(\Time\Unit-c\big)
\big\}
\end{equation}
\end{theorem}
}

\newcommand\mytitle{Bagged filters for partially observed interacting systems}

\newcommand\MainFigureBmScaling{1}

\newcommand\MainFigureMeaslesScaling{3}
\newcommand\MainFigureMeaslesSlice{4}
\newcommand\MainFigureMeaslesProfile{5}

\newcommand\SuppSecGeneralization{S1}
\newcommand\SuppSecAdaptedSimulation{S2}
\newcommand\SuppSecThmI{S3}
\newcommand\SuppSecThmII{S4}
\newcommand\SuppSecBM{S5}
\newcommand\SuppSecMeasles{S6}
\newcommand\SuppSecMeaslesNbhd{S7}
\newcommand\SuppSecLorenz{S8}
\newcommand\SuppSecMemoryEfficient{S9}
\newcommand\SuppSecLatentStates{S10}
\newcommand\SuppSecParameterEst{S11}
\newcommand\SuppSecIvsJ{S12}

\newcommand\filts{k}
\newcommand\Filts{K}
\newcommand\filtFunc{\phi}
\newcommand\filtFuncSet{\Phi}

\newcommand\comma{{\hspace{-0.25mm},\hspace{-0.2mm}}}

\newcommand\amplitude{a}
\newcommand\meanBeta{{\bar\beta}}

\newcommand\siOnly[1]{} 
\newcommand\msOnly[1]{#1} 

\newcommand\inputSpace{\rule[-1.5mm]{0mm}{4.7mm}}
\newcommand\firstLineSpace{\rule[0mm]{0mm}{4mm}}
\newcommand\lastLineSpace{\rule[-2.5mm]{0mm}{4mm}}

\newcommand\gravity{G}

\newcommand\ABF{ABF}
\newcommand\ABFIR{ABF-IR}
\newcommand\UBF{UBF}

\newcommand\centersum{\Delta}
\newcommand\mediumint{\resizebox{2.8mm}{!}{$\displaystyle \int$}}
\newcommand\Bsize{b}

\newcommand\UnitSet{1{\mycolon}\Unit}
\newcommand\ObsTimeSet{1{\mycolon}\Time}

\newcommand\spaceTime{\UnitSet\times\ObsTimeSet}
\newcommand\unitTimeSubset{B}
\newcommand\adapted{\gamma}
\newcommand\myeqref[1]{(\ref{#1})}
\newcommand\tif{{}}

\newcommand\altAltTime{m}

\newcounter{wwii}
\newcounter{asii}
\newcounter{girasif}
\newcommand\code[1]{\texttt{#1}}
\usepackage{url}

\newcommand\simulation{\mathrm{sim}}

\usepackage{color}
\usepackage[normalem]{ulem}
\definecolor{orange}{rgb}{1,0.5,0}

\definecolor{green}{rgb}{0,0.5,0}

\newcommand\plusStrut{\rule{0mm}{1.2mm}}

\newcommand\Xspace{{\mathbb X}}
\newcommand\Yspace{{\mathbb Y}}

\newcommand\data[1]{#1^*}

\newcommand\unit{u}
\newcommand\altUnit{\tilde u}
\newcommand\Unit{U}
\newcommand\rep{i}
\newcommand\Rep{\mathcal{I}}

\newcommand\Island{\Rep}

\usepackage[mathscr]{euscript}

\renewcommand\time{n}
\newcommand\myvec[1]{\boldsymbol{#1}}
\newcommand\altTime{\tilde n}
\newcommand\Time{N}
\newcommand\Np{J}
\newcommand\np{j}
\newcommand\altNp{k}

\newcommand\resampleIndex{r}

\newcommand\IF{\mathrm{A}}  
\newcommand\IP{\mathrm{P}}  
\newcommand\GR{\mathrm{IR}}  
\newcommand\GP{\mathrm{IP}}  
\newcommand\resample{\mathrm{R}} 
 
\newcommand\LCP{\mathrm{P}}  

\newcommand\Ninter{S} 
\newcommand\ninter{s}
\newcommand\guideFunc{g}
\newcommand\thetaToV{\stackrel{\rightarrow}{\mathrm{v}}\!}
\newcommand\VtoTheta{\stackrel{\leftarrow}{\mathrm{v}}\!}
\newcommand\npgir{\np}
\newcommand\Npgir{\Np}
\newcommand\Nguide{G}

\newcommand\prob{\mathbb{P}}
\newcommand\E{\mathbb{E}}

\newcommand\given{{\,\vert\,}}
\newcommand\equals{{{\,}={\,}}}

\newcommand\seq[2]{{#1}\!:\!{#2}}
\newcommand\mydot{{\,\cdot\,}}

\newcommand\giventh{{\hspace{0.5mm};\hspace{0.5mm}}}
\newcommand\normal{\mathcal{N}}

\newcommand\loglik{\lambda}

\newcommand\R{\mathbb{R}}

\newcommand\mycolon{{\hspace{0.6mm}:\hspace{0.6mm}}}
\newcommand\MC[1]{#1^{\,\mbox{\tiny MC}}}
\newcommand\EMC{{\E}}
\newcommand\Var{\mathrm{Var}}
\newcommand\var{\Var}
\newcommand\Cov{\mathrm{Cov}} 
\newcommand\cov{\Cov}

\newcommand\dist{d}

\def\loglik{\ell}

\usepackage{amsthm}

\newtheorem{theorem}{Theorem}

\newtheorem{assumption}{Assumption}
\newtheorem{assumptionB}{Assumption}

\newcommand\asp{\hspace{4mm}}
\newcommand\codeIndent{\hspace{4mm}}

\usepackage{amsmath,amssymb}
\usepackage{graphicx}
\usepackage{enumerate}
\usepackage{natbib}
\usepackage{url} 

\newcommand{\blind}{1}

\usepackage{fullpage}
\IfFileExists{upquote.sty}{\usepackage{upquote}}{}
\begin{document}

\def\spacingset#1{\renewcommand{\baselinestretch}%
{#1}\small\normalsize} \spacingset{1}

\date{}

\if1\blind
{
  \title{\vspace{-10mm}\bf \mytitle}
  \author{Edward L. Ionides\thanks{This work was supported by National Science Foundation grants DMS-1761603 and DMS-1646108, and National Institutes of Health grants 1-U54-GM111274 and 1-U01-GM110712.}\hspace{.2cm}\\
    Department of Statistics, University of Michigan\\
    Kidus Asfaw\\
    Department of Statistics, University of Michigan\\
    Joonha Park\\
    Department of Mathematics, University of Kansas\\
    Aaron A. King\\
    Department of Ecology and Evolutionary Biology \& \\ Center for the Study of Complex Systems, University of Michigan
}
  \maketitle
} \fi

\if0\blind
{
  \bigskip
  \bigskip
  \bigskip
  \begin{center}
    {\LARGE\bf \mytitle}
\end{center}
  \medskip
} \fi

\vspace{-10mm}

\begin{abstract}
Bagging (i.e., bootstrap aggregating) involves combining an ensemble of bootstrap estimators.
We consider bagging for inference from noisy or incomplete measurements on a collection of interacting stochastic dynamic systems.
Each system is called a unit, and each unit is associated with a spatial location.
A motivating example arises in epidemiology, where each unit is a city: the majority of transmission occurs within a city, with smaller yet epidemiologically important interactions arising from disease transmission between cities.
Monte~Carlo filtering methods used for inference on nonlinear non-Gaussian systems can suffer from a curse of dimensionality as the number of units increases.
We introduce bagged filter (BF) methodology which combines an ensemble of Monte Carlo filters, using spatiotemporally localized weights to select successful filters at each unit and time.
We obtain conditions under which likelihood evaluation using a BF algorithm can beat a curse of dimensionality, and we demonstrate applicability even when these conditions do not hold.
BF can out-perform an ensemble Kalman filter on a coupled population dynamics model describing infectious disease transmission.
A block particle filter also performs well on this task, though the bagged filter respects smoothness and conservation laws that a block particle filter can violate.

\end{abstract}


\noindent%

\pagebreak

\spacingset{1.45} 

\section{Introduction}

Bagging is a technique to improve numerically unstable estimators by combining an ensemble of replicated bootstrap calculations \citep{breiman96}.
In the context of nonlinear partially observed dynamic systems, the bootstrap filter of \citet{gordon93} has led to a variety of particle filter (PF) methodologies \citep{doucet01,doucet11};
Here, we consider algorithms combining an ensemble of replicated particle filters, which we term {\it bagged filter} algorithms.
Standard PF methods suffer from a curse of dimensionality (COD), defined as an exponential increase in computational requirement as the problem size grows, limiting its applicability to large systems \citep{bengtsson08,snyder15,rebeschini15}.
The COD presents empirically as numerical instability of the Monte Carlo algorithm for affordable numbers of particles.
Much previous research has investigated scalable approaches to filtering and inference with applications to spatiotemporal systems.
Our bagged filters are in the class of {\it plug-and-play} algorithms, meaning that they require as input a simulator for the latent dynamic process but not an evaluator of transition probabilities \citep{breto09,he10}.
This simulation-based approach, also known as {\it likelihood-free} \citep{brehmer20} or {\it equation-free} \citep{kevrekidis09}, facilitates application to a wide class of models.
The ensemble Kalman filter \citep{evensen96,lei10,katzfuss19} is a widely used plug-and-play method which uses simulations to construct a nonlinear filter that is exact for a linear Gaussian model.
Another plug-and-play approach to combat the COD is the block particle filter \citep{rebeschini15,ng02}.
Both ensemble Kalman filter and block particle filter methods construct trajectories that can violate smoothness and conservation properties of the dynamic model.
By contrast, our bagged filters are built using valid trajectories of the dynamic model, making localization approximations only when comparing these trajectories to data.

The replicated stochastic trajectories in a bagged filter form an ensemble of representations of the dynamic system.
Unlike the particles in a particle filter or ensemble Kalman filter, the bagged replicates are independent in a Monte Carlo sense.
Bagged filters therefore bear some resemblance to \textit{poor man's ensemble} forecasting methodology in which a collection of independently constructed forecasts is generated using different models and methods \citep{ebert01}. 
Poor man's ensembles have sometimes been found to have greater forecasting skill than any one forecast \citep{leutbecher08,palmer02,chandler13}.
One explanation for this phenomenon is that even a hypothetically perfect model cannot provide effective filtering using methodology afflicted by the COD.  
We show that bagged filter methodology can relieve this limitation.
From this perspective, the independence of the forecasts in the poor man's ensemble, rather than the diversity of model structures, may be the key to its success.

We first consider a simple bagged filter where each replicate is an independent simulation of the latent process model.
We call this the unadapted bagged filter ({\UBF}) since the replicates in the ensemble depend on the model but not on the data.
{\UBF} is described in Sec.~\ref{sec:ubf}, with a theoretical analysis presented in Sec.~\ref{sec:ubf:theory}.
Each {\UBF} replicate corresponds to a basic PF algorithm with a single particle.
We show that {\UBF} formally beats the COD under a weak mixing assumption, though {\UBF} can have poor numerical behavior if a very large number of replicates are needed to reach this happy asymptotic limit.
Subsequent empirical results show that {\UBF} may nevertheless be a useful algorithm in some situations.
In Sec.~\ref{sec:abf}, we generalize {\UBF} to construct an adapted bagged filter (\ABF) where each replicate tracks the data.
The price of adaptation is that {\ABF} no longer fully avoids the COD, a limitation that can be controlled in certain situations by supplementing {\ABF} with a technique called intermediate resampling, to obtain the {\ABFIR} algorithm.
Theoretical results for {\ABF} and {\ABFIR} algorithms are developed in Sec.~\ref{sec:abf:theory}.
The algorithms are demonstrated in action and compared with alternative approaches in Sec.~\ref{sec:examples}.

\section{The unadapted bagged filter (UBF)}
\label{sec:ubf}

Suppose the collection of units is indexed by the set $\{1,2,\dots,\Unit\}$, which is written as $1\mycolon\Unit$.
The latent Markov process is denoted by $\{\myvec{X}_{\time},\time\in 0\mycolon\Time\}$, with $\myvec{X}_{\time}=X_{1:\Unit,\time}$ taking values in a product space $\Xspace^\Unit$.
This discrete time process may arise from a continuous time Markov process $\{\myvec{X}(t), t_0\le t\le t_{\Time} \}$ observed at times $t_{1:\Time}$, and in this case we set $\myvec{X}_\time=\myvec{X}(t_\time)$.
The initial value $\myvec{X}_{0}$ may be stochastic or deterministic.
Observations are made on each unit, modeled by an observable process $\{\myvec{Y}_{\time}=Y_{1:\Unit,\time},\time\in 1\mycolon\Time\}$ which takes values in a product space $\Yspace^\Unit$.
Observations are modeled as being conditionally independent given the latent process.
The conditional independence of measurements applies over both time and the unit structure, so the collection $\big\{Y_{\unit\comma\time},\unit\in\seq{1}{\Unit},\time\in\seq{1}{\Time}\big\}$ is conditionally independent given
$\big\{X_{\unit\comma\time},\unit\in\seq{1}{\Unit},\time\in\seq{1}{\Time}\big\}$.
The unit structure for the observation process is not necessary for all that follows (see Sec.~\SuppSecGeneralization).
We suppose the existence of a joint density $f_{\myvec{X}_{0:\Time},\myvec{Y}_{1:\Time}}$  of $X_{1:\Unit,1:\Time}$ and $Y_{1:\Unit,1:\Time}$ with respect to some appropriate measure, following a notational convention that the subscripts of $f$ denote the joint or conditional density under consideration.
The data are $\data{y}_{\unit\comma\time}$ for unit $\unit$ at time $\time$.
This model is a special case of a  partially observed Markov process  \citep[POMP,][]{breto09}, also known as a state space model or hidden Markov model.
The additional unit structure, not generally required for a POMP, is appropriate for modeling interactions between units characterized by a spatial location, and so we call the model a SpatPOMP.
In the following, we use a lexicographical ordering on the set of observations;
Specifically, we define the set of observations preceding unit $\unit$ at time $\time$ as
\begin{equation}\label{eq:setA}
A_{\unit\comma\time}=\big\{(\tilde\unit,\tilde\time): 1 \leq \tilde\time<\time \mbox{ or } (\tilde\time=\time \mbox{ and } \tilde\unit <\unit)\big\}.
\end{equation}
The ordering of the spatial locations in \myeqref{eq:setA} might seem artificial, and indeed densities such as
$f_{X_{\unit\comma\time}|X_{A_{\unit\comma\time}}}$ will frequently be hard to compute or simulate from.
The bagged filter algorithms we study do not evaluate or simulate such transition densities but only compute the measurement model on neighborhoods, unlike the filter of \citet{beskos17} built on a similar factorization. 
If sufficiently distant units are approximately independent, we say the system is {\it weakly coupled}.
In this case, we suppose there is a neighborhood $B_{\unit\comma\time}\subset A_{\unit\comma\time}$ such that the latent process on 
$A_{\unit\comma\time} \setminus B_{\unit\comma\time}$ 
is approximately conditionally independent of $X_{\unit\comma\time}$ given data on $B_{\unit\comma\time}$. 

Our primary interest is estimation of the log likelihood for the data given the model, $\loglik=\log f_{\myvec{Y}_{1:\Time}}(\data{\myvec{y}}_{1:\Time})$, which is of fundamental importance in both Bayesian and non-Bayesian statistical inference.
A general filtering problem is to evaluate $\E\big[h(X_{\unit,\time}) \given Y_{A_{\unit,\time}}\equals \data{y}_{A_{\unit,\time}}\big]$ for some function $h:\Xspace\to\R$.
Taking $h(x)= f_{Y_{\unit,\time}| X_{\unit,\time}}\big(\data{y}_{\unit,\time}\given x\big)$ gives a filtering representation of the likelihood evaluation problem.
Further discussion on bagged filtering for other filtering problems is given in Sec.~{\SuppSecLatentStates}.
For likelihood-based inference, maximization plays an important role in point estimation, confidence interval construction, hypothesis testing and model selection. 
An extension of bagged filtering to likelihood maximization is demonstrated in Sec.~\ref{sec:profile} following the approach described in Sec.~{\SuppSecParameterEst}.

Pseudocode for a {\UBF} algorithm for likelihood evaluation is given below.
The prediction weight  $w^P_{u,n,i}$ gives an appropriate weighting for replicate $i$ for predicting $\data{y}_{u,n}$ based on the most relevant data, $\data{y}_{B_{u,n}}$.
Conditional log likelihoods are estimated using an approximation
\begin{eqnarray*}
    \ell_{u,n} &=& \log f_{Y_{u,n}|Y_{A_{u,n}}}\big(\data{y}_{u,n}\given \data{y}_{A_{u,n}}\big)
    = \log \left( \int f_{Y_{u,n}|X_{u,n}}(\data{y}_{u,n}\given x) \, f_{X_{u,n}|Y_{A_{u,n}}}(x \given \data{y}_{A_{u,n}}) \, dx \right)
\\
  &\approx& \log \left( \int f_{Y_{u,n}|X_{u,n}}(\data{y}_{u,n}\given x) \, f_{X_{u,n}|Y_{B_{u,n}}}(x\given \data{y}_{B_{u,n}}) \, dx \right).
\end{eqnarray*}
The choice of $B_{u,n}$ is determined empirically, with a bias-variance trade-off used to compare small neighborhoods such as $B_{u,n} = \{ (u,n-1),(u-1,n)\}$ or $B_{u,n} = \{ (u,n-1),(u,n-2)\}$ against larger neighborhoods.
The plug-and-play property is evident because {\UBF} requires as input a simulator for the latent coupled dynamic process but not an evaluator of transition probabilities.
The pseudocode for {\UBF} adopts a convention that implicit loops are carried out over all free indices, meaning indices with values that are not explicitly specified.
For example, the construction of $w^P_{\unit,\time,\rep}$ in {\UBF} has an implicit loop over $\unit$, $\time$ and $\rep$.
However, the summation constructing $\MC{\loglik}_{\unit,\time}$ does not have an implicit loop over $\rep$ since the summation index $\rep$ is specified explicitly and so is not a free index.


\begin{center}
\noindent\begin{tabular}{l}
\hline
\inputSpace {\bf {\UBF}. Unadapted bagged filter.}\\
\hline
\inputSpace {\bf input:}
\\ \codeIndent
simulator for $f_{\myvec{X}_0}(\myvec{x}_0)$ and $f_{\myvec{X}_{\time}|\myvec{X}_{\time-1}}(\myvec{x}_{\time}\given \myvec{x}_{\time-1})$
\\ \codeIndent
evaluator for $f_{{Y}_{\unit\comma\time}|{X}_{\unit\comma\time}}({y}_{\unit\comma\time}\given {x}_{\unit\comma\time})$
\\ \codeIndent
number of replicates, ${\Rep}$
\\ \codeIndent
neighborhood structure, $B_{\unit\comma\time}$
\\ \codeIndent
data, $\data{\myvec{y}}_{\time}$
\\
\inputSpace {\bf implicit loops:}
\\ \codeIndent
$\unit \mbox{ in } \seq{1}{\Unit}$, 
$\; \time \mbox{ in } \seq{1}{\Time}$, 
$\; \rep \mbox{ in } \seq{1}{\Rep}$
\\
\inputSpace {\bf algorithm:}
\\ \codeIndent
simulate $\myvec{X}^{\tif}_{0:\Time,\rep}\sim f_{\myvec{X}_{0:\Time}}(\myvec{x}_{0:\Time})$
\\ \codeIndent 
measurement weights,
$w^M_{\unit,\time,\rep}=f_{Y_{\unit,\time}|X_{\unit,\time}}(\data{y}_{\unit,\time}\given X^{\tif}_{\unit,\time,\rep})$
\\ \codeIndent
prediction weights, 
$w^P_{\unit,\time,\rep}=\prod_{(\tilde \unit,\tilde n)\in B_{\unit,\time}}
w^M_{\tilde\unit,\tilde n,\rep}$
\\ \codeIndent
$\MC{\loglik}_{\unit,\time}= 
\log\left(
  \sum_{\rep=1}^\Rep w^M_{\unit,\time,\rep}w^P_{\unit,\time,\rep}
\right)
-\log\left(
  \sum_{\rep=1}^\Rep w^P_{\unit,\time,\rep}
\right)
$
\\
\inputSpace {\bf output:}
\\ \codeIndent
log likelihood estimate, $\MC{\loglik}= \sum_{\time=1}^\Time\sum_{\unit=1}^\Unit \MC{\loglik}_{\unit\comma\time}$\\
\lastLineSpace
\\
\hline
\end{tabular}
\end{center}

\subsection{{\UBF} theory}
\label{sec:ubf:theory}

A dataset $\data{\myvec{y}}_{1:\Time}$ with $\Unit$ units is modeled via a joint density  $f_{\myvec{X}_{0:\Time},\myvec{Y}_{1:\Time}}$.
We consider non-asymptotic bounds that apply for all values of $\Unit$ and $\Time$. 
To impose a requirement that distant regions of space-time behave similarly and have only weak dependence, we assert the following conditions which define constants $\eone$, $\etwo$ and $Q$ used to bound the bias and variance in Theorem~\ref{thm:tif}.
Stronger bounds are obtained when the conditions hold for small $\eone$, $\etwo$ and $Q$.

\Ai  

\Aii  

\Aiii   

\Aiv    

The two mixing conditions in Assumptions~\ref{A1} and~\ref{A:unconditional:mix} are subtly different. 
Assumption~\ref{A1} describes a conditional mixing property dependent on the data, whereas~\ref{A:unconditional:mix} asserts a form of unconditional mixing.
Although both capture a similar concept of weak coupling, conditional and unconditional mixing properties do not readily imply one another.
Assumption~\ref{A2} is a compactness condition of a type that has proved useful in the theory of particle filters despite the rarity of its holding exactly.
Theorem~\ref{thm:tif} shows that these conditions let {\UBF} compute the likelihood with a Monte Carlo variance of order $\Unit\Time\Rep^{-1}$ with a bias of order $\Unit\Time\epsilon$.

\TheoremI 

\begin{proof}
A complete proof is given in Sec.~\SuppSecThmI. 
Briefly, the assumptions imply a multivariate central limit theorem for $\{\MC{\loglik}_{\unit\comma\time}, (\unit,\time)\in \UnitSet{\times}\ObsTimeSet\}$ as $\Rep\to\infty$.
The limiting variances and covariances are uniformly bounded, using Assumptions~\ref{A1b} and~\ref{A2}. 
Assumption~\ref{A1} provides a uniform bound on the discrepancy between ${\loglik}_{\unit\comma\time}$ and mean of the Gaussian limit.
This is enough to derive \myeqref{th1:lik:bound}.
Assumption~\ref{A:unconditional:mix} gives a stronger bound on covariances between sufficiently distant units, leading to \myeqref{th1:lik:bound2}.
\end{proof}

Theorem~\ref{thm:tif} does not guarantee uniformity over $\Unit$ and $\Time$ of the rate of convergence as $\Rep\to\infty$.
However, it does guarantee that the polynomial bounds in (\ref{th1:lik:bound}) and (\ref{th1:lik:bound2}) hold for sufficiently large $\Rep$.
The COD is characterized by exponential bounds, and so Theorem~\ref{thm:tif} shows a specific sense in which {\UBF} can avoid COD.
Uniformity of the central limit convergence in Theorem~\ref{thm:tif} may be expected to hold via a Berry-Esseen theorem, but extension of existing Berry-Esseen results for dependent processes \citep{bentkus97,jirak16} is beyond the scope of this article.

The approximation error for {\UBF} can be divided into two sources: a localization bias due to conditioning on a finite neighborhood, and Monte Carlo error. 
The localization bias does not disappear in the limit as Monte Carlo effort increases.
It does become small as the conditioning neighborhood increases, but the Monte Carlo effort grows exponentially in the size of this neighborhood.
Although the filtering inference is carried out using localization, the simulation of the process is carried out globally which avoids the introduction of additional boundary effects and ensures that the simulations comply with any constraints satisfied by the model for the latent process.

\section{Adaptation and intermediate resampling}
\label{sec:abf}

Theorem~\ref{thm:tif} shows that {\UBF} can beat COD.
However, {\UBF} can perform poorly on long time series unless weak temporal dependence allows simulated sample paths to remain relevant over the course of a long time series. 
For example, we will find that {\UBF} performs well on an epidemiological model (Sec.~\ref{sec:discussion}) but less well on a geophysical model (Sec.~{\SuppSecLorenz}).
It is sometimes necessary to select simulations consistent with the data, much as standard PF algorithms do.
We look for approaches that build on the basic insight of {\UBF} while having superior practical performance.

Whereas the full global filtering problem of drawing from $f_{\myvec{X}_{\time}|\myvec{Y}_{1:\time}}$ may be intractable via importance sampling methods, a version of this problem localized in space and time may nevertheless be feasible.
The conditional density, $f_{\myvec{X}_{\time}|\myvec{Y}_{\time},\myvec{X}_{\time-1}}$, is called the {\it adapted density}, and simulating from this density is called {\it adapted simulation}.
For models where $\myvec{X}_{\time-1}$ is highly informative about $\myvec{X}_{\time}$, importance sampling for adapted simulation may be much easier than the full filter calculation.
The following adapted bagged filter ({\ABF}) is constructed under a hypothesis that the adapted simulation problem is tractable, and it is applicable when the number of units is prohibitive for Monte Carlo sampling from the full filter distribution but not for sampling from the adapted distribution.
In {\ABF}, the adapted simulations are reweighted in a neighborhood of each unit and time point to construct a local approximation to the filtering problem which leads to an estimate of the likelihood.
The pseudocode for {\ABF}, below, reduces to {\UBF} when using a single particle per repicate, $\Np=1$.


\begin{center}
\noindent\begin{tabular}{l}
\hline
{\bf 
{\ABF}. Adapted bagged filter.}\inputSpace\\
\hline
\inputSpace {\bf input:} same as for {\UBF} plus
\\ \codeIndent
particles per replicate,  $\Np$
\\
\inputSpace {\bf implicit loops:}
\\ \codeIndent
$\unit \mbox{ in } \seq{1}{\Unit}$, 
$\; \time \mbox{ in } \seq{1}{\Time}$, 
$\; \rep \mbox{ in } \seq{1}{\Rep}$,
$\; \np \mbox{ in } \seq{1}{\Np}$
\\
\inputSpace {\bf algorithm:}
\\ \codeIndent
\firstLineSpace
Initialize adapted simulation: $\myvec{X}^{\IF}_{0,\rep} \sim f_{\myvec{X}_0}(\myvec{x}_0)$
\\ \codeIndent
For $\time\ \mathrm{in}\ \seq{1}{\Time}$
\\ \codeIndent
\asp  Proposals:
    $\myvec{X}_{\time,\rep,\np}^{\IP} \sim 
    f_{\myvec{X}_{\time}|X_{1:\Unit,\time-1}} 
    \big( \myvec{x}_{\time}\given \myvec{X}^{\IF}_{\time-1,\rep}\big)$
\\ \codeIndent
\asp Measurement weights:
  $w^M_{\unit,\time,\rep,\np} = 
    f_{Y_{\unit,\time}|X_{\unit\comma\time}} 
    \big (\data{y}_{\unit\comma\time}\given X^{\IP}_{\unit\comma\time,\rep,\np}\big)$
\\ \codeIndent
\asp  Adapted resampling weights:
  $w^{\IF}_{\time,\rep,\np} = 
    \prod_{\unit=1}^{\Unit} w^M_{\unit,\time,\rep,\np}$
\\ \codeIndent
\asp
      Resampling:
        $\prob\big[\resampleIndex({\rep})=a \big] = w^{\IF}_{\time,\rep,a}
  \Big( 
  \sum_{\altNp=1}^{\Np} w^{\IF}_{\time,\rep,\altNp}
  \Big)^{-1}$
\\ \codeIndent
\asp 
$\myvec{X}^{\IF}_{\time,\rep} = \myvec{X}^{\IP}_{\time,\rep,r(\rep)}$ 
\\ \codeIndent
\asp 
  $w^{\LCP}_{\unit,\time,\rep,\np}= \displaystyle
  \prod_{\altTime=1}^{\time-1}
  \Big[
    \frac{1}{\Np}\sum_{k=1}^{\Np}
    \hspace{1mm}
       \prod_{\altUnit:(\altUnit,\altTime)\in B_{\unit,\time}} 
    \hspace{-1mm}
        w^M_{\altUnit,\altTime,\rep,k}
  \Big] \prod_{\altUnit:(\altUnit,\time)\in B_{\unit,\time}} 
    \hspace{-1mm}
        w^M_{\altUnit,\time,\rep,\np}$
\\ \codeIndent
End for
\\ 
\inputSpace {\bf output:}
\\ \codeIndent
$\displaystyle \MC{\loglik}_{\unit,\time}= 
\log\Bigg(
\frac{
\sum_{\rep=1}^\Rep \sum_{\np=1}^{\Np} w^M_{\unit,\time,\rep,\np}w^P_{\unit,\time,\rep,\np}
}{
\sum_{\rep=1}^\Rep \sum_{\np=1}^{\Np} w^P_{\unit,\time,\rep,\np}
}
\Bigg)
$
\vspace{1mm}
\\
\hline
\end{tabular}
\end{center}

{\ABF} remedies a weakness of {\UBF} by making each boostrap filter adapted to the data.
However, this benefit carries a cost, since adapted simulation is not immune from the curse of dimensionality.
Therefore, we also consider an algorithm called {\ABFIR} which uses an intermediate resampling technique to carry out the adapted simulation.
Intermediate resampling involves assessing the satisfactory progress of particles toward the subsequent observation at a collection of times between observations.
This is well defined when the latent process has a continuous time representation, $\{\myvec{X}(t)\}$, with observation times $t_{1:\Time}$.
We write $\Ninter$ intermediate resampling times as 
\begin{equation}
\nonumber
t_{\time-1}=t_{\time,0}<t_{\time,1}<\dots<t_{\time,\Ninter}=t_{\time}.
\end{equation}
Carrying out an intermediate resampling procedure can have favorable scaling properties when $\Ninter$ is proportional to $\Unit$ \citep{park20}.
In the case $\Ninter=1$, {\ABFIR} reduces to {\ABF}.
Intermediate resampling was developed in the context of sequential Monte Carlo \citep{delmoral15,park20}; however, the same theory and methodology can be applied to the simpler and easier problem of adapted simulation. 
{\ABFIR} employs a guide function to gauge the compatibility of each particle with future data.
This is a generalization of the popular auxiliary particle filter \citep{pitt99}.
Only an ideal guide function fully addresses COD \citep{park20} and on nontrivial problems this is not available.
However, practical guide functions can nevertheless improve performance.

The implementation in the {\ABFIR} pseudocode constructs the guide $g_{\time,\ninter,\rep,\np}$ using a simulated moment method proposed by \citet{park20}.
The quantities  $\myvec{X}_{\time,\rep,\npgir}^{G}$, $V_{\unit,\time,\rep}$, $\myvec{\mu}^{\GP}_{\time,\ninter,\rep,\np}$, $V^{\mathrm{meas}}_{\unit,\time,\ninter,\rep,\np}$, $V^{\mathrm{proc}}_{\unit,\time,\ninter,\rep}$ and $\theta_{\unit,\time,\ninter,\rep,\np}$ constructed in {\ABFIR} are used only to construct $g_{\time,\ninter,\rep,\np}$.
Heuristically, we use guide simulations to approximate the variance of the increment in each particle between time points, and we augment the measurement variance to account for both dynamic variability and measurement error.
The guide function affects numerical performance of the algorithm but not its correctness: it enables a computationally convenient approximation to improve performance on the intractable target problem.
Our guide function supposes the availability of a deterministic function approximating evolution of the mean of the latent process, written as
\begin{equation}
\nonumber
\myvec{\mu}(\myvec{x},s,t)\approx \E\big[\myvec{X}(t) \given \myvec{X}(s)=\myvec{x}\big]. 
\end{equation}
Further, the guide requires that the measurement model has known conditional mean and variance as a function of the model parameter vector $\theta$, written as
\begin{eqnarray}
\nonumber
h_{\unit\comma\time}(x_{\unit\comma\time})
  &=& \E\big[Y_{\unit\comma\time}\given X_{\unit\comma\time}=x_{\unit\comma\time}\big]
\\
\nonumber
{\thetaToV}_{\unit\comma\time}(x_{\unit\comma\time},\theta)
  &=& \var\big(Y_{\unit\comma\time}\given X_{\unit\comma\time}=x_{\unit\comma\time} \giventh \theta\big)
\end{eqnarray}
Also required for {\ABFIR} is an inverse function ${\VtoTheta}_{\unit\comma\time}$ such that
\begin{equation}
\nonumber
{\thetaToV}_{\unit\comma\time} \big( x_{\unit\comma\time},\VtoTheta_{\unit\comma\time}(V,x_{\unit\comma\time},\theta) \big) = V.
\end{equation}


\newcommand\fitVspace{\vspace{-1mm}}
\begin{center} 
\noindent\begin{tabular}{l}
\hline 
{\bf {\ABFIR}. Adapted bagged filter with intermediate resampling.} 
\inputSpace\\
\hline
\inputSpace {\bf input:} same as for {\ABF} plus \firstLineSpace \fitVspace
\\ \codeIndent \fitVspace
number of intermediate timesteps, $\Ninter$
\\ \codeIndent  \fitVspace
measurement variance parameterizations, ${\VtoTheta}_{\unit\comma\time}$ and ${\thetaToV}_{\unit\comma\time}$
\\ \codeIndent  \fitVspace
approximate process and observation mean functions, $\myvec{\mu}$ and $h_{\unit\comma\time}$
\\  \fitVspace
\inputSpace {\bf implicit loops:}
\\ \codeIndent  \fitVspace
$\unit \mbox{ in } \seq{1}{\Unit}$, 
$\; \time \mbox{ in } \seq{1}{\Time}$, 
$\; \rep \mbox{ in } \seq{1}{\Rep}$,
$\; \np \mbox{ in } \seq{1}{\Np}$,
$\; \npgir \mbox{ in } \seq{1}{\Npgir}$
\\  \fitVspace
\inputSpace {\bf algorithm:}
\\ \codeIndent  \fitVspace
\firstLineSpace
Initialize adapted simulation: $\myvec{X}^{\IF}_{0,\rep} \sim f_{\myvec{X}_0}(\myvec{x}_0)$
\\ \codeIndent  \fitVspace
For $\time\ \mathrm{in}\ \seq{1}{\Time}$
\\ \codeIndent  \fitVspace
\asp Guide simulations:
    $\myvec{X}_{\time,\rep,\npgir}^{G} \sim 
    f_{\myvec{X}_{\time}|\myvec{X}_{\time-1}} 
    \big( \myvec{x}_{\time}\given \myvec{X}^{\IF}_{\time-1,\rep} \big)$
\\ \codeIndent  \fitVspace
\asp Guide sample variance: $V_{\unit,\time,\rep}=
      \var \big\{
        h_{\unit\comma\time}\big( {X}_{\unit,\time,\rep,\npgir}^{G}\big), \npgir \mbox{ in } \seq{1}{\Npgir}
      \big\}$ 
\\ \codeIndent  \fitVspace
\asp $\guideFunc^{\resample}_{\time,0,\rep,\np}=1 \; \; $ and
$\; \myvec{X}_{\time,0,\rep,\np}^{\GR}=\myvec{X}^{\IF}_{\time-1,\rep}$
\\ \codeIndent  \fitVspace
\asp For $\ninter  \,\, \mathrm{in} \,\, \seq{1}{\Ninter}$
\\ \codeIndent  \fitVspace
\asp\asp Intermediate proposals:
        ${\myvec{X}}_{\time,\ninter,\rep,\np}^{\GP}
          \sim {f}_{{\myvec{X}}_{\time,\ninter}|{\myvec{X}}_{\time,\ninter-1}}
          \big(\mydot|{\myvec{X}}_{\time,\ninter-1,\rep,\np}^{\GR}\big)$ 
\\ \codeIndent  \fitVspace
\asp\asp 
        $\myvec{\mu}^{\GP}_{\time,\ninter,\rep,\np} 
           = \myvec{\mu}\big( \myvec{X}^{\GP}_{\time,\ninter,\rep,\np},t_{\time,\ninter},t_{\time} \big)$
\\ \codeIndent  \fitVspace
\asp\asp      
        $V^{\mathrm{meas}}_{\unit,\time,\ninter,\rep,\np}
           = \thetaToV_{\unit}(\theta,\mu^{\GP}_{\unit,\time,\ninter,\rep,\np})$
\\ \codeIndent  \fitVspace
\asp\asp  
        $V^{\mathrm{proc}}_{\unit,\time,\ninter,\rep}
          = V_{\unit,\time,\rep} \,
          \big(t_{\time}-t_{\time,\ninter}\big) \Big/
          \big(t_{\time}-t_{\time,0}\big)$ 
\\ \codeIndent  \fitVspace
\asp\asp
        $\theta_{\unit,\time,\ninter,\rep,\np}= 
          \VtoTheta_{\unit}\big(
            V^{\mathrm{meas}}_{\unit,\time,\ninter,\rep,\np} + V^{\mathrm{proc}}_{\unit,\time,\ninter,\rep}, 
            \, \mu^{\GP}_{\unit,\time,\ninter,\rep,\np}
          \big)$
\\ \codeIndent  \fitVspace
\asp\asp  
        $
\guideFunc_{\time,\ninter,\rep,\np}=
          \prod_{\unit=1}^{\Unit}
          f_{Y_{\unit,\time}|X_{\unit,\time}}
          \big(
            \data{y}_{\unit,\time}\given \mu^{\GP}_{\unit,\time,\ninter,\rep,\np} \giventh \theta_{\unit,\time,\ninter,\rep,\np} 
          \big)$
\\ \codeIndent  \fitVspace
\asp\asp Guide weights:
$w^G_{\time,\ninter,\rep,\np}= \guideFunc^{}_{\time,\ninter,\rep,\np}
         \big/ \guideFunc^{\resample}_{\time,\ninter-1,\rep,\np}$
\\ \codeIndent  \fitVspace
\asp\asp
      Resampling:
        $\prob\big[\resampleIndex({\rep,\np})=a \big] = w^G_{\time,\ninter,\rep,a}
\Big( \sum_{\altNp=1}^{\Np}w^G_{\time,\ninter,\rep,\altNp}\Big)^{-1}$
\\ \codeIndent  \fitVspace
\asp\asp
        $\myvec{X}_{\time,\ninter,\rep,\np}^{\GR}=\myvec{X}_{\time,\ninter,\rep,\resampleIndex({\rep,\np})}^{\GP}\; \; $ and
        $\; \guideFunc^{\resample}_{\time,\ninter,\rep,\np}= \guideFunc^{}_{\time,\ninter,\rep,\resampleIndex({\rep,\np})}\,$
\\ \codeIndent  \fitVspace
\asp
End For
\\ \codeIndent  \fitVspace
\asp
  Set $\myvec{X}^{\IF}_{\time,\rep}=\myvec{X}^{\GR}_{\time,\Ninter,\rep,1}$ 
\\ \codeIndent  \fitVspace
\asp Measurement weights:
  $w^M_{\unit,\time,\rep,\npgir} = 
    f_{Y_{\unit,\time}|X_{\unit,\time}} 
    \big (\data{y}_{\unit,\time}\given X^{G}_{\unit,\time,\rep,\npgir} \big)$
\\ \codeIndent  \fitVspace
\asp 
  $w^{\LCP}_{\unit,\time,\rep,\npgir}= \displaystyle
  \prod_{\altTime=1}^{\time-1}
  \Big[
    \frac{1}{\Npgir}\sum_{a=1}^{\Npgir}
    \hspace{1mm}
       \prod_{\altUnit:(\altUnit,\altTime)\in B_{\unit,\time}} 
    \hspace{-1mm}
        w^M_{\altUnit,\altTime,\rep,a}
  \Big] \prod_{\altUnit:(\altUnit,\time)\in B_{\unit,\time}} 
    \hspace{-1mm}
        w^M_{\altUnit,\time,\rep,\npgir}$
\\ \codeIndent  \fitVspace
End for
\\  \fitVspace
\inputSpace {\bf output:}
\\ \codeIndent  \fitVspace
$\displaystyle \MC{\loglik}_{\unit,\time}= 
\log\Bigg(
\frac{
\sum_{\rep=1}^\Rep \sum_{\npgir=1}^{\Npgir} w^M_{\unit,\time,\rep,\npgir}w^P_{\unit,\time,\rep,\npgir}
}{
\sum_{\rep=1}^\Rep \sum_{\npgir=1}^{\Npgir} w^P_{\unit,\time,\rep,\npgir}
}
\Bigg)
$
\vspace{2mm}
\\
\hline
\end{tabular}
\end{center}

\clearpage

This guide function is applicable to spatiotemporal versions of a broad range of population and compartment models used to model dynamic systems in ecology, epidemiology, and elsewhere. 
Other guide functions could be developed and inserted into the {\ABFIR} algorithm, including other constructions considered by \citet{park20}.

One might wonder why it is appropriate to keep many particle representations at intermediate timesteps while resampling down to a single representative at each observation time.
An answer is that adaptive simulation can fail to track the observation sequence when one resamples down to a single particle too often (Sec.~\SuppSecAdaptedSimulation).



\subsection{{\ABFIR} theory}
\label{sec:abf:theory}

We  start by considering a deterministic limit for infinite Monte Carlo effort and explaining why the {\ABF} and {\ABFIR} algorithms approximately target the likelihood function, subject to suitable mixing behavior.
Subsequently, we consider the scaling properties as Monte Carlo effort increases.
We adopt a convention that densities involving $Y_{\unit\comma\time}$ are implicitly evaluated at the data, $\data{y}_{\unit\comma\time}$, and densities involving $X_{\unit\comma\time}$ are implicitly evaluated at $x_{\unit\comma\time}$ unless otherwise specified.
We write $A^{+}_{\unit\comma\time}=A_{\unit\comma\time}\cup (\unit,\time)$, matching the defintion $B^{+}_{\unit\comma\time}=B_{\unit\comma\time}\cup (\unit,\time)$.
The essential ingredient in all the algorithms is a localization of the likelihood, which may be factorized sequentially as
\begin{equation}
\nonumber
f_{Y_{1:\Unit,1:\Time}}
=
\prod_{\time=1}^\Time\prod_{\unit=1}^\Unit 
f_{Y_{\unit\comma\time}|Y_{A_{\unit\comma\time}}}
= \prod_{\time=1}^\Time\prod_{\unit=1}^\Unit 
\frac{f_{Y^{}_{A^+_{\unit\comma\time}}}}{f_{Y^{}_{A^{}_{\unit\comma\time}}}}.
\end{equation}
In particular, the approximations assume that the full history $A_{\unit\comma\time}$ can be well approximated by a neighborhood $B_{\unit\comma\time}\subset A_{\unit\comma\time}$.
{\UBF} approximates $f_{Y_{\unit\comma\time}|Y_{A_{\unit\comma\time}}}$ by
\begin{equation}
\nonumber
f_{Y_{\unit\comma\time}|Y_{B_{\unit\comma\time}}} 
=
\frac{f^{}_{Y^{}_{B^+_{\unit\comma\time}}}}{f^{}_{Y^{}_{B^{}_{\unit\comma\time}}}}
=
\frac{\int f_{Y_{B^+_{\unit\comma\time}}|X_{B^+_{\unit\comma\time}}} f_{X_{B^+_{\unit\comma\time}}}\, dx_{B^+_{\unit\comma\time}} }
{\int f_{Y_{B_{\unit\comma\time}}|X_{B_{\unit\comma\time}}} f_{X_{B_{\unit\comma\time}}}\, dx_{B_{\unit\comma\time}} }.
\end{equation}
For $\unitTimeSubset \subset \seq{1}{\Unit}\times\seq{1}{\Time}$, define $\unitTimeSubset^{[m]}=\unitTimeSubset \cap \big(\UnitSet\times \{m\}\big)$.
{\ABF} and {\ABFIR} build on the following identity, 
\begin{equation}
\nonumber
\hspace{-1mm}
f_{Y_{A_{\unit\comma\time}}}{=} \int 
\! \!
f_{\myvec{X}_0} 
\! \!
\left[
\prod_{\altAltTime=1}^{\time}
f_{\myvec{X}_{\altAltTime}|\myvec{X}_{\altAltTime-1},\myvec{Y_{\altAltTime}}}
f_{Y_{A^{[\altAltTime]}_{\unit\comma\time}}|\myvec{X}_{\altAltTime-1}}
\right]
\! d\myvec{x}_{0:\time},
\end{equation}
where
$f_{\myvec{X}_{\altAltTime}|\myvec{X}_{\altAltTime-1},\myvec{Y}_{\altAltTime}}$ is called the adapted transition density.
The adapted process (i.e., a stochastic process following the adapted transition density) can be interpreted as a one-step greedy procedure using the data to guide the latent process.
Let 
$g^{}_{\myvec{X}_{0:\Time},\myvec{X}^P_{1:\Time}}(\myvec{x}_{0:\Time},\myvec{x}^P_{1:\Time})$ 
be the joint density of the adapted process and the proposal process, 
\begin{eqnarray}
\nonumber
\hspace{-3mm} g^{}_{\myvec{X}_{0:\Time},\myvec{X}^P_{1:\Time}}(\myvec{x}_{0:\Time},\myvec{x}^P_{1:\Time})
&=& f_{\myvec{X}_0}(\myvec{x}_0) \times 
\\
\label{eq:g}
&& \hspace{-42mm}
\prod_{\time=1}^{\Time} 
f_{\myvec{X}_{\time}|\myvec{X}_{\time-1},\myvec{Y}_{\time}} 
  \big( 
    \myvec{x}_{\time} \given \myvec{x}_{\time-1},\data{\myvec{y}}_{\time} 
  \big)
\,\,
f_{\myvec{X}_{\time}|\myvec{X}_{\time-1}}
  \big(
    \myvec{x}^P_{\time} \given \myvec{x}_{\time-1}
  \big).
\end{eqnarray}
Using the convention that an empty density $f_{Y_{\emptyset}}$ evaluates to 1, we define
\begin{equation}
\nonumber
\adapted^{}_{\unitTimeSubset}
= 
\prod_{m=1}^{\Time} f_{Y_{\unitTimeSubset^{[m]}}|\myvec{X}^{}_{m-1}} \big( \data{y}_{\unitTimeSubset^{[m]}} \given \myvec{X}^{}_{m-1} \big).
\end{equation}
Denoting $\E_{g}$ for expectation for $(\myvec{X}_{0:\Time},\myvec{X}^P_{1:\Time})$ having density $g_{\myvec{X}_{0:\Time},\myvec{X}^P_{1:\Time}}$, we have $ f_{Y_{A_{u,n}}} = \E_g \big[ \adapted^{}_{A^{}_{\unit\comma\time}}\big]$ and thus
\begin{equation}
\nonumber
f_{Y_{\unit\comma\time}|Y_{A_{\unit\comma\time}}}
= \frac{\E_{g}\big[\adapted^{}_{A^+_{\unit\comma\time}}\big] }{\E_{g}\big[\adapted^{}_{A^{}_{\unit\comma\time}}\big] }.
\end{equation}
Estimating this ratio by Monte Carlo sampling from $g$ is problematic due to the growing size of $A_{\unit\comma\time}$.
Thus, {\ABF} and {\ABFIR} make a localized approximation,
\begin{equation}
\label{eq:AB:ratio2}
 \frac{\E_{g}\big[\adapted^{}_{A^+_{\unit\comma\time}}\big] }{\E_{g}\big[\adapted^{}_{A^{}_{\unit\comma\time}}\big] }
\approx \frac{\E_{g}\big[\adapted^{}_{B^+_{\unit\comma\time}}\big] }{\E_{g}\big[\adapted^{}_{B^{}_{\unit\comma\time}}\big] }.
\end{equation}
The conditional log likelihood estimate $\MC{\loglik}_{\unit\comma\time}$ in {\ABF} and {\ABFIR} come from replacing the expectations on the right hand side of \myeqref{eq:AB:ratio2} with averages over 
Monte Carlo replicates of simulations from the adapted process. 
To see that we expect the approximation in \myeqref{eq:AB:ratio2} to hold when dependence decays across spatiotemporal distance, we can write 
\begin{eqnarray}
\nonumber
\adapted^{}_{A^{}_{\unit\comma\time}} 
&=&
 \adapted^{}_{B^{}_{\unit\comma\time}}  \hspace{1mm} \adapted^{}_{B^{c}_{\unit\comma\time}} 
\\
\nonumber
\adapted^{}_{A^+_{\unit\comma\time}}
&=&
 \adapted^{}_{B^{+}_{\unit\comma\time}}  \hspace{1mm}  \adapted^{}_{B^{c}_{\unit\comma\time}} ,
\end{eqnarray}
where $B^c_{\unit\comma\time}$ is the complement of $B_{\unit\comma\time}$ in $A_{\unit\comma\time}$.
Under our assumptions, the term corresponding to $\adapted_{B^c_{\unit\comma\time}}$ approximately cancels in the numerator and denominator of the right hand side of \myeqref{eq:AB:ratio2}.

Since {\ABF} is {\ABFIR} with $\Ninter=1$, we focus attention on {\ABFIR}.
At a conceptual level, the localized likelihood estimate in {\ABFIR} has the same structure as its {\UBF} counterpart.
However, {\ABFIR} additionally requires the capability to satisfactorily implement adapted simulation.
Adapted simulation is a local calculation, making it an easier task than the global operation of filtering.
Nevertheless, adapted simulation via importance sampling is vulnerable to COD for sufficiently large values of $\Unit$.
For a continuous time model, the use of $\Ninter>1$ is motivated by a result that guided intermediate resampling can reduce, or even remove, the COD in the context of a particle filtering algorithm \citep{park20}.
Assumptions~\ref{B1}--\ref{B:unconditional:mix} below are analogous to ~\ref{A1}--\ref{A:unconditional:mix} and are non-asymptotic assumptions involving $\ethree>0$, $\efourA>0$ and $Q>1$ which are required to hold uniformly over space and time.
Assumptions~\ref{B:temporal:mix}--\ref{B:XG_XA_ind} control the Monte~Carlo error arising from adapted simulation.
~\ref{B:temporal:mix} is a stability property which asserts that the effect of the latent process on the future of the adapted process decays over time. 
Assumption~\ref{B:girf} is a non-asymptotic bound on Monte~Carlo error for a single step of adapted simulation.
The scaling of the constant $\BvConstant$ with  $\Unit$,  $\Time$ and $\Ninter$ in  Assumption~\ref{B:girf} has been studied by \citet{park20}, where it was established that setting $\Ninter=\Unit$ can lead to $\BvConstant$ being constant, when using an ideal guide function, or slowly growing with $\Unit$ otherwise.
The $\efive^{-3}$ error rate in Assumption~\ref{B:girf} follows from balancing the two sources of error defined in the statement of Theorem~2 of \citet{park20}.
Assumption~\ref{B:XG_XA_ind} can be guaranteed by the construction of the algorithm, if independently generated Monte~Carlo random variables are used for building the guide function and the one-step prediction particles.
The asymptotic limit in Theorem~\ref{thm:abf} arises as the number of replicates increases.

\Bi 

\Bii 

\Biii 

\BivA 

\BivB 

\Bv  

\Bvi    

\TheoremII 

\begin{proof}
 A full proof is provided in Sec.~\SuppSecThmII. The extra work to prove Theorem~\ref{thm:abf} beyond the argument for Theorem~\ref{thm:tif} is to bound the error arising from the importance sampling approximation to a draw from the adapted transition density. 
This bound is constructed using Assumptions~\ref{B:temporal:mix}, \ref{B:girf} and~\ref{B:XG_XA_ind}.
The remainder of the proof follows the same approach as Theorem~\ref{thm:tif}, with the adapted process replacing the unconditional latent process.
\end{proof}

The theoretical results foreshadow our empirical observations (Sec.~\ref{sec:examples}) that the relative performance of {\UBF}, {\ABF} and {\ABFIR} is situation-dependent.
Assumption~\ref{A:unconditional:mix} is a mixing assumption for the unconditional latent process, whereas Assumption~\ref{B:unconditional:mix} replaces this with a mixing assumption for the adapted process conditional on the data. 
For a non-stationary process, Assumption~\ref{A:unconditional:mix} may fail to hold uniformly in $\Unit$ whereas the adapted process may provide stable tracking of the latent process (Sec.~\SuppSecAdaptedSimulation).
When Assumption~\ref{A:unconditional:mix} holds, {\UBF} can benefit from not requiring Assumptions~\ref{B:temporal:mix}, \ref{B:girf} and~\ref{B:XG_XA_ind}. 
Adapted simulation is an easier problem than filtering, but nevertheless can become difficult in high dimensions, with the consequence that Assumption~\ref{B:girf} could require large $\BvConstant$.
The tradeoff between {\ABF} and {\ABFIR} depends on the effectiveness of the guide function for the problem at hand.
Intermediate resampling and guide function calculation require additional computational resources, which will necessitate smaller values of $\Rep$ and $\Np$.
In some situations, the improved scaling properties of {\ABFIR} compared to {\ABF}, corresponding to a lower value of $\BvConstant$, will outweigh this cost.

\medskip

\section{Examples}
\label{sec:examples}

We compare the performance of the three bagged filters ({\UBF}, {\ABF} and {\ABFIR}) against each other and against alternative plug-and-play approaches.
The plug-and-play property facilitates numerical implementation for general classes of models, and all the algorithms and models under consideration are implemented in the \code{spatPomp} R package \citep{asfaw21arxiv}.
Ensemble Kalman filter (EnKF) methods propagate the ensemble members by simulation from the dynamic model and then update the ensemble to assimilate observations using a Gaussian-inspired rule \citep{evensen96,lei10}.
The block particle filter \citep[BPF,][]{rebeschini15,ng02} partitions the latent space and combines independently drawn components from each partition.
BPF overcomes COD under weak coupling assumptions \citep{rebeschini15}.
Unlike these two methods, our bagged filters modify particles only according to the latent dynamics.
Thus, our methods respect conservation laws and continuity or smoothness conditions obeyed by the dynamic model.
We also compare with a guided intermediate resampling filter \citep[GIRF,][]{park19}, one of many variants of the particle filter designed to scale to larger numbers of units than are possible with a basic particle filter.

First, in Sec.~\ref{sec:bm}, we consider a spatiotemporal Gaussian process for which the exact likelihood is available via a Kalman filter. 
We see in Fig.~\ref{fig:bm_alt_plot} that {\ABFIR} can have a considerable advantages over {\UBF} and {\ABF} for problems with an intermediate level of coupling.
Then, in Sec.~\ref{sec:measles}, we develop a model for measles transmission within and between cities.
The measles model is weakly coupled, leading to successful performance for all three bagged filters.
This class of metapopulation models was the primary motivation for the development of these methodologies.
In Sec.~\ref{sec:profile} we demonstrate an extension from likelihood evaluation to likelihood maximization for the measles model.
Additionally, in  Sec.~\SuppSecLorenz, we compare performance on the Lorenz-96 model, a highly coupled system used to test inference methods for geophysical applications.

\subsection{Correlated Brownian motion}
\label{sec:bm}

Suppose $\myvec{X}(t) = \Omega \myvec{W}(t)$ where $\myvec{W}(t)=W_{1:\Unit}(t)$ comprises $\Unit$ independent standard Brownian motions, and $\Omega_{\unit,\altUnit}=\rho^{\dist(\unit,\altUnit)}$ with $\dist(\unit,\altUnit)$ being the circle distance,
\[
\dist(\unit,\altUnit) 
= \min\big(|\unit-\altUnit|, |\unit-\altUnit+\Unit|, |\unit-\altUnit-\Unit|\big).
\]
Set $t_\time=\time$ for $\time=0,1,\dots,\Time$ with initial value $\myvec{X}(0)=\myvec{0}$ and suppose measurement errors are independent and normally distributed, $Y_{\unit\comma\time}= X_{\unit\comma\time}+ \eta_{\unit\comma\time}$ with
$\eta_{\unit\comma\time}\sim \normal(0,\tau^2)$.
The parameter $\rho$ determines the strength of the spatial coupling.


\begin{knitrout}
\definecolor{shadecolor}{rgb}{1, 1, 1}\color{fgcolor}\begin{figure}

\includegraphics[width=4.5in]{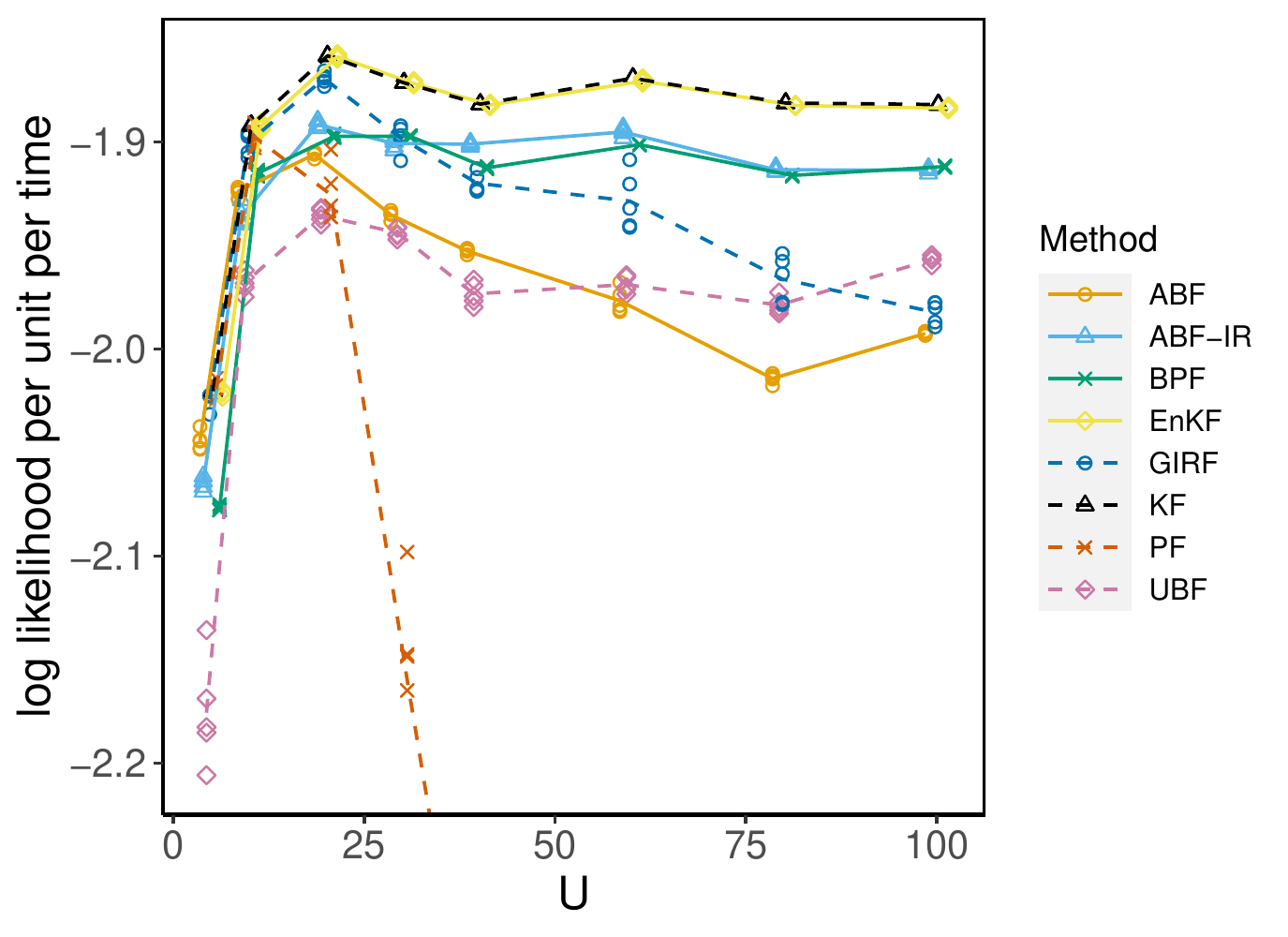} \hfill{}

\caption[log likelihood estimates for a correlated Brownian motion model of various dimensions]{log likelihood estimates for a correlated Brownian motion model of various dimensions. {\UBF}, {\ABF} and {\ABFIR} are compared with a guided intermediate resampling filter (GIRF), standard particle filter (PF), block particle filter (BPF) and ensemble Kalman filter (EnKF). The exact likelihood was computed via a Kalman filter (KF).}\label{fig:bm_alt_plot}
\end{figure}

\end{knitrout}

Fig.~\ref{fig:bm_alt_plot} shows how the bagged filters scale on this Gaussian model, compared to a standard particle filter (PF), a guided intermediate resampling filter (GIRF), a block particle filter (BPF), and an ensemble Kalman filter.
For our numerical results, we use $\tau=1$, $\rho=0.4$ and $\Time=50$.
The algorithmic parameters and run times are listed in Sec.~\SuppSecBM, together with a plot of the simulated data and supplementary discussion.
In this case, the exact likelihood is computable via the Kalman filter (KF). 
Since EnKF is based on a Gaussian approximation, it is also exact in this case, up to a small Monte Carlo error. 
The GIRF framework encompasses lookahead particle filter techniques, such as the auxiliary particle filter \citep{pitt99}, and intermediate resampling techniques \citep{delmoral17}. 
GIRF methods combining these techniques were found to perform better than either of these component techniques alone \citep{park20}.
Thus, GIRF here represents a state-of-the-art auxiliary particle filter that targets the complete joint filter density for all units.
We use the general-purpose, plug-and-play implementation of GIRF provided by the \texttt{spatPomp} R package \citep{asfaw21github}; for a Gaussian model, one can calculate an ideal guide function for GIRF but that was not used.
PF works well for small values of $\Unit$ in Fig.~\ref{fig:bm_alt_plot} and rapidly starts struggling as $\Unit$ increases.
GIRF behaves comparably to PF for small $\Unit$ but its performance is maintained for larger $\Unit$.
{\ABF} and {\ABFIR} have some efficiency loss, for small $\Unit$, relative to PF and GIRF due to the localization involved in the filter weighting, but for large $\Unit$ this cost is paid back by the benefit of the reduced Monte Carlo variability.
{\UBF} has a larger efficiency loss for small $\Unit$, but its favorable scaling properties lead it to overtake {\ABF} for larger $\Unit$.
BPF shows stable scaling and modest efficiency loss.
This linear Gaussian SpatPOMP model provides a simple scenario to demonstrate scaling behavior.
For filters that cannot take direct advantage of the Gaussian property of the model, we see that there is a tradeoff between efficiency at low $\Unit$ and scalability.
This is unavoidable, since there is no known algorithm that is simultaneously fully efficient (up to Monte Carlo error), scalable, and applicable to general SpatPOMP models.
We now explore this tradeoff empirically on to a more complex SpatPOMP exemplifying the nonlinear non-Gaussian models motivating our new filtering approach.


\subsection{Spatiotemporal measles epidemics}
\label{sec:measles}

Data analysis for spatiotemporal systems featuring nonlinear, nonstationary mechanisms and partial observability has been a longstanding open challenge for ecological and epidemiological analysis \citep{bjornstad01}.
A compartment modeling framework for spatiotemporal population dynamics divides the population at each spatial location into categories, called compartments, which are modeled as homogeneous. 
Spatiotemporal compartment models can be called patch models or metapopulation models in an ecological context.
Ensemble Kalman filter (EnKF) methods provide a state-of-the-art approach to inference for metapopulation models \citep{li20} despite concerns that the approximations inherent in the EnKF can be problematic for models that are highly nonlinear or non-Gaussian \citep{ades15}.
Our bagged filter methodologies have theoretical guarantees for arbitrarily nonlinear and non-Gaussian models, while having improved scaling properties compared to particle filters.

We consider a spatiotemporal model for disease transmission dynamics of measles within and between multiple cities, based on the model of \citet{park20} which adds spatial interaction to the compartment model presented by \cite{he10}.
The model compartmentalizes the population of each city into susceptible ($S$), exposed ($E$), infectious ($I$), and recovered/removed ($R$) categories.
The number of individuals in each compartment city $\unit$ at time $t$ are denoted by integer-valued random variables $S_\unit(t)$, $E_\unit(t)$, $I_\unit(t)$, and $R_\unit(t)$.
The population dynamics are written in terms of counting processes $N_{\bullet\bullet,\unit}(t)$ enumerating cumulative transitions in city $\unit$, up to time $t$, between compartments identified by the subscripts.
We model the $\Unit$ largest cities in the UK, ordered in decreasing size so that $\unit=1$ corresponds to London.
We vary $\Unit$ to test methodologies on a hierarchy of filtering challenges.
Our model is described by the following system of stochastic differential equations, for $\unit=1,\dots, \Unit$,
\begin{equation}
\nonumber
\begin{array}{lllllll}
\displaystyle dS_\unit(t) &=& dN_{BS,\unit}(t) &-& dN_{SE,\unit}(t) &-& dN_{SD,\unit}(t) \\
\displaystyle dE_\unit(t) &=& dN_{SE,\unit}(t) &-& dN_{EI,\unit}(t) &-& dN_{ED,\unit}(t) \\
\displaystyle dI_\unit(t) &=& dN_{EI,\unit}(t) &-& dN_{IR,\unit}(t) &-& dN_{ID,\unit}(t) 
\end{array}
\end{equation}
Here,  $N_{BS,\unit}(t)$ models recruitment into the susceptible population, and $N_{\bullet D,\unit}(t)$ models emigration and death. 
The total population $P_\unit(t)=S_\unit(t)+E_\unit(t)+I_\unit(t)+R_\unit(t)$ is calculated by smoothing census data and is treated as known.
The number of recovered individuals $R_\unit(t)$ in city $\unit$ is therefore defined implicitly.
$N_{SE,\unit}(t)$ is modeled as negative binomial death processes \citep{breto09,breto11}
with over-dispersion parameter $\sigma_{SE}$, and rate given by
\begin{eqnarray}
\nonumber
\mathbb{E} \big[ N_{SE,\unit}(t+dt) - N_{SE,\unit}(t) \big] 
&=& 
\beta(t) \, S_\unit(t) 
\Big[ 
  \left( \frac{I_\unit+\iota}{P_\unit} \right)^\alpha
\\
\label{eq:dEdt}
&& \hspace{-3.8cm}
 + \sum_{\altUnit \neq \unit} \frac{v_{\unit\altUnit}}{P_\unit} 
  \left\{ 
    \left(
      \frac{ I_{\altUnit} }{ P_{\altUnit} }
    \right)^\alpha - 
    \left(
      \frac{I_\unit}{P_\unit}  
    \right)^\alpha
  \right\}
\Big] dt + o(dt),\label{eqn:transmissionrate}
\end{eqnarray}
where $\beta(t)$ models seasonality driven by high contact rates between children at school, described by
\begin{equation}
\nonumber
  \beta(t)=\begin{cases}
\big(1+\amplitude(1-p)p^{-1} \big)\, \meanBeta & \mbox{ during school term},\\
\big( 1-\amplitude\big) \, \meanBeta& \mbox{ during vacation}
  \end{cases} \label{eq:term}
\end{equation}
with $p = 0.759$ being the proportion of the year taken up by the school terms, $\meanBeta$ is the mean transmission rate, and $\amplitude$ measures the reduction of transmission during school holidays.
In \myeqref{eq:dEdt}, $\alpha$ is a mixing exponent modeling inhomogeneous contact rates within a city, and $\iota$ models immigration of infected individuals which is appropriate when analyzing a subset of cities that cannot be treated as a closed system.
The number of travelers from city $\unit$ to $\altUnit$ is denoted by $v_{\unit\altUnit}$. 
Here, $v_{\unit\altUnit}$ is constructed using the gravity model of \cite{xia04}, 
\[
v_{\unit\altUnit} = \gravity \cdot \frac{\;\overline{\dist}\;}{\bar{P}^2} \cdot \frac{P_\unit \cdot P_{\altUnit}}{\dist(\unit,\altUnit)},
\]
where $\dist(\unit,\altUnit)$ denotes the distance between city $\unit$ and city $\altUnit$, $P_\unit$ is the average population for city $\unit$ across time, $\bar{P}$ is the average population across cities, and $\overline{\dist}$ is the average distance between a randomly chosen pair of cities.
Here, we model $v_{\unit\altUnit}$ as fixed through time and symmetric between any two arbitrary cities, though a natural extension would allow for temporal variation and asymmetric movement between two cities.
The transition processes $N_{EI,\unit}(t)$, $N_{IR,\unit}(t)$ and $N_{\bullet D,\unit}(t)$ are modeled as conditional Poisson processes with per-capita rates $\mu_{EI}$, $\mu_{IR}$ and $\mu_{\bullet D}$ respectively, and we fix $\mu_{\bullet D}=50 \mbox{ year}^{-1}$.
The birth process $N_{BS,\unit}(t)$ is an inhomogeneous Poisson processes with rate $\mu_{BS,\unit}(t)$, given by interpolated census data.

To complete the model specification, we must describe the measurement process.
Let $Z_{\unit\comma\time}=N_{IR\comma\unit}(t_\time)-N_{IR\comma\unit}(t_{\time-1})$ be the number of removed infected individuals in the $n$th reporting interval.
Suppose that cases are quarantined once they are identified, so that reported cases comprise a fraction $\rho$ of these removal events.
The case report $\data{y}_{\unit\comma\time}$ is modeled as a realization of a discretized conditionally Gaussian random variable $Y_{\unit\comma\time}$, defined for $y>0$ via
\begin{eqnarray}
\nonumber
\prob\big[Y_{\unit\comma\time}{=}y\mid Z_{\unit\comma\time}{=}z\big] &=& \Phi\big(y+0.5; \rho z,\rho(1-\rho)z+\psi^2\rho^2z^2\big)
\\
&&\hspace{-20mm}
- \Phi\big(y-0.5; \rho z,\rho(1-\rho)z+\psi^2\rho^2z^2\big)
\label{eq:obs}
\end{eqnarray}
where $\Phi(\cdot;\mu,\sigma^2)$ is the $\normal(\mu,\sigma^2)$ cumulative distribution function, and $\psi$ models overdispersion relative to the binomial distribution.
For $y=0$, we replace $y-0.5$ by $-\infty$ in \myeqref{eq:obs}.


\begin{knitrout}
\definecolor{shadecolor}{rgb}{1, 1, 1}\color{fgcolor}\begin{figure}

\includegraphics[width=6.5in]{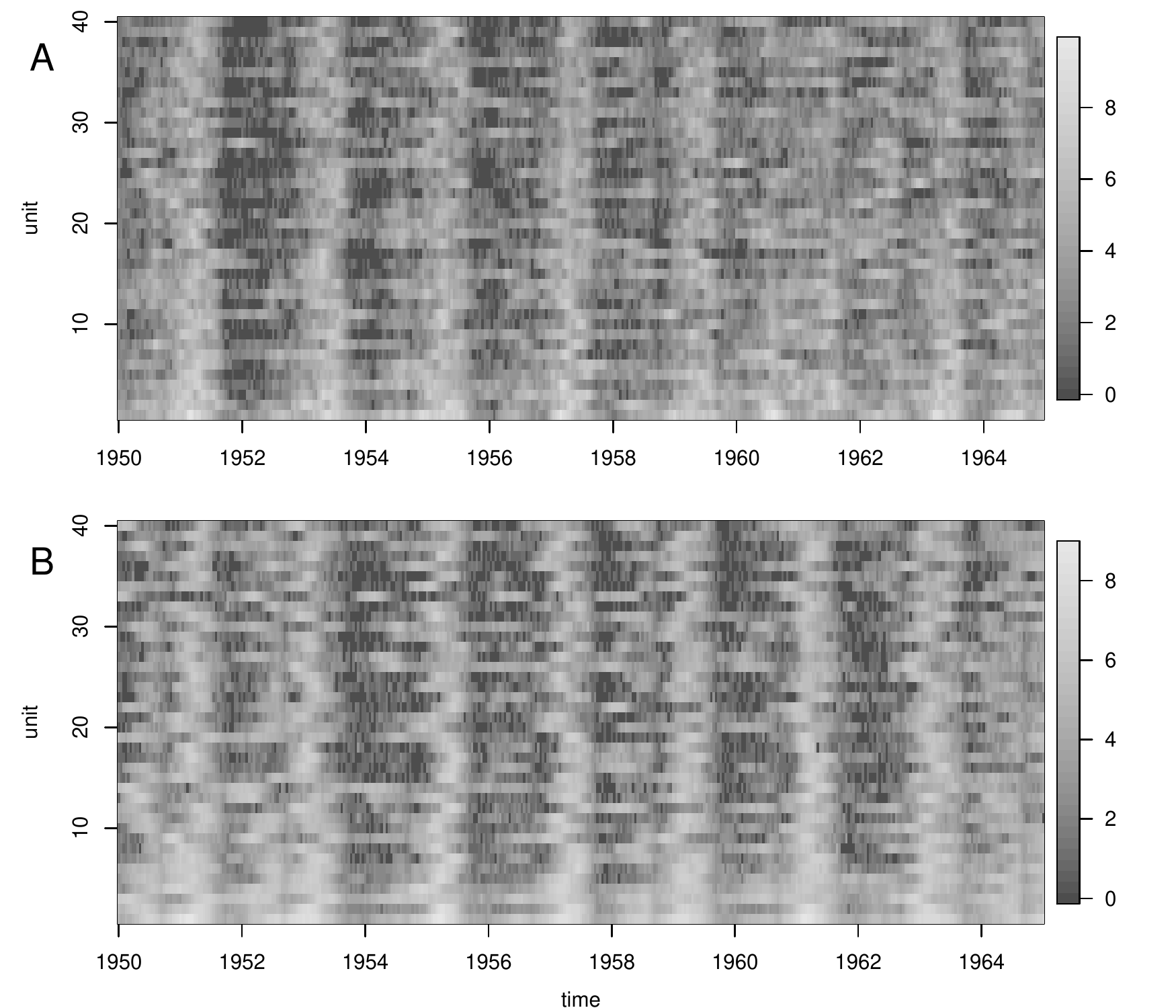} \hfill{}

\caption[Log(reported cases $+$ 1) for (A) the measles simulation used for the likelihood slice]{Log(reported cases $+$ 1) for (A) the measles simulation used for the likelihood slice; (B) the corresponding UK measles data. The simulation shares the biennial pattern, with most but not all cities locked in phase most of the time.}\label{fig:measles_image_plot}
\end{figure}

\end{knitrout}

\begin{knitrout}
\definecolor{shadecolor}{rgb}{1, 1, 1}\color{fgcolor}\begin{figure}

\includegraphics[width=4in]{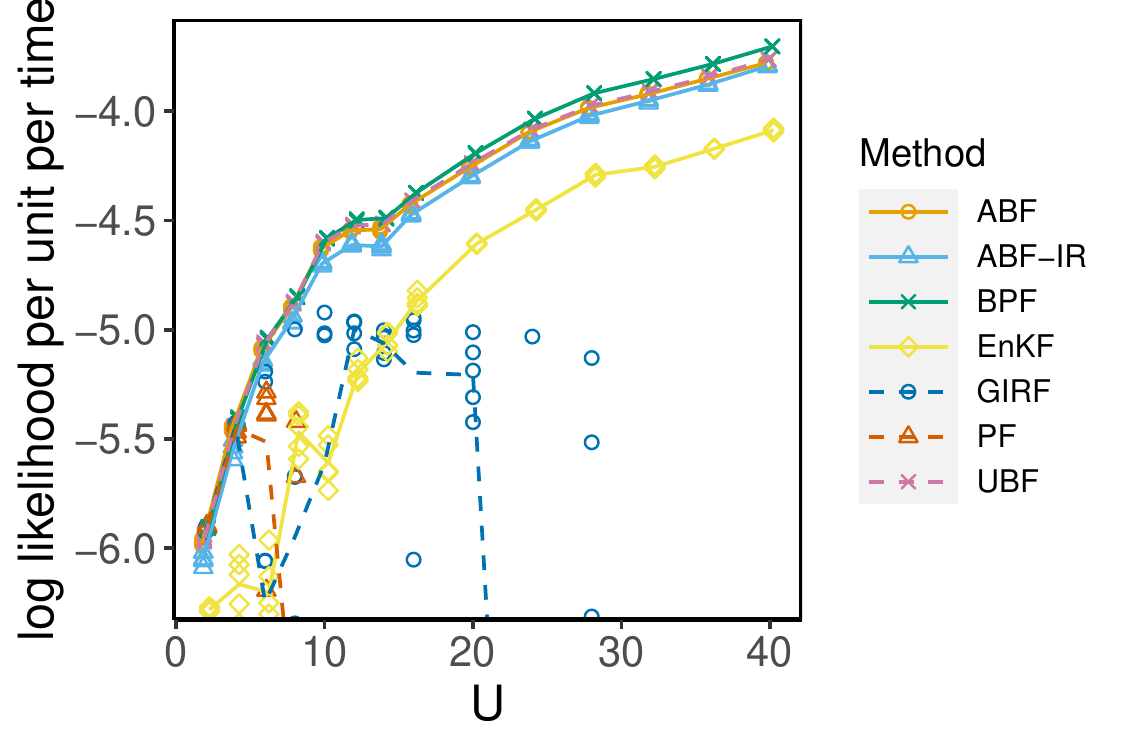} \hfill{}

\caption[log likelihood estimates for simulated data from the measles model of various dimensions]{log likelihood estimates for simulated data from the measles model of various dimensions. {\UBF}, {\ABF} and {\ABFIR} are compared with a guided intermediate resampling filter (GIRF), a standard particle filter (PF), a block particle filter (BPF) and an ensemble Kalman filter (EnKF).}\label{fig:mscale_loglik_plot}
\end{figure}

\end{knitrout}

This model includes many features that have been proposed to be relevant for understanding measles transmission dynamics \citep{he10}.
Our plug-and-play methodology permits consideration of all these features, and readily extends to the investigation of further variations. 
Likelihood-based inference via plug-and-play methodology therefore provides a framework for evaluating which features of a dynamical model are critical for explaining the data \citep{king08}.
By contrast, \cite{xia04} developed a linearization for a specific spatiotemporal measles model which is numerically convenient but not readily adaptable to assess alternative model choices.
Fig.~\ref{fig:measles_image_plot} shows a simulation from our model, showing that trajectories from this model can capture some features of the system that have been hard to understand: how can it be that disease transmission dynamics between locations have important levels of interaction yet are not locked in synchrony \citep{becker20}?
Here, we are testing statistical tools rather than engaging directly in the scientific debate so we test methods on the simulated data.

We first assess the scaling properties of the filters on the measles model by evaluating the likelihood over varying numbers of units, $\Unit$, for fixed parameters.
The results are given in Fig.~\ref{fig:mscale_loglik_plot}, with additional information about timing, algorithmic choices, parameter values and a plot of the data provided in Sec.~\SuppSecMeasles.
In Fig.~\ref{fig:mscale_loglik_plot}, the log likelihood per unit per time increases with $\Unit$ because city size decreases with $\Unit$. 
Smaller cities have fewer measles cases, resulting in a narrower and taller probability density function.
Fig.~\ref{fig:mscale_loglik_plot} shows a rapid decline in the performance of the particle filter (PF) beyond $\Unit=4$.
This is a challenging filtering problem, with dynamics including local fadeouts and high stochasticity in each city stabilized at the metapopulation level by the coupling.
In this example, GIRF performs poorly suggesting that the simulated moment guide function is less than successful.
We used the general-purpose implementation of GIRF in the \code{spatPomp} package, and there might be room for improvement by developing a model-specific guide function.
{\ABFIR} uses the same guide function, and this may explain why {\ABFIR} performs worse than {\ABF} here, though {\ABFIR} is much less sensitive than GIRF to the quality of the guide.
{\ABF} and {\UBF} are competing with BPF as winners on this challenge.
The bagged filters and BPF have substantial advantages compared to EnKF, amounting to more than 0.2 log likelihood units per observation.
We suspect that the limitations of EnKF on this problem are due to the nonlinearity, non-Gaussianity, and discreteness of fadeout and reintroduction dynamics.
All the algorithms have various tuning parameters that could influence the results. 
Some investigations of alternatives are presented in Secs.~\SuppSecMeasles,~\SuppSecMeaslesNbhd and~\SuppSecIvsJ.
Generalizable conclusions are hard to infer from numerical comparisons of complex algorithms on complex models.
Experimentation with different methods, and their tuning parameters, is recommended when investigating a new model.

\begin{knitrout}
\definecolor{shadecolor}{rgb}{1, 1, 1}\color{fgcolor}\begin{figure}

\includegraphics[width=\maxwidth,height=7in]{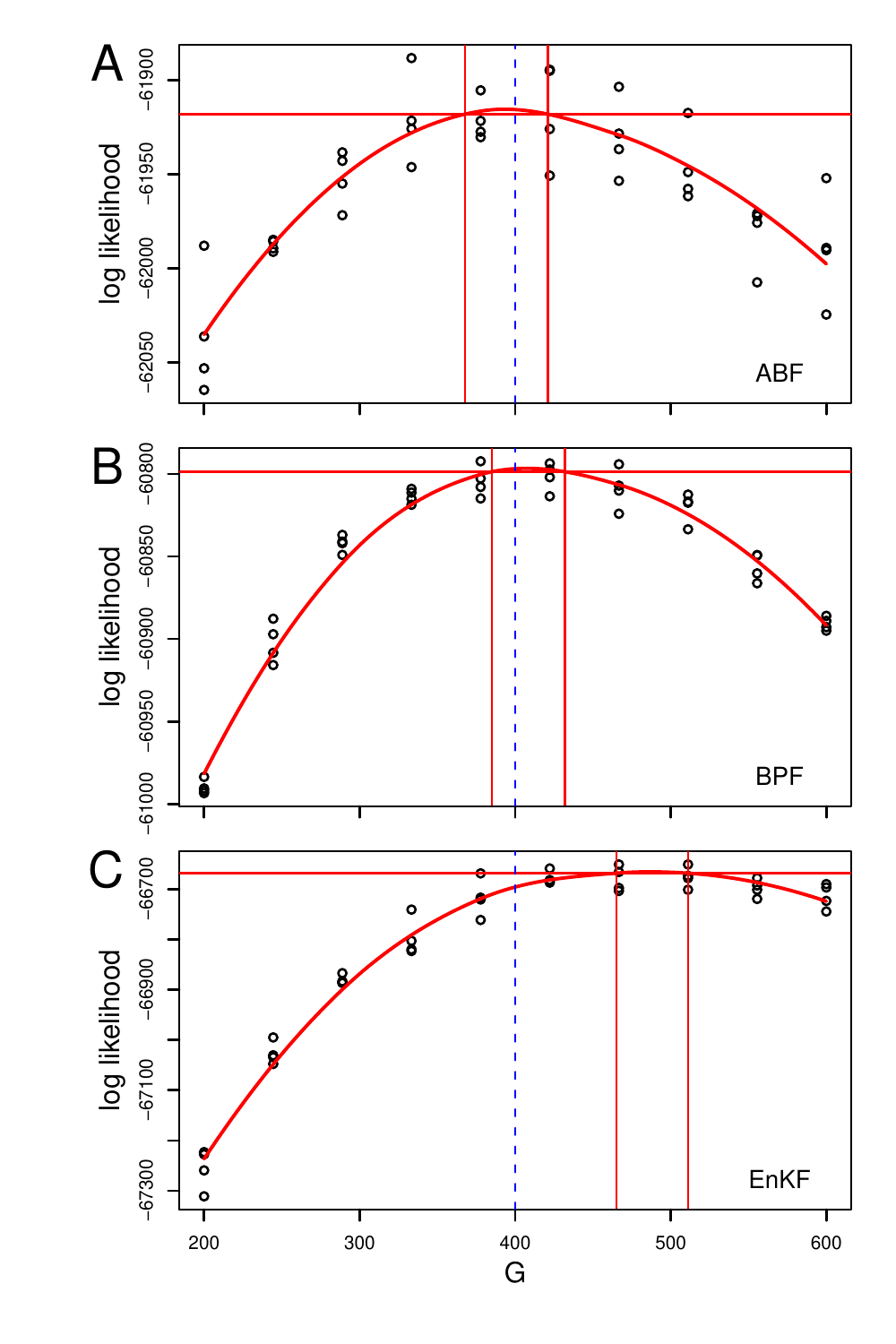} \hfill{}

\caption{Likelihood slices varying the coupling parameter, for the measles model with $U=40$ cities, computed via (A) {ABF}; (B) BPF; (C) EnKF. The solid perpendicular lines construct 95\% Monte Carlo adjusted confidence intervals \citep{ionides17}. The true parameter value is identified by a blue dashed line.}\label{fig:slice_combined_plot}
\end{figure}

\end{knitrout}

Fig.~\ref{fig:slice_combined_plot}(A) demonstrates an application of {\ABF} to the task of computing a slice of the likelihood function over the coupling parameter, $\gravity$, for simulated data.
This slice varies $\gravity$ while fixing the other parameters at the values used for the simulation.
Fig.~\ref{fig:slice_combined_plot}(B) shows a similar plot calculated using BPF with comparable computational effort.
Both ABF and BPF are successful here, though BPF is more computationally efficient.
By contrast, Fig.~\ref{fig:slice_combined_plot}(C) shows that EnKF has substantial bias in estimating $\gravity$, as well as considerably lower likelihood. 
Likelihood slices have less inferential value than likelihood profiles, but provide a computationally and conceptually simpler setting that can be insightful.
Scientifically, the slices in Fig.~\ref{fig:slice_combined_plot} give an upper bound on the identifiability of $\gravity$ from such data, since the likelihood slice provides statistically efficient inference when all other parameters are known.

%
%


\subsection{Likelihood maximization and profile likelihood}
\label{sec:profile}
\begin{knitrout}
\definecolor{shadecolor}{rgb}{1, 1, 1}\color{fgcolor}\begin{figure}

\includegraphics[width=4in]{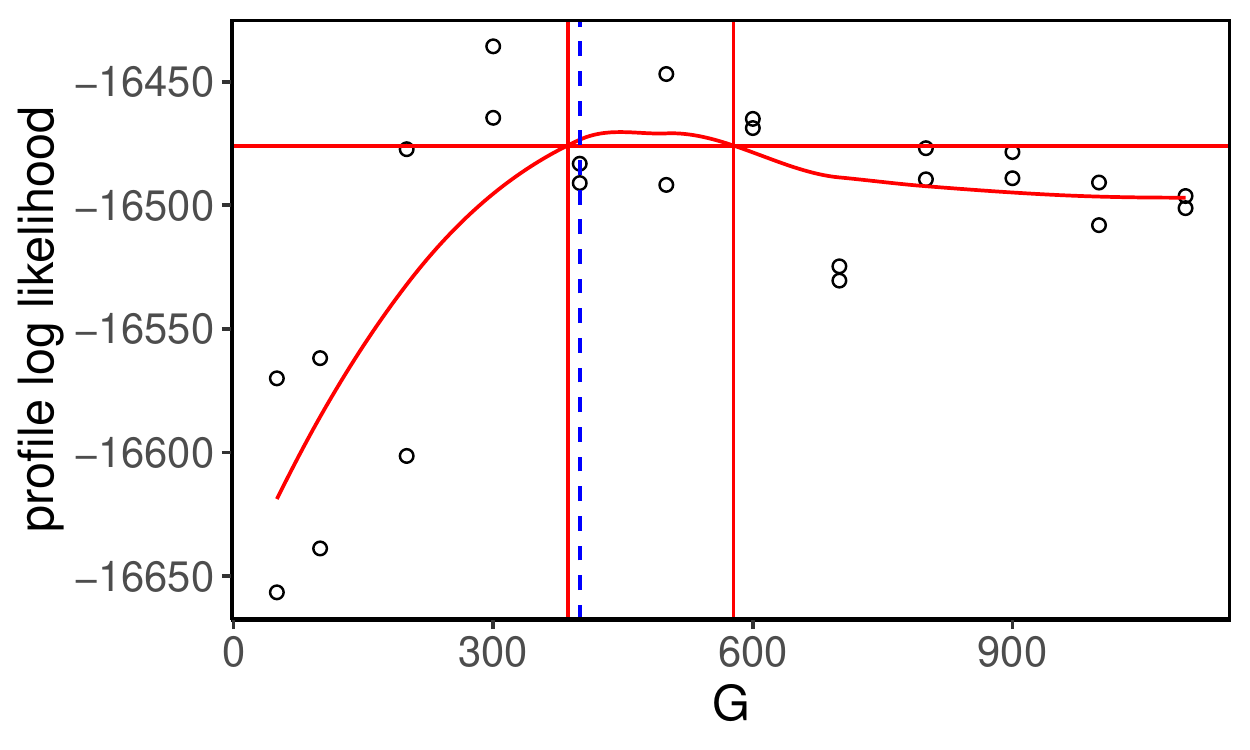} \hfill{}

\caption[An iterated bagged filter used to maximize the likelihood, compute a profile likelihood, and hence construct a confidence interval]{An iterated bagged filter used to maximize the likelihood, compute a profile likelihood, and hence construct a confidence interval. The profiling is carried out over the coupling parameter, $G$.}\label{fig:profile_plot}
\end{figure}

\end{knitrout}

Likelihood evaluation via filtering does not by itself enable parameter estimation for POMP models, however it provides a foundation for Bayesian and likelihood-based inference.
In particular, filtering algorithms can be modified to carry out likelihood maximization by stochastically perturbing parameters in a sequence of filtering operations with decreasing perturbation variance \citep{ionides15}.
We demonstrate this for the measles model in Fig.~\ref{fig:profile_plot} using an iterated bagged filter algorithm which is fully described in Sec.~{\SuppSecParameterEst}.

Monte Carlo methods for computing and maximizing the log likelihood suffer from  bias and variance, both of which can be considerable for large datasets and complex models.
Appropriate inference methodology, such as Monte Carlo adjusted profile (MCAP) confidence intervals, can accommodate substantial Monte Carlo variance so long as the bias is slowly varying across the statistically plausible region of the parameter space \citep{ionides17,ning21}.
Fig.~\ref{fig:profile_plot} constructs an MCAP 95\% confidence interval for the coupling parameter, $G$, using an iterated unadapted bagged filter to maximize over the parameters, $a$, $\bar\beta$, $\sigma_{SE}$, $\psi$, $\mu_{EI}$ and $\mu_{IR}$.
This simulation study, carried out with $\Unit=20$ and $\Time=208$, shows that $G$ is identifiable via likelihood-based inference in the absence of assumptions about these parameters.

\section{Discussion}
\label{sec:discussion}

The pseudocode presented for the bagged filters describes how the outputs are calculated given the inputs, but does not prescribe details of how these quantities are calculated. 
There is scope for implementations to trade off memory, computation and communication by varying decisions on how the loops defined in the pseudocode are coded, including decisions on memory over-writing and parallelization.
This article focuses on  the logical structure of the algorithms, leaving room for future research on implementation-specific considerations, though some supplementary discussion of memory-efficient implementation is given in Sec.~\SuppSecMemoryEfficient.

Plug-and-play inference based on sequential Monte Carlo likelihood evaluation has proved successful for investigating highly nonlinear partially observed dynamic systems of low dimension arising in analysis of epidemiological and ecological population dynamics \citep{breto18statSci,pons-salort18,decelles18,marino18}.
This article develops a methodological extension motivated by the analysis of interacting biological populations.
Similar challenges related to nonlinear non-Gaussian dynamic models arise in geophysical modeling.
Relative to biological systems, geophysical applications are characterized by a greater number of spatial locations, better mathematical understanding of the underlying processes, and lower stochasticity.
From this literature, the locally weighted particle filter of \citet{poterjoy16} is perhaps closest to our approach, but the local weights of \citet{poterjoy16} are used to construct a localized Kalman gain which is motivated by a Gaussian approximation comparable to EnKF. 
EnKF arose originally via geophysical research \citep{evensen96} and has since become used more widely for inference on SpatPOMP models \citep{katzfuss19,lei10}.
However, EnKF can fail entirely even on simple POMP models if the structure is sufficiently non-Gaussian.
For example, let $X_n$ be a one-dimensional Gaussian random walk, and let $Y_n$ given $X_n=x_n$ be normally distributed with mean $0$ and variance $x_n^2$.
The linear filter rule used by EnKF to update the estimate of $X_n$ given $Y_n$ has mean zero for any value of $X_n$, since $X_n$ and $Y_n$ are uncorrelated.
Therefore, the EnKF filter estimate of the latent process remains essentially constant regardless of the data.
Models of this form are used in finance to describe stochastic volatility.
EnKF could be applied more successfully by modifying model, such as replacing $Y_n$ by $|Y_n|$, but for complex models it may be unclear whether and where such problems are arising.
Our results show that there is room for improvement over EnKF on a spatiotemporal epidemiology model, though in our example there is no clear advantage for BF methods over BPF.

Latent state trajectories constructed in our BF algorithms are all generated from the model simulator, appropriately reweighted and resampled, and so they are necessarily valid sample paths of the model.
For example, spatial smoothness properties of the model through space, or conservation properties where some function of the system remains unchanged through time, are maintained in the BF trajectories.
This is not generally true for the block particle filter (since resampling blocks can lead to violations at block boundaries) or for EnKF (since the filter procedure perturbs particles using a linear update rule that cannot respect nonlinear relationships).
The practical importance of smoothness and conservation considerations will vary with the system under investigation, but this property of BF gives the scientific investigator one less thing to worry about.

The algorithms {\UBF}, {\ABF}, {\ABFIR}, GIRF, PF, BPF, and EnKF compared in this article all enjoy the plug-and-play property, facilitating their implementations in general-purpose software.
The numerical results for this paper use the \code{abf}, \code{abfir}, \code{girf}, \code{pfilter}, \code{bpfilter} and \code{enkf} functions via the open-source R package \code{spatPomp} \citep{asfaw21arxiv} that provides a spatiotemporal extension of the R package \code{pomp} \citep{king16}.
{\UBF} was implemented using \code{abf} with $\Np=1$ particles per replicate.
The source code for this paper will be contributed to an open-source scientific archive upon acceptance for publication.

\bibliography{bib-iif}

\begin{thebibliography}{}

\bibitem[Ades and Van~Leeuwen, 2015]{ades15}
Ades, M. and Van~Leeuwen, P.~J. (2015).
\newblock The equivalent-weights particle filter in a high-dimensional system.
\newblock {\em Quarterly Journal of the Royal Meteorological Society},
  141(687):484--503.

\bibitem[Asfaw et~al., 2021a]{asfaw21github}
Asfaw, K., Ionides, E.~L., and King, A.~A. (2021a).
\newblock \texttt{spatPomp}: R package for statistical inference for
  spatiotemporal partially observed {M}arkov processes.
\newblock {\em \upshape{\texttt{https://github.com/kidusasfaw/spatPomp}}}.

\bibitem[Asfaw et~al., 2021b]{asfaw21arxiv}
Asfaw, K., Park, J., Ho, A., King, A.~A., and Ionides, E.~L. (2021b).
\newblock Statistical inference for spatiotemporal partially observed {M}arkov
  processes via the {R} package spatpomp.
\newblock {\em arXiv:2101.01157}.

\bibitem[Becker et~al., 2020]{becker20}
Becker, A.~D., Zhou, S.~H., Wesolowski, A., and Grenfell, B.~T. (2020).
\newblock Coexisting attractors in the context of cross-scale population
  dynamics: Measles in {L}ondon as a case study.
\newblock {\em Proceedings of the Royal Society of London, Series B},
  287(1925):20191510.

\bibitem[Bengtsson et~al., 2008]{bengtsson08}
Bengtsson, T., Bickel, P., and Li, B. (2008).
\newblock Curse-of-dimensionality revisited: Collapse of the particle filter in
  very large scale systems.
\newblock In Speed, T. and Nolan, D., editors, {\em Probability and Statistics:
  Essays in Honor of {D}avid {A}. {F}reedman}, pages 316--334. Institute of
  Mathematical Statistics, Beachwood, OH.

\bibitem[Bentkus et~al., 1997]{bentkus97}
Bentkus, V., G{\"o}tze, F., and Tikhomoirov, A. (1997).
\newblock Berry--{E}sseen bounds for statistics of weakly dependent samples.
\newblock {\em Bernoulli}, 3(3):329--349.

\bibitem[Beskos et~al., 2017]{beskos17}
Beskos, A., Crisan, D., Jasra, A., Kamatani, K., and Zhou, Y. (2017).
\newblock A stable particle filter for a class of high-dimensional state-space
  models.
\newblock {\em Advances in Applied Probability}, 49(1):24--48.

\bibitem[Bj{\o}rnstad and Grenfell, 2001]{bjornstad01}
Bj{\o}rnstad, O.~N. and Grenfell, B.~T. (2001).
\newblock Noisy clockwork: Time series analysis of population fluctuations in
  animals.
\newblock {\em Science}, 293:638--643.

\bibitem[Brehmer et~al., 2020]{brehmer20}
Brehmer, J., Louppe, G., Pavez, J., and Cranmer, K. (2020).
\newblock Mining gold from implicit models to improve likelihood-free
  inference.
\newblock {\em Proceedings of the National Academy of Sciences of the USA},
  117(10):5242--5249.

\bibitem[Breiman, 1996]{breiman96}
Breiman, L. (1996).
\newblock Bagging predictors.
\newblock {\em Machine learning}, 24(2):123--140.

\bibitem[Bret{\'o}, 2018]{breto18statSci}
Bret{\'o}, C. (2018).
\newblock Modeling and inference for infectious disease dynamics: A
  likelihood-based approach.
\newblock {\em Statistical Science}, 33(1):57--69.

\bibitem[Bret\'{o} et~al., 2009]{breto09}
Bret\'{o}, C., He, D., Ionides, E.~L., and King, A.~A. (2009).
\newblock Time series analysis via mechanistic models.
\newblock {\em Annals of Applied Statistics}, 3:319--348.

\bibitem[Bret\'{o} and Ionides, 2011]{breto11}
Bret\'{o}, C. and Ionides, E.~L. (2011).
\newblock Compound {M}arkov counting processes and their applications to
  modeling infinitesimally over-dispersed systems.
\newblock {\em Stochastic Processes and their Applications}, 121:2571--2591.

\bibitem[Chandler, 2013]{chandler13}
Chandler, R.~E. (2013).
\newblock Exploiting strength, discounting weakness: Combining information from
  multiple climate simulators.
\newblock {\em Philosophical Transactions of the Royal Society A: Mathematical,
  Physical and Engineering Sciences}, 371(1991):20120388.

\bibitem[de~Cell{\`e}s et~al., 2018]{decelles18}
de~Cell{\`e}s, M.~D., Magpantay, F.~M., King, A.~A., and Rohani, P. (2018).
\newblock The impact of past vaccination coverage and immunity on pertussis
  resurgence.
\newblock {\em Science Translational Medicine}, 10(434):eaaj1748.

\bibitem[Del~Moral et~al., 2017]{delmoral17}
Del~Moral, P., Moulines, E., Olsson, J., and Verg{\'e}, C. (2017).
\newblock Convergence properties of weighted particle islands with application
  to the double bootstrap algorithm.
\newblock {\em Stochastic Systems}, 6(2):367--419.

\bibitem[Del~Moral and Murray, 2015]{delmoral15}
Del~Moral, P. and Murray, L.~M. (2015).
\newblock Sequential {M}onte {C}arlo with highly informative observations.
\newblock {\em Journal on Uncertainty Quantification}, 3:969--997.

\bibitem[Doucet et~al., 2001]{doucet01}
Doucet, A., de~Freitas, N., and Gordon, N.~J. (2001).
\newblock {\em Sequential \mbox{M}onte \mbox{C}arlo Methods in Practice}.
\newblock Springer, New York.

\bibitem[Doucet and Johansen, 2011]{doucet11}
Doucet, A. and Johansen, A. (2011).
\newblock A tutorial on particle filtering and smoothing: Fifteen years later.
\newblock In Crisan, D. and Rozovsky, B., editors, {\em Oxford Handbook of
  Nonlinear Filtering}. Oxford University Press.

\bibitem[Ebert, 2001]{ebert01}
Ebert, E.~E. (2001).
\newblock Ability of a poor man's ensemble to predict the probability and
  distribution of precipitation.
\newblock {\em Monthly Weather Review}, 129(10):2461--2480.

\bibitem[Evensen and van Leeuwen, 1996]{evensen96}
Evensen, G. and van Leeuwen, P.~J. (1996).
\newblock Assimilation of geostat altimeter data for the \mbox{A}gulhas
  \mbox{C}urrent using the ensemble \mbox{K}alman filter with a
  quasigeostrophic model.
\newblock {\em Monthly Weather Review}, 124:58--96.

\bibitem[Gordon et~al., 1993]{gordon93}
Gordon, N., Salmond, D.~J., and Smith, A. F.~M. (1993).
\newblock Novel approach to nonlinear/non-\mbox{G}aussian \mbox{B}ayesian state
  estimation.
\newblock {\em IEE Proceedings--F}, 140(2):107--113.

\bibitem[He et~al., 2010]{he10}
He, D., Ionides, E.~L., and King, A.~A. (2010).
\newblock Plug-and-play inference for disease dynamics: Measles in large and
  small towns as a case study.
\newblock {\em Journal of the Royal Society Interface}, 7:271--283.

\bibitem[Ionides et~al., 2017]{ionides17}
Ionides, E.~L., Breto, C., Park, J., Smith, R.~A., and King, A.~A. (2017).
\newblock Monte {C}arlo profile confidence intervals for dynamic systems.
\newblock {\em Journal of the Royal Society Interface}, 14:1--10.

\bibitem[Ionides et~al., 2015]{ionides15}
Ionides, E.~L., Nguyen, D., Atchad\'{e}, Y., Stoev, S., and King, A.~A. (2015).
\newblock Inference for dynamic and latent variable models via iterated,
  perturbed {B}ayes maps.
\newblock {\em Proceedings of the National Academy of Sciences of the USA},
  112(3):719–--724.

\bibitem[Jirak, 2016]{jirak16}
Jirak, M. (2016).
\newblock Berry--{E}sseen theorems under weak dependence.
\newblock {\em The Annals of Probability}, 44(3):2024--2063.

\bibitem[Katzfuss et~al., 2020]{katzfuss19}
Katzfuss, M., Stroud, J.~R., and Wikle, C.~K. (2020).
\newblock Ensemble {K}alman methods for high-dimensional hierarchical dynamic
  space-time models.
\newblock {\em Journal of the American Statistical Association},
  115(530):866--885.

\bibitem[Kevrekidis and Samaey, 2009]{kevrekidis09}
Kevrekidis, I.~G. and Samaey, G. (2009).
\newblock Equation-free multiscale computation: Algorithms and applications.
\newblock {\em Annual Review of Physical Chemistry}, 60:321--344.

\bibitem[King et~al., 2008]{king08}
King, A.~A., Ionides, E.~L., Pascual, M., and Bouma, M.~J. (2008).
\newblock Inapparent infections and cholera dynamics.
\newblock {\em Nature}, 454:877--880.

\bibitem[King et~al., 2016]{king16}
King, A.~A., Nguyen, D., and Ionides, E.~L. (2016).
\newblock Statistical inference for partially observed {M}arkov processes via
  the {R} package pomp.
\newblock {\em Journal of Statistical Software}, 69:1--43.

\bibitem[Lei et~al., 2010]{lei10}
Lei, J., Bickel, P., and Snyder, C. (2010).
\newblock Comparison of ensemble {K}alman filters under non-{G}aussianity.
\newblock {\em Monthly Weather Review}, 138(4):1293--1306.

\bibitem[Leutbecher and Palmer, 2008]{leutbecher08}
Leutbecher, M. and Palmer, T.~N. (2008).
\newblock Ensemble forecasting.
\newblock {\em Journal of Computational Physics}, 227(7):3515--3539.

\bibitem[Li et~al., 2020]{li20}
Li, R., Pei, S., Chen, B., Song, Y., Zhang, T., Yang, W., and Shaman, J.
  (2020).
\newblock Substantial undocumented infection facilitates the rapid
  dissemination of novel coronavirus ({SARS}-{CoV}-2).
\newblock {\em Science}, 368(6490):489--493.

\bibitem[Marino et~al., 2019]{marino18}
Marino, J.~A., Peacor, S.~D., Bunnell, D.~B., Vanderploeg, H.~A., Pothoven,
  S.~A., Elgin, A.~K., Bence, J.~R., Jiao, J., and Ionides, E.~L. (2019).
\newblock Evaluating consumptive and nonconsumptive predator effects on prey
  density using field times series data.
\newblock {\em Ecology}, 100(3):e02583.

\bibitem[Ng et~al., 2002]{ng02}
Ng, B., Peshkin, L., and Pfeffer, A. (2002).
\newblock Factored particles for scalable monitoring.
\newblock {\em Proceedings of the 18th Conference on Uncertainty and Artificial
  Intelligence}, pages 370--377.

\bibitem[Ning et~al., 2021]{ning21}
Ning, N., Ionides, E.~L., and Ritov, Y. (2021).
\newblock Scalable {M}onte {C}arlo inference and rescaled local asymptotic
  normality.
\newblock {\em Bernoulli}, pre-published online.

\bibitem[Palmer, 2002]{palmer02}
Palmer, T.~N. (2002).
\newblock The economic value of ensemble forecasts as a tool for risk
  assessment: From days to decades.
\newblock {\em Quarterly Journal of the Royal Meteorological Society},
  128(581):747--774.

\bibitem[Park and Ionides, 2019]{park19}
Park, J. and Ionides, E.~L. (2019).
\newblock Inference on high-dimensional implicit dynamic models using a guided
  intermediate resampling filter.
\newblock {\em Arxiv:1708.08543v3}.

\bibitem[Park and Ionides, 2020]{park20}
Park, J. and Ionides, E.~L. (2020).
\newblock Inference on high-dimensional implicit dynamic models using a guided
  intermediate resampling filter.
\newblock {\em Statistics \& Computing}, 30:1497--1522.

\bibitem[Pitt and Shepard, 1999]{pitt99}
Pitt, M.~K. and Shepard, N. (1999).
\newblock Filtering via simulation: Auxillary particle filters.
\newblock {\em Journal of the American Statistical Association}, 94:590--599.

\bibitem[Pons-Salort and Grassly, 2018]{pons-salort18}
Pons-Salort, M. and Grassly, N.~C. (2018).
\newblock Serotype-specific immunity explains the incidence of diseases caused
  by human enteroviruses.
\newblock {\em Science}, 361(6404):800--803.

\bibitem[Poterjoy, 2016]{poterjoy16}
Poterjoy, J. (2016).
\newblock A localized particle filter for high-dimensional nonlinear systems.
\newblock {\em Monthly Weather Review}, 144(1):59--76.

\bibitem[Rebeschini and van Handel, 2015]{rebeschini15}
Rebeschini, P. and van Handel, R. (2015).
\newblock Can local particle filters beat the curse of dimensionality?
\newblock {\em The Annals of Applied Probability}, 25(5):2809--2866.

\bibitem[Snyder et~al., 2015]{snyder15}
Snyder, C., Bengtsson, T., and Morzfeld, M. (2015).
\newblock Performance bounds for particle filters using the optimal proposal.
\newblock {\em Monthly Weather Review}, 143(11):4750--4761.

\bibitem[Xia et~al., 2004]{xia04}
Xia, Y., Bj{\o}rnstad, O.~N., and Grenfell, B.~T. (2004).
\newblock Measles metapopulation dynamics: A gravity model for epidemiological
  coupling and dynamics.
\newblock {\em American Naturalist}, 164(2):267--281.

\end{thebibliography}


\begin{thebibliography}{}

\bibitem[Ades and Van~Leeuwen, 2015]{ades15}
Ades, M. and Van~Leeuwen, P.~J. (2015).
\newblock The equivalent-weights particle filter in a high-dimensional system.
\newblock {\em Quarterly Journal of the Royal Meteorological Society},
  141(687):484--503.

\bibitem[Asfaw et~al., 2021]{asfaw21arxiv}
Asfaw, K., Park, J., Ho, A., King, A.~A., and Ionides, E.~L. (2021).
\newblock Statistical inference for spatiotemporal partially observed {M}arkov
  processes via the {R} package spatpomp.
\newblock {\em arXiv:2101.01157}.

\bibitem[Evensen and van Leeuwen, 1996]{evensen96}
Evensen, G. and van Leeuwen, P.~J. (1996).
\newblock Assimilation of geostat altimeter data for the \mbox{A}gulhas
  \mbox{C}urrent using the ensemble \mbox{K}alman filter with a
  quasigeostrophic model.
\newblock {\em Monthly Weather Review}, 124:58--96.

\bibitem[Ionides et~al., 2011]{ionides11}
Ionides, E.~L., Bhadra, A., Atchad\'{e}, Y., and King, A.~A. (2011).
\newblock Iterated filtering.
\newblock {\em Annals of Statistics}, 39:1776--1802.

\bibitem[Ionides et~al., 2006]{ionides06-pnas}
Ionides, E.~L., Bret\'{o}, C., and King, A.~A. (2006).
\newblock Inference for nonlinear dynamical systems.
\newblock {\em Proceedings of the National Academy of Sciences of the USA},
  103:18438--18443.

\bibitem[Ionides et~al., 2017]{ionides17}
Ionides, E.~L., Breto, C., Park, J., Smith, R.~A., and King, A.~A. (2017).
\newblock Monte {C}arlo profile confidence intervals for dynamic systems.
\newblock {\em Journal of the Royal Society Interface}, 14:1--10.

\bibitem[Ionides et~al., 2015]{ionides15}
Ionides, E.~L., Nguyen, D., Atchad\'{e}, Y., Stoev, S., and King, A.~A. (2015).
\newblock Inference for dynamic and latent variable models via iterated,
  perturbed {B}ayes maps.
\newblock {\em Proceedings of the National Academy of Sciences of the USA},
  112(3):719–--724.

\bibitem[Liu, 2001]{liu01}
Liu, J.~S. (2001).
\newblock {\em Monte \mbox{C}arlo Strategies in Scientific Computing}.
\newblock Springer, New York.

\bibitem[Lorenz, 1996]{lorenz96}
Lorenz, E.~N. (1996).
\newblock Predictability: A problem partly solved.
\newblock {\em Proceedings of the Seminar on Predictability}, 1:1--18.

\bibitem[Ning et~al., 2021]{ning21}
Ning, N., Ionides, E.~L., and Ritov, Y. (2021).
\newblock Scalable {M}onte {C}arlo inference and rescaled local asymptotic
  normality.
\newblock {\em Bernoulli}, pre-published online.

\bibitem[Park and Ionides, 2020]{park20}
Park, J. and Ionides, E.~L. (2020).
\newblock Inference on high-dimensional implicit dynamic models using a guided
  intermediate resampling filter.
\newblock {\em Statistics \& Computing}, 30:1497--1522.

\bibitem[van Kekem and Sterk, 2018]{vankekem18}
van Kekem, D.~L. and Sterk, A.~E. (2018).
\newblock Travelling waves and their bifurcations in the {L}orenz-96 model.
\newblock {\em Physica D: Nonlinear Phenomena}, 367:38--60.

\end{thebibliography}

\end{document}



\date{\today}
\title{Supplement to ``{\mytitle}''}
\author{E. L. Ionides, K. Asfaw, J. Park and A. A. King}

\newcommand{\blind}{1}

\if1\blind
{
\maketitle
}
\fi

\if0\blind
{
  \bigskip
  \bigskip
  \bigskip
  \begin{center}
    {\LARGE\bf \mytitle}
\end{center}
  \bigskip
  \bigskip
}\fi




\tableofcontents

\newpage

\section{\secTitleSpace A generalization to models without latent unit structure}

Variations of the algorithms in the main text apply when there is no latent unit structure. 
In this case, the observation vector $\myvec{Y}_{\time}=(Y_{1,\time},\dots,Y_{\Unit,\time})$ consists of a collection of measurements on a general latent vector ${X}_{\time}$.
We may have the structure that $Y_{1,\time},\dots,Y_{\Unit,\time}$ are conditionally independent given ${X}_{\time}$, but even this is not essential to the approach.
This is most readily seen in the context of the unadapted bagged filter, giving rise to the generalized unadapted bagged filter (G-{\UBF}) algorithm defined as follows.

\begin{center}
\noindent\begin{tabular}{l}
\hline
{\bf Algorithm~G-{\UBF} (Generalized unadapted bagged filter).}\rule[-1.5mm]{0mm}{6mm}\\
\hline
{\bf input:}\rule[-1.5mm]{0mm}{6mm} \\
Simulator for $f_{{X}_{\time}|{X}_{\time-1}}({x}_{\time}|{x}_{\time-1})$\\
Evaluator for $f_{Y_{\unit,\time}|{X}_{\time}}(\data{y}_{\unit,\time}\given {x}_{\time})$\\
Number of bootstrap filters, $\Rep$\\
Neighborhood structure, $B_{\unit,\time}$, for $\unit\in 1\mycolon\Unit$ and $\time\in 1\mycolon\Time$\\
Data, $\data{y}_{\unit,\time}$ for $\unit\in 1\mycolon\Unit$ and $\time\in 1\mycolon\Time$ \\
{\bf output:}\rule[-1.5mm]{0mm}{5mm} \\
Log likelihood estimate, $\MC{\loglik}= \sum_{\time=1}^\Time\sum_{\unit=1}^\Unit \MC{\loglik}_{\unit,\time}$  \rule[-2mm]{0mm}{3mm}
\\
\hline
For $\rep$ in $1\mycolon\Rep$ \rule[-1.5mm]{0mm}{6mm}
\\
\asp simulate ${X}_{\time,\rep}^{\simulation}$ from the dynamic model, for $\time\in 1{\mycolon}\Time$\\
End For\\
Prediction weights, 
$w^P_{\unit,\time,\rep}= f_{Y_{B_{\unit,\time}}|{X}_{1:\time}}(\data{y}_{B_{\unit,\time}}\given {X}^{\simulation}_{1:\time,\rep})$\\
Measurement weights,
$w^M_{\unit,\time,\rep}=f_{Y_{\unit,\time}|X_{\time},Y_{B_{\unit,\time}}}(\data{y}_{\unit,\time}\given X^{\simulation}_{\time,\rep},\data{y}_{B_{\unit,\time}})$
\\
Conditional log likelihood estimate,
$\MC{\loglik}_{\unit,\time}= 
\log \left(
  \sum_{\rep=1}^\Rep w^M_{\unit,\time,\rep}\,   w^P_{\unit,\time,\rep}
\right)
-
\log \left(
    \sum_{\tilde \rep=1}^\Rep w^P_{\unit,\time,\tilde \rep}
\right)
$
\\
\hline
\end{tabular}
\end{center}

The algorithm G-{\UBF} operates on an arbitrary POMP model. 
G-{\UBF} therefore provides a potential approach to extending methodologies from SpatPOMP models to models that have some similarity to a SpatPOMP without formally meeting the definition.
For example, there may be collections of interacting processes at different spatial scales in a spatiotemporal system. 
Alternatively, the potential outcomes of the latent process may vary between spatial units, such as when modeling interactions between terrestrial and aquatic ecosystems.
We do not further explore G-{\UBF} here.

\clearpage


\section{\secTitleSpace Adapted simulation for an Euler approximation}

\newcommand\nextn{n+1}
\newcommand\thisn{n}

We investigate the adapted simulation process by considering a continuous-time limit where it becomes a diffusion process.
We find that adapted simulation can effectively track the latent process when the measurement error is on an appropriate scale.
However, when the measurement error is large compared to the latent process noise, adapted simulation can fail in situations where filtering succeeds.
We work with a one-dimensional POMP model having a latent process constructed as an Euler approximation,
\begin{eqnarray}\label{x:euler}
X_{\nextn}&=& X_{\thisn} + \mu(X_{\thisn})\delta + \sigma \sqrt{\delta} \epsilon_{\nextn}, 
\end{eqnarray}
which provides a numerical solution to a one-dimensional stochastic differential equation,
\begin{eqnarray}
\nonumber
dX(t)&=& \mu\big(X(t)\big)\, dt + \sigma\, dU(t) ,
\end{eqnarray}
where $\{U(t)\}$ is a standard Brownian motion. 
We will consider several different measurement processes.


\subsection{Measurement error on the same scale as the process noise}
\label{subsec:m1}


\noindent Here, we consider the measurement model
\begin{equation} \label{m1}
Y_{\nextn}= \mu(X_{\thisn})\delta + \sigma \sqrt{\delta} \epsilon_{\nextn} + \tau\sqrt{\delta}\,\eta_{\nextn}.
\end{equation}
This is an approximation to the increment $Y(t+\delta)-Y(t)$ of a continuous time measurement model
\begin{equation}
\label{m1:cts}
dY(t) = dX(t) + \tau\, dV(t),
\end{equation}
where $\{V(t)\}$ is a standard Brownian motion independent of $\{U(t)\}$.
The measurement model (\ref{m1:cts}) makes inference on $X(t)$ given $Y(t)$ a continuous time version of the filtering problem.
A feature of this model is that $Y(t)$ does not directly track the level of the state, since the solution with initial conditions $Y(t_0)=X(t_0)$ and $V(t_0)=0$ is
\begin{equation}
\nonumber
Y(t) = X(t) + \tau V(t).
\end{equation}
The measurement error, $\tau V(t)$, has variance $\tau^2 t$ that increases with $t$. 
However, under appropriate conditions, information on changes in $\{X(t)\}$ obtained via $\{Y(t)\}$ are enough to track $X(t)$ indirectly via the filtering equations.
For the POMP given by (\ref{x:euler}) and (\ref{m1}), we can calculate exactly the adapted simulation distribution $f_{X_{\nextn}|Y_{\nextn},X_{\thisn}}$.
It is convenient to work conditionally on $X_{\thisn}$, allowing us to treat $X_{\thisn}$ and $\mu(X_{\thisn})$ as constants, with $X_{\nextn}$ and $Y_{\nextn}$ therefore being jointly normally distributed.
A Gaussian distribution calculation then gives the conditional moments.
First, we find
\begin{eqnarray*}
\E\big[ X_{\nextn}|Y_{\nextn},X_{\thisn} \big]
&=& 
X_{\thisn} + \mu(X_{\thisn})\delta + \E\big[{\sigma \sqrt{\delta} \epsilon_{\nextn} \, \big| \, \sigma \sqrt{\delta}\epsilon_{\nextn}+\tau{\sqrt{\delta}}\eta_{\nextn}}\big]
\\
&=&
X_{\thisn} + \mu(X_{\thisn})\delta + \frac{\sigma^2}{\sigma^2 +\tau^2}
\big(
  \sigma \sqrt{\delta}\epsilon_{\nextn}+\tau{\sqrt{\delta}}\eta_{\nextn}
\big)
\\
&=& X_{\thisn} + \mu(X_{\thisn})\delta + \frac{\sigma^2}{\sigma^2 +\tau^2} 
  \big( Y_{\nextn}- \mu(X_{\thisn})\delta \big).
\end{eqnarray*} 
Then,
\begin{eqnarray*}
\var\big[{X_{\nextn} \, \big| \, Y_{\nextn},X_{\thisn}}\big]
&=& 
\var\big[
   \sigma \sqrt{\delta} \epsilon_{\nextn} \, \big| \,
   \sigma \sqrt{\delta}\epsilon_{\nextn}+\tau{\sqrt{\delta}}\eta_{\nextn}
\big]
\\
&=& \sigma^2\delta - \frac{\sigma^4\delta^2}{\sigma^2\delta+\tau^2\delta}
\\
&=& \delta\frac{\sigma^2\tau^2}{\sigma^2+\tau^2}.
\end{eqnarray*}
Call the adapted simulation process $\{A_{\time},\time=1,2,\dots\}$, defined conditionally on $\{Y_{\time},\time=1,2,\dots\}$. 
We see from the above calculation that $A_{\time}$ can be constructed by the recursion
\begin{eqnarray*}
A_{\nextn} &=& A_{\thisn} + \mu(A_{\thisn})\delta +
  \frac{\sigma^2}{\sigma^2 +\tau^2} 
  \Big( 
    \mu(X_{\thisn})\delta + \sigma \sqrt{\delta} \, \epsilon_{\nextn}
    + \tau\sqrt{\delta} \, \eta_{\nextn}  -\mu(A_{\thisn})\, \delta
  \Big) 
\\
&& 
\hspace{4cm} + 
  \frac{\sigma\tau}{\sqrt{\sigma^2+\tau^2}} \sqrt{\delta} \, \zeta_{\nextn}
\end{eqnarray*}
where $\{\zeta_n\}$ is an iid standard normal sequence independent of $\{\epsilon_n,\eta_n\}$.
To study how well the adapted simulation tracks $\{X_n\}$, we subtract $X_{\nextn}$ from both sides to get
\begin{eqnarray*}
\hspace{-8mm} [A_{\nextn}-X_{\nextn}] &=& [A_{\thisn} -X_{\thisn}] + [\mu(A_{\thisn}) - \mu(X_{\thisn})]\delta - \sigma \sqrt{\delta} \, \epsilon_{\nextn} 
\\
&& \hspace{1mm}
+ 
  \frac{\sigma^2}{\sigma^2 +\tau^2} 
  \Big( 
     [\mu(X_{\thisn})-\mu(A_{\thisn})]\delta + \sigma \sqrt{\delta} \, \epsilon_{\nextn} + \tau\sqrt{\delta} \, \eta_{\nextn}  
   \Big) 
   + \frac{\sigma\tau}{\sqrt{\sigma^2+\tau^2}} \sqrt{\delta} \, \zeta_{\nextn}
\\
&=&   [A_{\thisn} -X_{\thisn}] +  \frac{2\sigma^2+\tau^2}{\sigma^2 +\tau^2}[\mu(X_{\thisn})-\mu(A_{\thisn})]\delta
\\
&& \hspace{3mm}
+   \frac{\sigma^2\tau\sqrt{\delta}\eta_{\nextn} - \sigma\tau^2\sqrt{\delta}\epsilon_{\nextn}}{\sigma^2 +\tau^2} 
   + \frac{\sigma\tau}{\sqrt{\sigma^2+\tau^2}} \sqrt{\delta} \, \zeta_{\nextn}.
\end{eqnarray*}
$A_{\time}$ tracks $X_{\time}$ when the process $\{A_{\time}-X_{\time},\time=1,2,\dots\}$ is stable.
This happens when $\mu(x)-\mu(y)$ is negative when $x$ is sufficiently larger than $y$. 
For example, a stable autoregressive process with $\mu(x)=-ax$ gives a stable adapted filter process.

\subsection{Independent measurement error on a scale that gives a finite limiting amount of information about $X(t)$ from measurements on a unit time interval}
\label{subsec:m2}

\noindent We now consider the measurement model
\begin{eqnarray} 
\nonumber
Y_{\nextn}&=& X_{\nextn}+ \frac{\tau}{\sqrt{\delta}}\, \eta_{\nextn}
\\
  &=& X_{\thisn} + \mu(X_{\thisn}) \, \delta + \sigma \sqrt{\delta}  \, \epsilon_{\nextn}+\frac{\tau}{\sqrt{\delta}}\eta_{\nextn},
\label{e0}
\end{eqnarray}
where $\{\epsilon_n,\eta_n\}$ is a collection of independent standard normal random variables.
The conditional mean is now
\begin{eqnarray}
\nonumber
\E\big[{X_{\nextn}|Y_{\nextn},X_{\thisn}}\big].
&=&
X_{\thisn} + \mu(X_{\thisn})\delta +
\E\Big[
  \sigma \sqrt{\delta} \epsilon_{\nextn}\,\Big| \, \sigma \sqrt{\delta}\epsilon_{\nextn}+\frac{\tau}{\sqrt{\delta}} \eta_{\nextn}\Big]
\\
&=&
X_{\thisn} + \mu(X_{\thisn})\delta + \frac{\sigma^2\delta}{\sigma^2\delta +\tau^2/\delta}
\big(
  \sigma \sqrt{\delta}\epsilon_{\nextn}+\tau{\sqrt{\delta}} \, \eta_{\nextn}
\big)
\label{m2.2}
\end{eqnarray} 
Using (\ref{e0}) and (\ref{m2.2}) gives
\begin{eqnarray*}
\E\big[{X_{\nextn}|Y_{\nextn},X_{\thisn}}\big] &=& 
X_{\thisn} + \mu(X_{\thisn})\delta + \frac{\sigma^2\delta^2}{\sigma^2\delta^2+\tau^2} 
\Big(
  Y_{\nextn}- X_{\thisn} - \mu(X_{\thisn})\, \delta
\Big).
\end{eqnarray*} 
In the limit as $\delta\to 0$, the contribution from the measurement is order $\delta^2$ and is therefore negligible.
Although the observation process is meaningfully informative about the latent process, the adapted simulation fails to track the latent process in this limit. 
Intuitively, this is because the adapted simulation is trying to track differences in the latent process, but for this model the signal to noise ratio for the difference in each interval of length $\delta$ tends to zero.

\subsection{Independent measurements of the latent process with measurement error on a scale that gives a useful adapted process as $\delta\to 0$}
\label{subsec:m3}

\noindent We now consider the measurement model
\begin{eqnarray} 
\nonumber
Y_{\nextn}&=& X_{\nextn}+ \tau\eta_{\nextn}
\\
  &=& X_{\thisn} + \mu(X_{\thisn})\delta + \sigma \sqrt{\delta} \epsilon_{\nextn}+\tau\eta_{\nextn}.
\label{m3:e0}
\end{eqnarray}
The conditional mean is now
\begin{eqnarray}
\nonumber
\E\big[{X_{\nextn}|Y_{\nextn},X_{\thisn}}\big].
&=&
X_{\thisn} + \mu(X_{\thisn})\, \delta +
\E\big[
  \sigma \sqrt{\delta} \epsilon_{\nextn}\,\big| \,
  \sigma \sqrt{\delta}\epsilon_{\nextn}+\tau \eta_{\nextn}
\big]
\\
&=&
X_{\thisn} + \mu(X_{\thisn})\delta + \frac{\sigma^2\delta}{\sigma^2\delta +\tau^2}
\big(
  \sigma \sqrt{\delta}\epsilon_{\nextn}+\tau \eta_{\nextn}
\big)
\label{m3.2}
\end{eqnarray} 
Using (\ref{m3:e0}) and (\ref{m3.2}) gives
\begin{eqnarray*}
\E\big[ X_{\nextn}|Y_{\nextn},X_{\thisn}\big] &=& 
X_{\thisn} + \mu(X_{\thisn})\delta + \frac{\sigma^2\delta}{\sigma^2\delta+\tau^2} 
\Big(
  Y_{\nextn}- X_{\thisn} - \mu(X_{\thisn})\delta
  \Big)
\\
&=& X_n + \mu(X_n)\delta + \frac{\sigma^2}{\tau^2} \delta
\Big(
  Y_{\nextn}- X_{\thisn} - \mu(X_{\thisn}) \, \delta
  \Big)
+ o(\delta)
\end{eqnarray*} 
In the limit as $\delta\to 0$, the adapted simulation has a diffusive drift toward the value of the latent process.

For disease models, incidence data can arguably be considered as noisy measurements of the change of a state variable (number of susceptibles) that is not directly measured. 
This could correspond to a situation where the measurement error is on the same scale as the process noise (Subsection~\ref{subsec:m1}). 
Alternatively, we could think of weekly aggregated incidence as a noisy measurement of the infected class, in which case the measurement error could match the scaling in Subsection~\ref{subsec:m3}.

The model in Subsection~\ref{subsec:m2} is a cautionary tale, warning us against carrying out adapted simulation on short time intervals.
An interpretation is that one should not carry out adapted simulation unless a reasonable amount of information has accrued.
When each observation has low information, a particle filter may enable solution to the filtering problem without particle depletion.
It is when the data are highly informative that the curse of dimensionality makes basic particle filters ineffective, opening up demand for alternative methods.

We are now in a better position to understand why it may be appropriate to keep many particle representations at intermediate timesteps while resampling down to a single representative at each observation time, as {\ABF} and {\ABFIR} do.
We have seen that adaptive simulation can fail when observations occur frequently.
Resampling down to a single particle too often can lose the ability for the adapted process to track the latent process.
This implies that adapted simulation should not be relied upon more than necessary to ameliorate the curse of dimensionality: once proper importance sampling for filtering problem becomes tractable in a sufficiently small spatiotemporal neighborhood, one should maintain weighted particles on this spatiotemporal scale rather than resorting to adapted simulation.


\section{\secTitleSpace {\UBF} convergence: Proof of Theorem~1}

We consider a collection of models $f_{\myvec{X}_{0:\Time},\myvec{Y}_{1:\Time}}$ and data $\data{\myvec{y}}_{1:\Time}$ defined for each $(\Unit,\Time)$.
These models and datasets are not required to have any nesting relationship, so we do not insist that $X_{1,1}$ or $\data{y}_{1,1}$ should be the same for $(\Unit,\Time)=(10,10)$ as for $(\Unit,\Time)=(100,100)$.
Formally, we define probability and expectation on a product space of the stochastic model and Monte Carlo outcomes. 
Monte Carlo quantities such as the output of the {\UBF}, {\ABF} and {\ABFIR} algorithms depend on the data but not on any random variables constructed under the model.
As a consequence, we can use $\E$ to correspond both to expectations over Monte Carlo stochasticity (for Monte Carlo quantities) and model stochasticity (for random variables constructed under the model).
We suppress discussion of measurability by assuming that all functions considered have appropriate measurability properties.
We restate the assumptions and statement for Theorem~1.

\Ai  

We use the total variation bound in Assumption~~\ref{A1} via the following Proposition~\ref{weak_coupling:lemma}, which replaces conditioning on $X_{B^c_{\unit,\time}}$ with conditioning on $Y_{B^c_{\unit,\time}}$. The bound in \myeqref{eq:implication:of:A1} could be used in place of Assumption~\ref{A1}.

\begin{prop}\label{weak_coupling:lemma}
Under the conditions of Assumption~\ref{A1}, \myeqref{eq:weak_coupling2} implies
\begin{equation}\label{eq:implication:of:A1}
\bigg| \int h(x_{\unit,\time}) f_{X_{\unit,\time}|Y_{A_{\unit,\time}}}(x_{\unit,\time}\given \data{y}_{A_{\unit,\time}}) \, dx_{\unit,\time}
- \int h(x_{\unit,\time})f_{X_{\unit,\time}|Y_{B_{\unit,\time}}}(x_{\unit,\time}\given \data{y}_{B_{\unit,\time}})
\, dx_{\unit,\time} \, 
\bigg| < \eone
\end{equation}
\end{prop}
\begin{proof} For notational compactness, we suppress the arguments $x_{\unit,\time}$, $x_{B^c_{\unit,\time}}$, $\data{y}_{A_{\unit,\time}}$, $\data{y}_{B_{\unit,\time}}$ matching the subscripts of conditional densities.
Using the conditional independence of the measurements given the latent process, we calculate
\begin{eqnarray}\nonumber
&& \hspace{-15mm}\bigg| \int h(x_{\unit,\time}) f_{X_{\unit,\time}|Y_{A_{\unit,\time}}} \, dx_{\unit,\time}
- \int h(x_{\unit,\time})f_{X_{\unit,\time}|Y_{B_{\unit,\time}}} \, dx_{\unit,\time}
\bigg|
\\
\nonumber
&=& \bigg| \int  \bigg\{
{\mediumint}  h(x_{\unit,\time}) f_{X^{}_{\unit,\time}|Y^{}_{B_{\unit,\time}},X_{B^c_{\unit,\time}
}} \, dx_{\unit,\time} -  {\mediumint}
 h(x_{\unit,\time})f_{X^{}_{\unit,\time}|Y^{}_{B_{\unit,\time}}}\, dx_{\unit,\time} \bigg\} \, 
 f_{X_{B^c_{\unit,\time}} \given Y_{A^{}_{\unit,\time}}}\, dx^{}_{B^c_{\unit,\time}} \bigg| 
\nonumber
\\
\nonumber
&\le& \int  \bigg| \,
{\mediumint}  h(x_{\unit,\time}) f_{X_{\unit,\time}|Y_{B_{\unit,\time}},X_{B^c_{\unit,\time}
}} \, dx_{\unit,\time} -  {\mediumint}
 h(x_{\unit,\time})f_{X_{\unit,\time}|Y_{B_{\unit,\time}}}\, dx_{\unit,\time} \, \bigg| \, 
 f_{X_{B^c_{\unit,\time}} \given Y_{A^{}_{\unit,\time}}}\, dx^{}_{B^c_{\unit,\time}} 
\\
\nonumber
& < & \int \eone \, f_{X_{B^c_{\unit,\time}} \given Y^{}_{A_{\unit,\time}}}\, dx^{}_{B^c_{\unit,\time}} \hspace{3mm} = \hspace{3mm} \eone.
\end{eqnarray}
\end{proof}

\Aii  

\Aiii   

\Aiv    


\newcommand\extraSpace{\hspace{1mm}}

Assumption~\ref{A:unconditional:mix} is needed only to ensure that the variance bound in Theorem~\ref{thm:tif} is essentially $O(\Unit \Time)$ rather than $O(\Unit^2 \Time^2)$. 
Both these rates avoid the exponentially increasing variance characterizing the curse of dimensionality.
Lower variance than $O(\Unit \Time)$ cannot be anticipated for any sequential Monte Carlo method since the log likelihood estimate can be written as a sum of $\Unit\Time$ terms each of which involves its own sequential Monte Carlo calculation.

\TheoremI 

\begin{proof}
Suppose the quantities $w^M_{\unit,\time,\rep}$ and $w^P_{\unit,\time,\rep}$ constructed in Algorithm~{\UBF} are considered i.i.d.\ replicates of jointly defined random variables $w^M_{\unit,\time}$ and $w^P_{\unit,\time}$, for each $(\unit,\time)\in \spaceTime$.
Also, write 
\begin{equation}
\nonumber
\centersum^{MP}_{\unit,\time}= \frac{1}{\sqrt{\Rep}}\sum_{\rep=1}^\Rep\big(w^M_{\unit,\time,\rep}w^P_{\unit,\time,\rep} - \EMC[w^M_{\unit,\time}w^P_{\unit,\time}]\big),
\hspace{2cm}
\centersum^{P}_{\unit,\time}= \frac{1}{\sqrt{\Rep}}\sum_{\rep=1}^\Rep\big(w^P_{\unit,\time,\rep} - \EMC[w^P_{\unit,\time}]\big),
\end{equation}
Then, using the delta method (e.g., Section 2.5.3 in \cite{liu01}) we find
\begin{eqnarray}
\nonumber
\MC{\loglik}_{\unit,\time} &=& \log\left( \frac{\sum_{\rep=1}^\Rep w^M_{\unit,\time,\rep}w^P_{\unit,\time,\rep}}{\sum_{\rep=1}^\Rep w^P_{\unit,\time,\rep}}\right)
\\
\nonumber
&=&  \log\Big(\EMC[w^M_{\unit,\time}w^P_{\unit,\time}] + \Rep^{-1/2}\centersum^{MP}_{\unit,\time}\Big) - \log\Big({\EMC[w^P_{\unit,\time}] + \Rep^{-1/2} \centersum^P_{\unit,\time}}\Big)
\\
\label{eq:likelihood:at:point}
&=& \log\left(\frac{\EMC[w^M_{\unit,\time}w^P_{\unit,\time}]}{\EMC[w^P_{\unit,\time}]}\right) +  \Rep^{-1/2}
\left(\frac{\centersum^{MP}_{\unit,\time}}{\EMC[w^M_{\unit,\time}w^P_{\unit,\time}]} - \frac{\centersum^{P}_{\unit,\time}}{\EMC[w^P_{\unit,\time}]} \right) + o_P\big(\Rep^{-1/2}\big)
\end{eqnarray}
The joint distribution of $\big\{ (\centersum^{MP}_{\unit,\time},\centersum^{P}_{\unit,\time}), (\unit,\time)\in \spaceTime \big\}$ follows a standard central limit theorem as $\Rep\to\infty$.
Each term has mean zero, with covariances uniformly bounded over $(\unit,\time,\altUnit,\altTime)$ due to Assumption~\ref{A2}.
Specifically,
\begin{equation}
\nonumber
\hspace{-3mm}
\var
\hspace{-1mm}
\left(
\hspace{-2mm}
\begin{array}{c}
\centersum^{MP}_{\unit,\time} \\
\centersum^{P}_{\unit,\time} \\
\centersum^{MP}_{\altUnit,\altTime} \\
\centersum^{P}_{\altUnit,\altTime} \\
\end{array} 
\hspace{-2mm}
\right)
\hspace{-1mm}
{=}
\hspace{-1mm}
\left(
\hspace{-2mm}
\begin{array}{cccc}
\var(w^{M}_{\unit,\time}w^{P}_{\unit,\time}) 
& \cov(w^{M}_{\unit,\time}w^{P}_{\unit,\time},w^{P}_{\unit,\time}) 
\hspace{-2mm}
& \cov(w^{M}_{\unit,\time}w^{P}_{\unit,\time},w^{M}_{\altUnit,\altTime}w^{P}_{\altUnit,\altTime}) 
\hspace{-2mm}
& \cov(w^{M}_{\unit,\time}w^{P}_{\unit,\time},w^{P}_{\altUnit,\altTime}) 
\\
\cov(w^{M}_{\unit,\time}w^{P}_{\unit,\time},w^{P}_{\unit,\time}) 
& \var(w^{P}_{\unit,\time}) 
& \cov(w^{P}_{\unit,\time},w^{M}_{\altUnit,\altTime}w^{P}_{\altUnit,\altTime}) 
& \cov(w^{P}_{\unit,\time},w^{P}_{\altUnit,\altTime}) 
\\
\cov(w^{M}_{\unit,\time}w^{P}_{\unit,\time},w^{M}_{\altUnit,\altTime}w^{P}_{\altUnit,\altTime}) 
\hspace{-2mm}
& \cov(w^{P}_{\unit,\time},w^{M}_{\altUnit,\altTime}w^{P}_{\altUnit,\altTime}) 
& \var(w^{M}_{\altUnit,\altTime}w^{P}_{\altUnit,\altTime}) 
& \cov(w^{M}_{\altUnit,\altTime}w^{P}_{\altUnit,\altTime},w^{P}_{\altUnit,\altTime}) 
\\
\cov(w^{M}_{\unit,\time}w^{P}_{\unit,\time},w^{P}_{\altUnit,\altTime}) 
& \cov(w^{P}_{\unit,\time},w^{P}_{\altUnit,\altTime}) 
& \cov(w^{M}_{\altUnit,\altTime}w^{P}_{\altUnit,\altTime},w^{P}_{\altUnit,\altTime}) 
& \var(w^{P}_{\altUnit,\altTime})
\end{array}
\hspace{-2mm}
\right)
\end{equation}
Note that
\begin{eqnarray*}
\nonumber
\log\left[ 
  \frac{\EMC[w^{M}_{\unit,\time}w^{P}_{\unit,\time}]}{\EMC[w^{P}_{\unit,\time}]} 
\right]
&=&\log \left[
\frac{
\int f_{Y_{\unit,\time}|X_{\unit,\time}}(\data{y}_{\unit,\time}\given x_{\unit,\time,\rep})\, f_{Y_{B_{\unit,\time}}|X_{B_{\unit,\time}}}(\data{y}_{B_{\unit,\time}}\given x_{B_{\unit,\time}})\,
f_{X_{B^+_{\unit,\time}}}(x_{B^+_{\unit,\time}}) \, dx_{B^+_{\unit,\time}}
}{
\int f_{Y_{B_{\unit,\time}}|X_{B_{\unit,\time}}}(\data{y}_{B_{\unit,\time}}\given x_{B_{\unit,\time}})\,
f_{X_{B_{\unit,\time}}}(x_{B_{\unit,\time}}) \, dx_{B_{\unit,\time}}
}
\right]
\\
\nonumber
&=& \log \big[ 
  f_{Y_{\unit,\time}|Y_{B_{\unit,\time}}}(\data{y}_{\unit,\time}\given \data{y}_{B_{\unit,\time}})
\big],
\end{eqnarray*}
where $B^+_{\unit,\time}=B_{\unit,\time}\cup (\unit,\time)$. Now, define
\begin{equation}
\nonumber
\centersum^{\loglik}_{\unit,\time}=
\left(\frac{\centersum^{MP}_{\unit,\time}}{\EMC[w^M_{\unit,\time}w^P_{\unit,\time}]} - \frac{\centersum^{P}_{\unit,\time}}{\EMC[w^P_{\unit,\time}]} \right)
\end{equation}
Summing over all $({\unit,\time})\in \spaceTime$, we get
\begin{equation}
\label{thm1:loglik:linearization}
\sqrt{\Rep}\left( \MC{\loglik} - \sum_{({\unit,\time})\in\, \spaceTime}
\log f_{Y_{\unit,\time}|Y_{B_{\unit,\time}}}(\data{y}_{\unit,\time}\given \data{y}_{B_{\unit,\time}})
\right)
= \sum_{({\unit,\time})\in \,  \spaceTime} \centersum^{\loglik}_{\unit,\time} + o(1).
\end{equation}
Now,
\begin{equation}
\nonumber
\cov\big(\centersum^{\loglik}_{\unit,\time},\centersum^{\loglik}_{\altUnit,\altTime}\big)
=\cov\left(
  \frac{w^M_{\unit,\time}w^P_{\unit,\time}}{\EMC[w^M_{\unit,\time}w^P_{\unit,\time}]}
  - \frac{w^P_{\unit,\time}}{\EMC[w^P_{\unit,\time}]},
  \frac{w^M_{\altUnit,\altTime}w^P_{\altUnit,\altTime}}{\EMC[w^M_{\altUnit,\altTime}w^P_{\altUnit,\altTime}]}
  - \frac{w^P_{\altUnit,\altTime}}{\EMC[w^P_{\altUnit,\altTime}]}
\right).
\end{equation}
Since 
\begin{equation}
\nonumber
\left|  \frac{w^M_{\unit,\time}w^P_{\unit,\time}}{\EMC[w^M_{\unit,\time}w^P_{\unit,\time}]}
  - \frac{w^P_{\unit,\time}}{\EMC[w^P_{\unit,\time}]} \right| <Q^{2\Bsize},
\end{equation}
we have 
\begin{equation}
\nonumber
\big| \cov\big(\centersum^{\loglik}_{\unit,\time},\centersum^{\loglik}_{\altUnit,\altTime}\big)
\big| < Q^{4\Bsize},
\end{equation}
implying that
\begin{equation}
\label{var:bound:without:assumption}
\var\left(\sum_{({\unit,\time})\in \spaceTime} \centersum^{\loglik}_{\unit,\time}\right)
< Q^{4\Bsize}\Unit^2\Time^2.
\end{equation}
If, in addition, $(\unit,\time)$ and $(\altUnit,\altTime)\in A_{\unit,\time}$ are sufficiently separated in the sense of Assumption~\ref{A:unconditional:mix}, then Lemma~\ref{lemma:covariance-bound} shows that Assumption~\ref{A:unconditional:mix} implies
\begin{equation}
\nonumber
\big| \cov\big(\centersum^{\loglik}_{\unit,\time},\centersum^{\loglik}_{\altUnit,\altTime}\big)
\big| < \etwo Q^{4\Bsize}.
\end{equation}
The number of insufficiently separated neighbors to $(\unit,\time)$ is bounded by $\altb$, and so we obtain
\begin{equation}
\label{var:bound:with:assumption}
\var\left(\sum_{({\unit,\time})\in S} \centersum^{\loglik}_{\unit,\time}\right)
< \ThmOneVarBound.
\end{equation}
Now we proceed to bound the bias in the Monte Carlo central limit estimator of $\loglik$.
Putting $h(x_{\unit,\time})=f_{Y_{\unit,\time}|X_{\unit,\time}}(\data{y}_{\unit,\time}\given x_{\unit,\time})$ into Assumption~\ref{A1}, using Assumption~\ref{A2}, gives
\begin{equation}
\nonumber
\big|
f_{Y_{\unit,\time}|Y_{B_{\unit,\time}}}(\data{y}_{\unit,\time}\given \data{y}_{B_{\unit,\time}})
- f_{Y_{\unit,\time}|Y_{A_{\unit,\time}}}(\data{y}_{\unit,\time}\given \data{y}_{A_{\unit,\time}})
\big|
< \eone Q.
\end{equation}
Noting that 
\begin{equation}
\label{LogInequality}
|a-b|<\delta, \hspace{3mm} a>Q^{-1} \mbox{ and } b>Q^{-1} \mbox{ implies }|\log(a)-\log(b)| < \delta Q, 
\end{equation}
we find
\begin{equation}
\label{eq:log:diff:identity}
\big|
\log f_{Y_{\unit,\time}|Y_{B_{\unit,\time}}}(\data{y}_{\unit,\time}\given \data{y}_{B_{\unit,\time}})
- \log f_{Y_{\unit,\time}|Y_{A_{\unit,\time}}}(\data{y}_{\unit,\time}\given \data{y}_{A_{\unit,\time}})
\big|
< \eone Q^2.
\end{equation}
Summing over $(\unit,\time)$, we get
\begin{equation}
\label{thm1:bias:bound}
\left| \loglik - 
\sum_{({\unit,\time})\in S}
\log f_{Y_{\unit,\time}|Y_{B_{\unit,\time}}}(\data{y}_{\unit,\time}\given \data{y}_{B_{\unit,\time}})
\right| 
< \eone Q^2 \Unit\Time.
\end{equation}
Together, the results in \myeqref{thm1:loglik:linearization}, \myeqref{var:bound:without:assumption}, \myeqref{var:bound:with:assumption} and \myeqref{thm1:bias:bound} confirm the assertions of the theorem.
\end{proof}


\newcommand\Xlemma{U}
\newcommand\Ylemma{V}
\newcommand\xlemma{u}
\newcommand\ylemma{v}

\begin{lemma}\label{lemma:covariance-bound}
Suppose $\Xlemma$ and $\Ylemma$ are random variables with joint density satisfying 
\begin{equation}
\big| f_{\Ylemma|\Xlemma}(\ylemma\given \xlemma)-f_{\Ylemma}(\ylemma) \big| < \epsilon f_{\Ylemma}(\ylemma).
\end{equation}
Suppose $|g(\Xlemma)| < a$ and $|h(\Ylemma)| < b$ for some real-valued function $g$ and $h$.
Then, $\cov\big(g(\Xlemma),h(\Ylemma)\big) <  ab\epsilon$.
\end{lemma}

\begin{proof}
The result is obtained by direct calculation, as follows.
\begin{eqnarray}
\nonumber
\cov\big(g(\Xlemma),h(\Ylemma)\big) &=& \E\bigg[ g(\Xlemma) \, \E\Big[ h(\Ylemma)-\E[h(\Ylemma)] \, \Big| \,  \Xlemma\Big] \bigg]
\\
\nonumber
&=& \int \left\{ \int g(\xlemma) h(\ylemma) \big( f_{\Ylemma|\Xlemma}(\ylemma \given \xlemma)-f_{\Ylemma}(\ylemma)\big)\, d{\ylemma} \right\} f_{\Xlemma}(\xlemma)\, d{\xlemma}
\\
\nonumber
&<& \int \left\{ \int a b \epsilon f_{\Ylemma}(\ylemma) \, d{\ylemma} \right\} f_{\Xlemma}(\xlemma)\, d{\xlemma}
\\
\nonumber
&=& ab\epsilon.
\end{eqnarray}
\end{proof}


\section{\secTitleSpace {\ABF} and {\ABFIR} convergence: Proof of Theorem~2}

Let 
$g^{}_{\myvec{X}_{0:\Time},\myvec{X}^P_{1:\Time}}(\myvec{x}_{0:\Time},\myvec{x}^P_{1:\Time})$ 
be the joint density of the adapted process and the proposal process, 
\begin{equation}
\label{eq:g}
g^{}_{\myvec{X}_{0:\Time},\myvec{X}^P_{1:\Time}}(\myvec{x}_{0:\Time},\myvec{x}^P_{1:\Time})
= f_{\myvec{X}_0}(\myvec{x}_0)
\prod_{\time=1}^{\Time} 
f_{\myvec{X}_{\time}|\myvec{X}_{\time-1},\myvec{Y}_{\time}} 
  \big( 
    \myvec{x}_{\time} \given \myvec{x}_{\time-1},\data{\myvec{y}}_{\time} 
  \big)
\,\,
f_{\myvec{X}_{\time}|\myvec{X}_{\time-1}}
  \big(
    \myvec{x}^P_{\time} \given \myvec{x}_{\time-1}
  \big).
\end{equation}
For 
$\unitTimeSubset \subset \UnitSet\times 1:\Time$, 
define 
$\unitTimeSubset^{[m]}=\unitTimeSubset \cap \big(\UnitSet\times \{m\}\big)$ 
and set
\begin{equation}
\label{eq:h_S}
\adapted^{}_{\unitTimeSubset}
= 
\prod_{m=1}^{\Time} f_{Y_{\unitTimeSubset^{[m]}}|\myvec{X}^{}_{m-1}} \big( \data{y}_{\unitTimeSubset^{[m]}} \given \myvec{X}^{}_{m-1} \big),
\end{equation}
using the convention that an empty density $f_{Y_{\emptyset}}$ evaluates to 1.
If we denoting $\E_{g}$ for expectation for $(\myvec{X}_{0:\Time},\myvec{X}^P_{1:\Time})$ having density $g_{\myvec{X}_{0:\Time},\myvec{X}^P_{1:\Time}}$, \eqref{eq:h_S} can be written as
\begin{equation}
\label{eq:h_S2}
\nonumber
\adapted^{}_{\unitTimeSubset}
= 
\E_g\left[
f_{Y_{\unitTimeSubset}|{X}^{}_{\unitTimeSubset}} 
\big( 
  \data{y}_{\unitTimeSubset} \given {X}^{P}_{\unitTimeSubset}
\big)
\, \Big| \, 
\myvec{X}^{}_{0:\Time}
\right],
\end{equation}
so we have
\[
\E_g \big[ \gamma_{B} \big] = \E_g \left[ f_{Y_{B}|X_{B}} \big(\data y_{B}| X_{B}^P \big) \right].
\]
Two useful identities are
\begin{eqnarray}
\nonumber
f_{X^{}_{\unit,\time}|Y_{A_{\unit,\time}}} 
\big( 
  x^{}_{\unit,\time}|\data{y}_{A_{\unit,\time}} 
\big) 
&=&
\frac{
  \E_{g}\Big[
    f_{Y^{}_{A_{\unit,\time}}|X^{}_{A_{\unit,\time}}} \big( \data y_{A_{\unit,\time}} \given X^P_{A^{}_{\unit,\time}} \big) \, 
    f^{}_{X_{\unit,\time}|X^{[\time]}_{A_{\unit,\time}},\myvec{X}_{\time-1}}\big( x^{}_{\unit,\time}|X^P_{A^{[\time]}_{\unit,\time}},\myvec{X}_{\time-1} \big)
  \Big]
}{
  \E_{g}\Big[
    f^{}_{Y_{A_{\unit,\time}}|X_{A_{\unit,\time}}}\big( \data{y}_{A_{\unit,\time}}|X^P_{A_{\unit,\time}} \big)
  \Big]
},
\\
\nonumber
f_{Y^{}_{\unit,\time}|Y_{A_{\unit,\time}}} \big( \data{y}_{\unit,\time}|\data{y}_{A_{\unit,\time}} \big) &=&
\frac{\E_{g}\big[\adapted^{}_{A^+_{\unit,\time}} 
\big]
}{\E_{g}\big[\adapted^{}_{A_{\unit,\time}} 
\big]}.
\end{eqnarray}

\Bi 

\Bii 

\Biii 

\begin{prop}\label{prop:ABF_AB}
Setting $h(x)= f_{Y_{\unit,\time}|X_{\unit,\time}}(\data{y}_{\unit,\time}|x)$, assumptions~\ref{B1} and~\ref{B2} imply
\begin{equation}
\Bigg| \,
\frac{\E_{g}\big[\adapted^{}_{A^+_{\unit,\time}}\big]}{\E_{g}\big[\adapted^{}_{A_{\unit,\time}}]}
- 
\frac{\E_{g}\big[\adapted^{}_{B^+_{\unit,\time}}\big]}{\E_{g}\big[\adapted^{}_{B_{\unit,\time}}]}
\,
\Bigg|
< Q \epsilon.
\end{equation}
\end{prop}
\begin{proof}
Using the non-negativity of all terms to justify interchange of integral and expectation,
\begin{equation}
\begin{split}
 & \int h(x) \E_g \Big[ f_{Y_{B_{u,n}}|X_{B_{u,n}}}(\data y_{B_{u,n}} | X_{B_{u,n}}^P) f_{X_{u,n}|X_{B_{u,n}^{[n]}},\myvec X_{n-1}}(x | X_{B_{u,n}^{[n]}}^P, \myvec X_{n-1}) \Big] dx \\
 & = \E_g \Big[ \int f_{Y_{u,n}|X_{u,n}}(\data y_{u,n}|x) f_{X_{u,n}|X_{B_{u,n}^{[n]}},\myvec X_{n-1}}(x | X_{B_{u,n}^{[n]}}^P, \myvec X_{n-1}) dx \cdot f_{Y_{B_{u,n}}|X_{B_{u,n}}}(\data y_{B_{u,n}}|X_{B_{u,n}}^P) \Big]
\end{split}
\label{eq:int_h_fubini}
\end{equation}
But by the construction of $g$,
\[
\begin{split}
f_{X_{u,n}|X_{B_{u,n}^{[n]}}, \myvec X_{n-1}} (x| X_{B_{u,n}^{[n]}}^P, \myvec X_{n-1})
&= g_{X_{u,n}^P|X_{B_{u,n}^{[n]}}^P, \myvec X_{n-1}}(x | X_{B_{u,n}^{[n]}}^P, \myvec X_{n-1}) \\
&= g_{X_{u,n}^P|X_{B_{u,n}}^P, \myvec X_{n-1}} (x|X_{B_{u,n}}^P, \myvec X_{n-1}).
\end{split}
\]
Thus \eqref{eq:int_h_fubini} becomes
\[
\begin{split}
  &\E_g \Big[ \int f_{Y_{u,n}|X_{u,n}}(\data y_{u,n}|x) g_{X_{u,n}^P|X_{B_{u,n}^P}, \myvec X_{n-1}} (x|X_{B_{u,n}}^P, \myvec X_{n-1}) dx \cdot f_{Y_{B_{u,n}}|X_{B_{u,n}}}(\data y_{B_{u,n}}|X_{B_{u,n}}^P) \Big]\\
  &= \E_g \left[ \E_g \Big[ f_{Y_{u,n}|X_{u,n}}(\data y_{u,n} | X_{u,n}^P) | X_{B_{u,n}}^P, \myvec X_{n-1} \Big] \cdot f_{Y_{B_{u,n}}|X_{B_{u,n}}}(\data y_{B_{u,n}} | X_{B_{u,n}}^P) \right]\\
  &= \E_g \Big[ f_{Y_{B_{u,n}^+} | X_{B_{u,n}^+}} (\data y_{B_{u,n}^+} | X_{B_{u,n}^+}^P) \Big]\\
  &= \E_g \gamma_{B_{u,n}^+}.
\end{split}
\]
Applying the same argument for the special case of $B_{u,n}=A_{u,n}$, we substitute into Assumption~\ref{B1} to complete the proof with the fact that $h<Q$.
\end{proof}

\BivA 

The mixing of the adapted process in Assumption~\ref{B:unconditional:mix} replaces the mixing of the unconditional process in  Assumption~\ref{A:unconditional:mix}.
Though mixing of the adapted process may be hard to check, one may suspect that the adapted process typically mixes more rapidly than the unconditional process.
Assumption~\ref{B:unconditional:mix} is needed only to ensure that the variance bound in Theorem~\ref{thm:abf} is essentially $O(\Unit \Time)$ rather than $O(\Unit^2 \Time^2)$. 
Either of these rates avoids the exponentially increasing variance characterizing the curse of dimensionality.
Lower variance than $O(\Unit \Time)$ cannot be anticipated for any sequential Monte Carlo method since the log likelihood estimate can be written as a sum of $\Unit\Time$ terms each of which involves its own sequential Monte Carlo calculation.
The following Proposition~\ref{prop:cov:mix} gives an implication of Assumption~\ref{B:unconditional:mix}.


\begin{prop} \label{prop:cov:mix}
  Assumption~\ref{B:unconditional:mix} implies that, if $(\altUnit, \altTime) \notin \altB_{\unit\comma\time}$,
\begin{equation}
\label{supp:eq:cov:g:bound}
\cov_{g_{}}\big(\adapted^{}_{B^{}_{\unit,\time}},\adapted^{}_{B^{}_{\altUnit\comma\altTime}} \big) < \efourA Q^{|B^{}_{\unit,\time}|+|B^{}_{\altUnit\comma\altTime}|}.
\end{equation}
\end{prop}
\begin{proof}
Write 
$\gamma=\adapted^{}_{B^{}_{\unit,\time}}
$ and 
$\tilde\gamma=\adapted^{}_{B^{}_{\altUnit\comma\altTime}}
$. 
Also, write $B=B^{}_{\unit,\time}$, $\tilde B=B^{}_{\altUnit,\altTime}$ and $f^{}_{B}(x^P_B)=f^{}_{Y_B|X_B}(\data{y}_B\given x_B^P)$.
Then,
\begin{eqnarray}
\nonumber
\hspace{-5mm} \E[\gamma\tilde\gamma]&=&
\int
  \left[
    \int f^{}_{B}(x^P_B)
         g^{}_{X^P_B|\myvec{X}_{0:\Time}}(x^P_B\given \myvec{x}_{0:\Time})\, dx^P_B
  \right]
\\
\nonumber
&& \hspace{15mm} 
\times
  \left[
    \int f^{}_{{\tilde B}}(x^P_{\tilde B})
         g^{}_{X^P_{\tilde B}|\myvec{X}_{0:\Time}}(x^P_{\tilde B}\given \myvec{x}_{0:\Time})\, dx^P_{\tilde B}
  \right]
  g^{}_{\myvec{X}_{0:\Time}}(\myvec{x}_{0:\Time})\, d\myvec{x}_{0:\Time}
\\
\nonumber
&=& \int \int f^{}_{B}(x^P_B)\, f^{}_{{\tilde B}}(x^P_{\tilde B})\,  \bigg\{ \int g^{}_{X^P_B|\myvec{X}_{0:\Time}}(x^P_B\given \myvec{x}_{0:\Time}) 
\\
\label{eq:prop:long:integral}
&& \hspace{15mm} 
\times \, 
g^{}_{X^P_{\tilde B}|\myvec{X}_{0:\Time}}(x^P_{\tilde B}\given \myvec{x}_{0:\Time}) \, g^{}_{\myvec{X}_{0:\Time}}(\myvec{x}_{0:\Time}) \, d\myvec{x}_{0:\Time} \bigg\}
\, dx^P_B \, dx^P_{\tilde B} 
\end{eqnarray}
Putting the approximate conditional independence requirement of Assumption~\ref{B:unconditional:mix} into \myeqref{eq:prop:long:integral}, we have
\begin{eqnarray}
\nonumber
&& \hspace{-15mm}
\Big|  
  \E[\gamma\tilde\gamma] -
  \int f^{}_{B}(x^P_B)\, f^{}_{{\tilde B}}(x^P_{\tilde B}) \, 
         g^{}_{X^P_{B}X^P_{\tilde B}|\myvec{X}_{0:\Time}}(x^{P}_{B},x^P_{\tilde B}\given \myvec{x}_{0:\Time})
\, g^{}_{\myvec{X}_{0:\Time}}(\myvec{x}_{0:\Time})
\, dx^P_B \, dx^P_{\tilde B} \, d\myvec{x}_{0:\Time}
\Big| 
\\
\nonumber
&& \hspace{40mm} < (1/2)\, \efourA \, Q^{|B|+|\tilde B|}.
\end{eqnarray}
This gives
\begin{equation}
\label{eq:prop:medium:integral}
\Big|  
  \E[\gamma\tilde\gamma] -
  \int f^{}_{B}(x^P_B)\, f^{}_{{\tilde B}}(x^P_{\tilde B}) \, 
         g^{}_{X^P_{B}X^P_{\tilde B}}(x^{P}_{B},x^P_{\tilde B})\, dx^P_B \, dx^P_{\tilde B} 
\Big| < (1/2) \, \efourA \, Q^{|B|+|\tilde B|}.
\end{equation}
Then, using the approximate unconditional independence requirement of Assumption~\ref{B:unconditional:mix}  combined with the triangle inequality, \myeqref{eq:prop:medium:integral} implies
\begin{equation}
\label{eq:prop:short:integral}
\Big| 
  \E[\gamma\tilde\gamma] -
  \int f^{}_{B}(x^P_B)\, f^{}_{{\tilde B}}(x^P_{\tilde B}) \, 
         g^{}_{X^P_{B}}(x^{P}_{B}) \,  g^{}_{X^P_{\tilde B}}(x^P_{\tilde B})\, dx^P_B \, dx^P_{\tilde B} 
\Big| < \efourA \, Q^{|B|+|\tilde B|}.
\end{equation}
We can rewrite \myeqref{eq:prop:short:integral} as
\begin{equation}
\label{eq:prop:without:integral}
\big|  \, 
  \E[\gamma\tilde\gamma] - \E[\gamma] \, \E[\tilde\gamma] 
\, \big|
< \efourA \, Q^{|B|+|\tilde B|},
\end{equation}
proving the proposition.
\end{proof}

\BivB 

Assumption~\ref{B:temporal:mix} is needed to ensure the stability of the Monte Carlo approximation to the adapted process. 
It ensures that any error due to finite Monte Carlo sample size has limited consequences at sufficiently remote time points.
One could instead propose a bound that decreases exponentially with $K$, but that is not needed for the current purposes.
The following Proposition~\ref{lemma:E1E2} is useful for taking advantage of Assumption~\ref{B:temporal:mix}.
  
\begin{prop}\label{lemma:E1E2}
  Suppose that $f$ is a non-negative function and that for some $\epsilon>0$,
  \[|f(x) - f(x')| < \epsilon f(x')\]
  holds for all $x,x'$.
  Then for any two probability distributions where the expectations are denoted by $\E_1$ and $\E_2$ and for any random variable $X$, we have
  \[
  |\E_1 f(X) - \E_2 f(X) | \leq 
\epsilon \, \E_2 f(X).
  \]
\end{prop}
\begin{proof}
  Let the two probability laws be denoted by $P_1$ and $P_2$. We have
  \[\begin{split}
  |\E_1 f(X) - \E_2 f(X)| &= \left| \int f(x) P_1(dx) - \int f(x') P_2(dx') \right|\\
  &\leq \left| \int \int f(x) P_1(dx) P_2(dx') - \int\int f(x') P_1(dx) P_2(dx') \right|\\
  &\leq \int\int | f(x) - f(x') | P_1(dx) P_2(dx')\\
  &\leq  \int 
\epsilon
f(x') P_2(dx') = 
\epsilon \, 
\E_2 f(X).
  \end{split}.\]
\end{proof}

\Bv  

Assumption~\ref{B:girf} controls the Monte Carlo error for a single time interval on a single bootstrap replicate. 
In the case $\Ninter=1$, {\ABFIR} becomes {\ABF} and this assumption is one of many alternatives for bounding error from importance sampling.  
The purpose behind the selection of Assumption~\ref{B:girf} is to draw on the results of \citet{park20} for intermediate resampling, and our assumption is a restatement of their Theorem~2.
When $\Ninter=1$, the curse of dimensionality for importance sampling has the consequence that $C_0$ grows exponentially with $\Unit$.
However, \citet{park20} showed that setting $\Ninter=\Unit$ can lead to situations where $C_0(\Unit,\Time,\Ninter)$ in Assumption~\ref{B:girf} grows polynomially with $\Unit$.
Here, we do not place requirements concerning the dependence of $C_0$ on $\Unit$, $\Time$ and $\Ninter$ since our immediate concern is a limit where $\Rep$ and $\Np$ increase. 
Nevertheless, the numerical results are consistent with the theoretical and empirical results obtained for intermediate resampling in the context of particle filtering by \cite{park20}.

\Bvi    

The Monte Carlo conditional independence required by Assumption~\ref{B:XG_XA_ind} would hold for {\ABFIR} if the guide variance $V_{\unit,\time,i}$ were calculated using an independent set of guide simulations to those used for evaluating the measurement weights $w^M_{\unit,\time,i,j}$.
For numerical efficiency, the {\ABFIR} algorithm implemented here constructs a shared pool of simulations for both purposes rather than splitting the pool up between them, in the expectation that the resulting minor violation of Assumption~\ref{B:XG_XA_ind} has negligible impact.

\TheoremII 


\begin{proof}   
First, we set up some notation. For $\unitTimeSubset_{\unit\comma\time}$ and $w^M_{\unit,\time,\rep,\np}$ constructed by {\ABFIR}, define
\begin{equation}\label{eq:gamma:def}
\adapted^{MC,\rep}_{\unitTimeSubset_{\unit\comma\time}}
=
  \prod_{m=1}^{\time}
  \left[
    \frac{1}{\Np}\sum_{\np=1}^{\Np}
    \hspace{1mm}
       \prod_{(\altUnit,m)\in \unitTimeSubset^{[m]}_{\unit\comma\time}} 
    \hspace{-1mm}
        w^M_{\altUnit,m,\rep,\np}
  \right]
\hspace{5mm}
\mbox{and}
\hspace{5mm}
\bar\adapted^{MC}_{\unitTimeSubset_{\unit\comma\time}}
=\frac{1}{\Rep} \sum_{\rep=1}^{\Rep} \adapted^{MC,\rep}_{\unitTimeSubset_{\unit\comma\time}}.
\end{equation}
The Monte Carlo conditional likelihoods output by {\ABFIR} can be written as
\begin{equation}
\MC{\loglik}_{\unit,\time} = \log\MC{\bar\adapted}_{B^+_{\unit,\time}} - \log\MC{\bar\adapted}_{B^{\plusStrut}_{\unit,\time}}.
\end{equation}
We proceed with a similar argument to the proof of Theorem~\ref{thm:tif}. 
Since $\adapted^{MC,\rep}_{\unitTimeSubset_{\unit\comma\time}}$ are i.i.d. for $\rep \in \seq{1}{\Rep}$, we can suppose they are replicates of a Monte Carlo random variable $\adapted^{MC}_{\unitTimeSubset_{\unit\comma\time}}$.
We define
\begin{equation}
\nonumber
\centersum^{+}_{\unit,\time}= \frac{1}{\sqrt{\Rep}}\sum_{\rep=1}^\Rep\big(
\adapted^{MC,\rep}_{B^+_{\unit,\time}} - \E\big[\adapted^{MC}_{B^+_{\unit,\time}}\big]\big),
\hspace{2cm}
\centersum^{}_{\unit,\time}= \frac{1}{\sqrt{\Rep}}\sum_{\rep=1}^\Rep\big(
\adapted^{MC,\rep}_{B^{}_{\unit,\time}} - \E\big[\adapted^{MC}_{B^{}_{\unit,\time}}\big]\big).
\end{equation}
The same calculation as \myeqref{eq:likelihood:at:point} gives
\begin{eqnarray}
\label{eq:likelihood:at:point:B}
\MC{\loglik}_{\unit,\time} 
&=& 
\log\left(\frac{\E\big[\adapted^{MC}_{B^+_{\unit,\time}}\big]}{\E\big[\adapted^{MC}_{B^{\plusStrut}_{\unit,\time}}\big]}\right) +  \Rep^{-1/2}
\left(\frac{\centersum^{+}_{\unit,\time}}{\E\big[\adapted^{MC}_{B^+_{\unit,\time}}\big]} - \frac{\centersum^{}_{\unit,\time}}{\E\big[\adapted^{MC}_{B^{}_{\unit,\time}}\big]} \right) + o_P\big(\Rep^{-1/2}\big)
\end{eqnarray}
The joint distribution of $\big\{ (\centersum^{+}_{\unit,\time},\centersum^{}_{\unit,\time}), (\unit,\time)\in \spaceTime \big\}$ follows a standard central limit theorem as $\Rep\to\infty$.
Each term has mean zero, with variances and covariances uniformly bounded over $(\unit,\time,\altUnit,\altTime)$ due to Assumption~\ref{B2}.
From Proposition~\ref{prop:ABF_AB}, using the same reasoning as \myeqref{eq:log:diff:identity},
\begin{equation}
\label{thm2:eq3}
\left|  
  \log \left(
    \frac{\E_{g}\big[\adapted^{}_{A^+_{\unit,\time}}
\big]}{\E_{g}\big[\adapted^{}_{A^{}_{\unit,\time}}
\big]} 
  \right)
  - 
  \log \left(
     \frac{\E_{g}\big[\adapted^{}_{B^+_{\unit,\time}}
\big]}{\E_{g}\big[\adapted^{}_{B^{}_{\unit,\time}}
\big]}
  \right)
\right|
< \ethree Q^2.
\end{equation}
Now we use  Lemma~\ref{lemma:AS_bias} and \eqref{LogInequality} to obtain
\begin{eqnarray}
\nonumber
&&\left|  
  \log \left(
     \frac{\E_{g}\big[\adapted^{}_{B^+_{\unit,\time}}
\big]}{\E_{g}\big[\adapted^{}_{B^{}_{\unit,\time}}
\big]}
  \right)
  -
  \log \left(
     \frac{\EMC\big[\adapted^{MC}_{B^+_{\unit,\time}}\big]}{\EMC\big[\adapted^{}_{B^{}_{\unit,\time}}\big]}
  \right)
\right|
\\
\nonumber
&& \le
  \left|  
    \log \E_{g}\big[\adapted^{}_{B^+_{\unit,\time}}
    \big]
    -
    \log \EMC\big[\adapted^{}_{B^{+}_{\unit,\time}}\big]
  \right|
  +
  \left|
    \log \E_{g}\big[\adapted^{}_{B^{}_{\unit,\time}}
    \big]
    -
    \log \EMC\big[\adapted^{}_{B^{}_{\unit,\time}}\big]
  \right|
\\
\label{thm2:eq3b}
&& < 2 Q^{2\Bsize}\big(\boundLemmaUniform\big).
\end{eqnarray}
The proof of the central limit result in \myeqref{th:abf:lik:bound} is completed by combining \myeqref{eq:likelihood:at:point:B}, \myeqref{thm2:eq3} and \myeqref{thm2:eq3b}.
To show \myeqref{th:abf:lik:bound2} we check that $\centersum^{}_{\unit,\time}$ and $\centersum^{}_{\altUnit,\altTime}$ are weakly correlated when $(\unit,\time)$ and $(\altUnit,\altTime)$ are sufficiently separated. 
By the same reasoning as the proof of Theorem~\ref{thm:tif}, it is sufficient to show that $\adapted^{MC}_{B^{}_{\unit\comma\time}}$ and $\adapted^{MC}_{B^{}_{\altUnit\comma\altTime}}$ are weakly correlated.
These Monte Carlo quantities approximate $\adapted^{}_{B^{}_{\unit,\time}}(\myvec{X}_{0:\time-1})$ and $\adapted^{}_{B^{}_{\altUnit\comma\altTime}}(\myvec{X}_{0:\altTime-1})$ with $\myvec{X}$ drawn from $g$.
Let us suppose $\time\ge\altTime$, and write $d_{\unit,\time}=\time-\inf_{(v,m)\in B_{\unit,\time}}m$.
First, we consider the situation $\time-\altTime>K+d_{\unit,\time}$, in which case we can use the Markov property to give
\begin{equation}
\label{supp:eq:cov:MC:bound}
\cov\big(\adapted^{MC}_{B^{}_{\unit,\time}},\adapted^{MC}_{B^{}_{\altUnit\comma\altTime}}\big)
< \EMC \big[ \adapted^{MC}_{B^{}_{\altUnit\comma\altTime}} \big]
\sup_{\myvec{x}}\left\{
\EMC \big[ \adapted^{MC}_{B^{}_{\unit\comma\time}} \big| \myvec{X}^A_{\time-d_{\unit,\time}-K,1}=\myvec{x} \big]
- \EMC \big[ \adapted^{MC}_{B^{}_{\unit\comma\time}} \big] \right\}
\end{equation}
Then, the triangle inequality followed by applications of Assumption~\ref{B:temporal:mix} and Lemma~\ref{lemma:AS_bias} gives
\newcommand\boundA{2\efourB + 2 \big( K+d_{\unit,\time}\big)\efive^{} }
\begin{eqnarray}
\nonumber
&&\hspace{-15mm} \left|
  \EMC \big[ \adapted^{MC}_{B^{}_{\unit\comma\time}} \big| \myvec{X}^A_{\time-d_{\unit,\time}-K,1}=\myvec{x} \big]
  - \EMC \big[ \adapted^{MC}_{B^{}_{\unit\comma\time}} \big] 
\right|
\\
\nonumber
&& 
\le 
  \Big|
    \E_g\big[ \adapted^{}_{B^{}_{\unit\comma\time}}
    \big| \myvec{X}_{\time-d_{\unit,\time}-K}=\myvec{x} \big]
    - \E_g \big[ \adapted^{}_{B^{}_{\unit\comma\time}}
    \big] 
  \Big|
\\
\nonumber
&& \hspace{15mm}
  +
  \Big|
    \EMC \big[ \adapted^{MC}_{B^{}_{\unit\comma\time}} \big| \myvec{X}^A_{\time-d_{\unit,\time}-K,1}=\myvec{x} \big]
    -    \E_g\big[ \adapted^{}_{B^{}_{\unit\comma\time}}
    \big| \myvec{X}_{\time-d_{\unit,\time}-K}=\myvec{x} \big]
  \Big|
\\
\nonumber
&& \hspace{15mm}
  +
  \Big| 
    \EMC \big[ \adapted^{MC}_{B^{}_{\unit\comma\time}} \big] 
   - \E_g\big[ \adapted^{}_{B^{}_{\unit\comma\time}}
    \big] 
  \Big|
\\
\label{supp:eq:cov:MC:triangle}
&& 
\le
Q^b 
\Big(
  \boundA
\Big) 
\end{eqnarray}
Putting \eqref{supp:eq:cov:MC:triangle} into \eqref{supp:eq:cov:MC:bound}, we get
\begin{equation}
\cov
\big(
  \adapted^{MC}_{B^{}_{\unit,\time}},\adapted^{MC}_{B^{}_{\altUnit\comma\altTime}}
\big) 
< 
Q^{2\Bsize} 
\Big(
  \boundA
\Big).
\end{equation}
Now we address the situation $\time-\altTime\le K+d_{\unit,\time}$. 
We apply Lemma~\ref{lemma:AS_bias} on the union $B^{}_{\unit\comma\time} \cup B^{}_{\altUnit\comma\altTime}$ for which the temporal depth is bounded by 
$d \le K+d_{\unit,\time}+d_{\altUnit,\altTime}$.
This gives
\newcommand\boundB{(2K+d_{\unit,\time}+d_{\altUnit,\altTime})\efive^{} +\efourB}
\begin{equation}
\label{application:of:thm2:eq2}
\Big| \,
  \EMC \big[\adapted^{MC}_{\unitTimeSubset_{\unit\comma\time}}\adapted^{MC}_{\unitTimeSubset_{\altUnit\comma\altTime}}\big] - \E_{g_{}}\big[\adapted^{}_{\unitTimeSubset_{\unit\comma\time}}
   \adapted^{}_{\unitTimeSubset_{\altUnit\comma\altTime}}
   \big]
\, \Big|
< Q^{2\Bsize}
\Big( 
\boundB
\Big).
\end{equation}
From
Proposition~\ref{prop:cov:mix}, if $(\altUnit\comma\altTime) \notin \altB_{\unit\comma\time}$,
\begin{equation}
\label{supp:eq:cov:g:bound}
\cov_{g_{}}\big(\adapted^{}_{B^{}_{\unit,\time}}
  ,\adapted^{}_{B^{}_{\altUnit\comma\altTime}}
  \big) < \efourA Q^{2\Bsize}.
\end{equation}
Now, we establish that 
$\cov\big(\adapted^{MC}_{\unitTimeSubset_{\unit\comma\time}},
    \adapted^{MC}_{\unitTimeSubset_{\altUnit\comma\altTime}}\big)$ 
is close to  
$\cov_g\big(\adapted^{}_{\unitTimeSubset_{\unit\comma\time}},
     \adapted^{}_{\unitTimeSubset_{\altUnit\comma\altTime}}\big)$.
\begin{eqnarray}
\nonumber
&& \hspace{-15mm} 
\Big| 
  \cov\big(\adapted^{MC}_{\unitTimeSubset_{\unit\comma\time}},
    \adapted^{MC}_{\unitTimeSubset_{\altUnit\comma\altTime}}\big)
  - 
  \cov_g\big(\adapted^{}_{\unitTimeSubset_{\unit\comma\time}},
     \adapted^{}_{\unitTimeSubset_{\altUnit\comma\altTime}}\big) 
\Big| 
\\
\nonumber
&&\le   
\Big| \,
  \EMC \big[\adapted^{MC}_{\unitTimeSubset_{\unit\comma\time}}\adapted^{MC}_{\unitTimeSubset_{\altUnit\comma\altTime}}\big] - \E_{g}\big[\adapted^{}_{\unitTimeSubset_{\unit\comma\time}}
  \adapted^{}_{\unitTimeSubset_{\altUnit\comma\altTime}}
  \big]
\, \Big|
\\
\nonumber
&& \hspace{10mm}
+ 
\Big|
  \EMC \big[\adapted^{MC}_{\unitTimeSubset_{\unit\comma\time}}\big]
  \big(
    \EMC \big[\adapted^{MC}_{\unitTimeSubset_{\altUnit\comma\altTime}}\big]
    -  \E_g\big[\adapted^{}_{\unitTimeSubset_{\altUnit\comma\altTime}}
  \big]
  \big)
\Big|
\\
\nonumber
&& \hspace{10mm}
+ 
\Big|
  \EMC \big[\adapted^{}_{\unitTimeSubset_{\altUnit\comma\altTime}}
  \big]
  \big(
    \EMC \big[\adapted^{MC}_{\unitTimeSubset_{\unit\comma\time}}\big]
    -  \E_g\big[\adapted^{}_{\unitTimeSubset_{\unit\comma\time}}
    \big]
  \big)
\Big|
\\
\nonumber
&&
\label{supp:eq:lots_of_K}
< 
Q^{2\Bsize}
\Big(
  \boundB + 2(\boundLemmaUniform)
\Big).
\\
&&
<
Q^{2\Bsize}
\Big(
\PartOfBoundOffDiagonal
\Big)
\end{eqnarray}
Using \myeqref{supp:eq:lots_of_K} together with \myeqref{supp:eq:cov:g:bound} to bound the $\Unit\Time(\Unit\Time-\altb)$ off-diagonal covariance terms completes the derivation of \myeqref{th:abf:lik:bound2}.

\end{proof}


\begin{lemma} \label{lemma:AS_bias}
Suppose Assumptions~\ref{B2}, \ref{B:temporal:mix}, \ref{B:girf} and \ref{B:XG_XA_ind}. 
Suppose the number of particles $\Np$ exceeds the requirement for~\ref{B:girf}.
If we write $d_B=\max_{(u_1,n_1), (u_2,n_2) \in B} |n_1-n_2|$ for $B \subset 1\mycolon U \times 1\mycolon N$,
then for any $B$,
\begin{equation*}
\Big| \,
  \EMC\big[\adapted^{MC}_{\unitTimeSubset} \big| \myvec X_{n-d_B-K,1}^A = \myvec x \big] - \E_{g}\big[\adapted^{}_{\unitTimeSubset}
 \big| \myvec X_{n-d_B-K}=\myvec x \big]
\, \Big|
< Q^{|B|}(K+d_B) \epsilon_{\mathrm{B6}}^{}, \quad \forall \myvec x \in \mathbb X^U,
\end{equation*}
and
\begin{equation}
\label{thm2:eq2}
\Big| \,
  \EMC\big[\adapted^{MC}_{\unitTimeSubset} \big] - \E_{g}\big[\adapted^{}_{\unitTimeSubset}
  \big]
\, \Big|
< Q^{|B|}(\epsilon_{\mathrm{B5}}^{} + (K+d_B) \epsilon_{\mathrm{B6}}^{}).
\end{equation}
\end{lemma}
\begin{proof}
Suppose that $\max_{(u',n')\in B} n' = n$.
Define $\eta_\time(\myvec{x}_{\time})=1$ and, for $0\le m \le \time-1$,
\begin{equation}
\label{eq:eta}
\eta_{m}(\myvec{x}_{m})
= \E_{g} \! \left[ \, \prod_{k=m+1}^{\time} \adapted^{}_{B^{[k]}} 
  \, \Big| \myvec{X}_{m}=\myvec{x}_{m} \right].
\end{equation}
We have a recursive identity
\begin{equation}
\label{eq:thm2:recursion}
\eta_m(\myvec{X}_m)=\E_{g} \! \left[ \adapted^{}_{B^{[m+1]}}
  \, \eta_{m+1}(\myvec{X}_{m+1}) \Big| \myvec{X}_m
\right].
\end{equation}
By taking the expectation of \eqref{eq:eta}, we have
\begin{equation}
\E_{g}\big[\eta_0(\myvec{X}_0)\big]=\E_{g}\big[\adapted^{}_{B}
  \big].
\end{equation}
Note that $g$ has marginal density $f_{\myvec{X}_{0}}$ for $\myvec{X}_0$.
We analyze an {\ABFIR} approximation to \myeqref{eq:thm2:recursion}. 
The function $\eta_{m+1}(\myvec{x})$ is not in practice computationally available for evaluation via {\ABFIR}, but the recursion nevertheless leads to a useful bound.
Let $\myvec{X}^{\IF}_{m+1}[\np](\myvec{x}_{m})$ correspond to the variable $\myvec{X}^{IR}_{m+1,\Ninter,1,\np}$ constructed by {\ABFIR} conditional on $\myvec{X}^{\IF}_{m,1}=\myvec{x}_{m}$.
Equivalently,  $\myvec{X}^{\IF}_{m+1}[\np](\myvec{x}_{m})$ matches the variable $\myvec{X}^{\IF}_{m+1,1}$ in {\ABFIR} if the assignment $\myvec{X}^{\IF}_{m+1,1}=\myvec{X}^{IR}_{m+1,\Ninter,1,1}$ is replaced by $\myvec{X}^{\IF}_{m+1,1}=\myvec{X}^{IR}_{m+1,\Ninter,1,\np}$ conditional on $\myvec{X}^{\IF}_{m,1}=\myvec{x}_{m}$.
We define an approximation error $e_m(\myvec{x}_m)$ by
\begin{equation}
\label{thm2:eq7}
\eta_m(\myvec{x}_m) =  
\frac{1}{\Np}\sum_{\np=1}^{\Np}
f_{Y_{B^{[m+1]}}|\myvec{X}_m}(\data{y}_{B^{[m+1]}}\given \myvec{x}_m)
 \, \eta_{m+1}\big( \myvec{X}_{m+1}^{\IF}[\np](\myvec{x}_{m})\big) + e_m(\myvec{x}_{m}).
\end{equation}
From Assumptions~\ref{B2} and~\ref{B:girf}, $\EMC\big|e_m(\myvec{x}_m)\big|<\efive^{} \, Q^{\big|B^{[m+1:n]}\big|}$ uniformly over $\myvec{x}_m$, 
Thus, setting $r_m = \EMC | e_m(\myvec{X}^A_{m,1}) |$, we have
\begin{equation}
\label{eq:lemma:r}
r_m < \efive^{} \, Q^{\big|B^{[m+1:n]}\big|}.
\end{equation} 
Now, setting $\K=K+d_{\unit,\time}$, we commence to prove inductively that, for $\time-\K\le m \le \time$,
\begin{equation}
\label{lemma:inductive:hypothesis}
\Bigg| \,
  \eta_{\time-\K}(\myvec{x}) - \EMC 
  \Big[ \eta_m(\myvec{X}^A_{m,1})
    \prod_{k=\time-\K+1}^m 
f_{Y_{B^{[k]}}|\myvec{X}_{k-1}}(\data{y}_{B^{[k]}}\given \myvec{X}^A_{k-1,1})
    \Big| \myvec{X}^A_{n-\K,1}=\myvec{x}
  \Big]
\, \Bigg|
< (m-n+\K)\efive \, Q^{|B|}.
\end{equation}
First, suppose that \myeqref{lemma:inductive:hypothesis} holds for $m$. 
From \myeqref{thm2:eq7} and \myeqref{eq:lemma:r},
\begin{equation}
\left| 
 \eta_m(\myvec x_m)
 - 
 \EMC\left[
  \frac{1}{\Np}\sum_{\np=1}^{\Np}
 f_{Y_{B^{[m+1]}}|\myvec{X}_m}(\data{y}_{B^{[m+1]}}\given \myvec{x}_m)
 \, \eta_{m+1}\big( \myvec{X}_{m+1}^{\IF}[\np](\myvec{x}_{m})\big)
 \right]
\right| 
< \efive \, Q^{\big|B^{[m+1:n]}\big|}.
\end{equation}
Since the particles are exchangeable, the expectation of the mean of $\Np$ particles can be replaced with the expectation of the first particle.
Plugging in $\myvec x_m = \myvec X_{m,1}^A$ gives us
\begin{equation}
\label{eq:lemma:eta:bound}
\left| 
 \eta_m(\myvec X_{m,1}^A)
 - 
  f_{Y_{B^{[m+1]}}|\myvec{X}_m}\big(\data{y}_{B^{[m+1]}}\given  \myvec{X}_{m,1}^{A}\big)
 \EMC\left[
   \eta_{m+1}\big(  \myvec{X}_{m+1,1}^{A} \big) \middle| \myvec X_{m,1}^A
 \right]
\right| 
< \efive \, Q^{\big| B^{[m+1:n]} \big| }
\end{equation}
Putting \myeqref{eq:lemma:eta:bound} into \myeqref{lemma:inductive:hypothesis}, for $m\le \time$, and taking an iterated expectation with respect to $\myvec{X}^A_{m,1}$, we find that \myeqref{lemma:inductive:hypothesis} holds also for $m+1$.
Since \myeqref{lemma:inductive:hypothesis} holds trivially for $m=\time-\K$, it holds for $\time-\K\le m\le \time$ by induction.
Then, noting $\eta_{\time}(\myvec{x})=1$, we have from \myeqref{lemma:inductive:hypothesis} that
\[
\left|
\eta_{n-\K}(x)
- \EMC \left[ \prod_{k=n-\K+1}^n 
  f_{Y_{B^{[k]}}|\myvec{X}_{k-1}}\big(\data{y}_{B^{[k]}}\given  \myvec{X}_{k-1,1}^{A}\big)
\middle \vert \myvec X_{n-\K, 1}^A = x \right]
\right|
< \K \efive \, Q^{|B|}.
\]
Integrating the above inequality over $\myvec x$ with respect to the law of $\myvec X_{n-\K,1}^A$, we obtain
\begin{equation}\label{eq:AS_bias_induction_result}
\left|
\EMC [\eta_{n-\K}(\myvec X_{n-\K,1}^A)]
- \EMC \left[ \prod_{k=n-\K+1}^n 
  f_{Y_{B^{[k]}}|\myvec{X}_{k-1}}\big(\data{y}_{B^{[k]}}\given  \myvec{X}_{k-1,1}^{A}\big)
\right]
\right|
< \K\efive \, Q^{| B |}.
\end{equation}
But under Assumption~\ref{B:XG_XA_ind}, we have
\begin{equation}\label{eq:gammaMC_underB6}
\EMC \left[ \prod_{k=n-\K+1}^n 
  f_{Y_{B^{[k]}}|\myvec{X}_{k-1}}\big(\data{y}_{B^{[k]}}\given  \myvec{X}_{k-1,1}^{A}\big)
\right]
= \EMC \big[ \gamma^{MC}_{B}\big].
\end{equation}
Assumption~\ref{B:temporal:mix} says
\begin{equation}
\label{eq:eta:bound}
| \eta_{n-\K}(\myvec x_{n-\K}^{(1)}) - \eta_{n-\K}(\myvec x_{n-\K}^{(2)}) |
< \efourB \, \eta_{n-\K}(\myvec x_{n-\K}^{(2)}).
\end{equation}
Application of Proposition~\ref{lemma:E1E2} to \myeqref{eq:eta:bound} gives
\begin{equation}\label{eq:Eg_EMC_eta}
\Big| \E_g \big[ \eta_{n-\K}(\myvec X_{n-\K})\big] - \EMC \big[\eta_{n-\K}(\myvec X_{n-\K,1}^A) \big] \Big|
< \efourB \, \E_g \big[ \eta_{n-\K}(\myvec X_{n-\K}) \big]
< \efourB \, Q^{| B |}.
\end{equation}
Combining \eqref{eq:AS_bias_induction_result}, \eqref{eq:gammaMC_underB6}, and \eqref{eq:Eg_EMC_eta} completes the proof of Lemma~\ref{lemma:AS_bias}.
\end{proof}

\clearpage

\clearpage

\section{\secTitleSpace The correlated Brownian motion example}


\begin{knitrout}
\definecolor{shadecolor}{rgb}{1, 1, 1}\color{fgcolor}\begin{figure}

\includegraphics[width=6.5in]{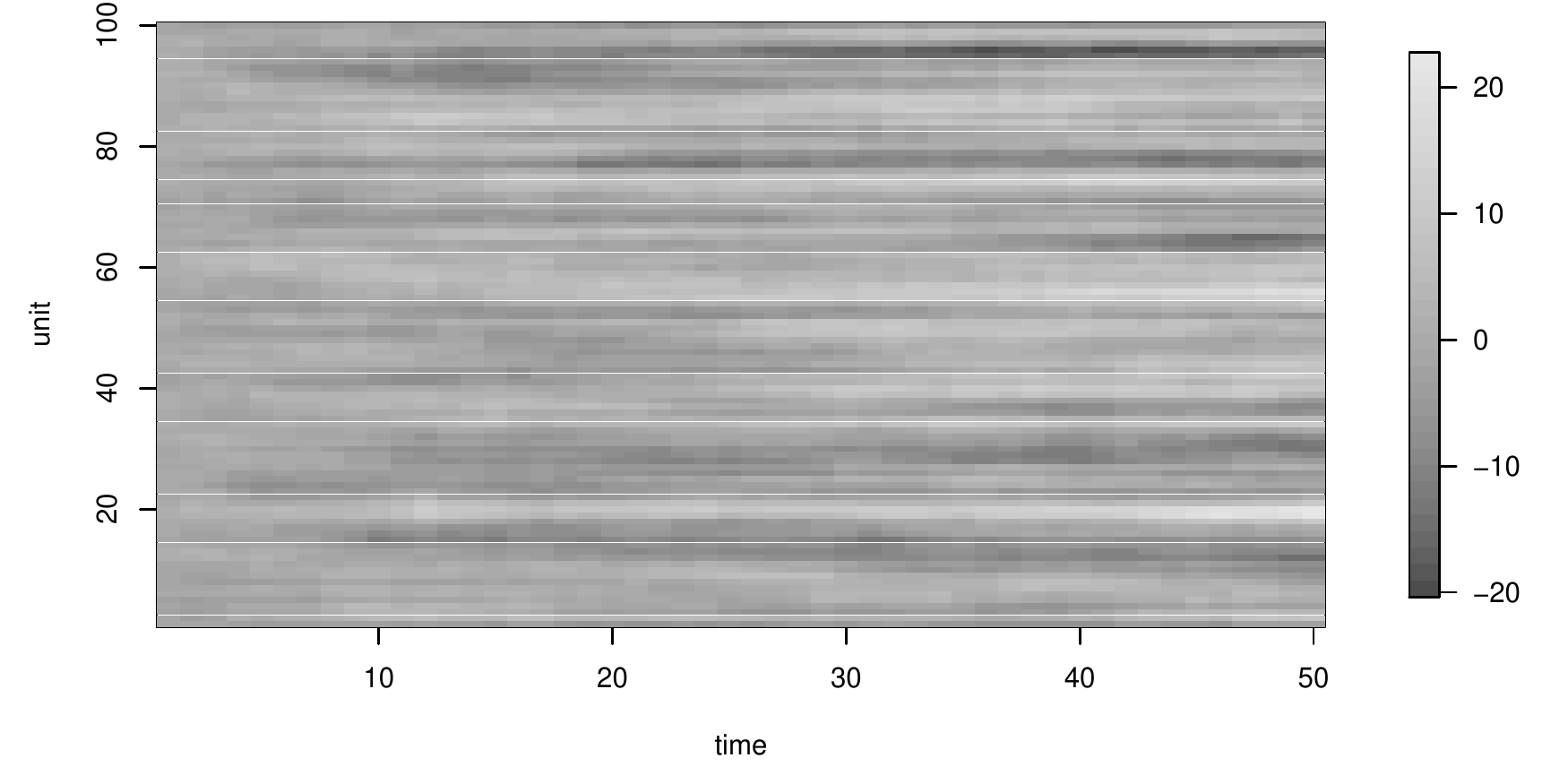} \hfill{}

\caption[correlated Brownian motion simulation used in the main text]{correlated Brownian motion simulation used in the main text}\label{fig:bm_image_plot}
\end{figure}

\end{knitrout}

To help visualize the correlated Brownian motion model, Fig.~\ref{fig:bm_image_plot} shows one of the simulations used for the results in Figure~{\MainFigureMeaslesSlice} of the main text.
Table~\ref{tab:bm} gives the algorithmic settings used for the filters and corresponding computational resource requirements.
Broadly speaking, $\Rep\Np$ for {\ABF} and {AIRSIF} should be compared with $\Rep$ for {\UBF}, $\Np$ for PF, and $\Np\Nguide$ for GIRF.

The computational effort allocated to each algorithm in Table~\ref{tab:bm} is given in core minutes.
{\UBF}, {\ABF} and {\ABFIR} parallelize readily, which is less true for PF and GIRF.
Therefore, the {\UBF}, {\ABF} and {\ABFIR} implementations run on all available cores (36 for this experiment)  whereas the PF and GIRF implementations run on a single core.
If sufficient replications are being carried out to utilize all available cores, comparison of core minute utilization is equivalent to comparison of total computation time.
However, a single replication of {\UBF}, {\ABF} or {\ABFIR} proceeds more quickly due to the parallelization. 

{\ABFIR} and GIRF have computational time scaling quadratically with $\Unit$ in this example, whereas the other methods scale linearly.
This is because the number of intermediate steps used, $\Ninter$, grows linearly with $\Unit$.

The main purpose of this example is not to provide a comparison between the functional capabilities of the methods on interesting scientific problems.
It is a toy example without the complexities that the methods are intended to address.
This simple example does show clearly the quick decline of PF and the slower declines of GIRF and {\ABF} as dimension increases. 

\begin{table}
\begin{center}
\begin{tabular}{|l|c|c|c|c|c|c|c|}
\hline &
  {\UBF}  &
  {\ABF} &
  {\ABFIR}  &
  GIRF  &
  PF &
  BPF &
  EnKF
\\ 
\hline 
\rule{0mm}{4.5mm}particles,
$\Np$ &
  --- &
  400 &
  200 &
  1000 &
  100000 &
  20000 &
  10000
\\
bootstrap replications, $\Rep$ &
  40000 &
  400 &
  200 &
  --- &
  --- &
  --- &
  ---
\\
guide simulations, $\Nguide$ &
  ---&
  ---&
  ---&
  50&
  --- &
  --- &
  ---
\\
lookahead lag, $L$ &
  ---&
  ---&
  ---&
  2 &
  --- &
  --- &
  ---
\\
intermediate steps, $S$ &
  ---&
  --- &
  $\Unit/2$ &
  $\Unit$ &
  --- &
  --- &
  ---
\\
\cline{2-4}
\hspace{-3mm} \begin{tabular}{l}
neighborhood, $B_{\unit\comma\time}$ \\
or block size
\end{tabular} &
\multicolumn{3}{c|}{
$ \rule[-4mm]{0mm}{11mm}
\begin{array}{c} 
$\big\{(\unit-1,\time),(\unit-2,\time),$ \\  $(\unit,\time-1),(\unit,\time-2)\big\}$
\end{array}$
} &
  --- &
  --- &
  3 &
  ---
\\
\cline{2-5}
forecast mean, $\myvec{\mu}(\myvec{x},s,t)$ &
  --- &
  --- &
  \multicolumn{2}{c|}{$\myvec{x}$} &
  --- &
  --- &
  ---
\\
measurement mean, $h_{\unit\comma\time}(x)$ &
  --- &
  --- &
  \multicolumn{2}{c|}{$x$} &
  --- &
  --- &
  $x$
\\
\cline{4-5}
$\tau={\VtoTheta}_{\unit\comma\time}(V,x)$ &
  --- &
  --- &
  \multicolumn{2}{c|}{\rule{0mm}{5mm}$\sqrt{V}$}  &
  --- &
  --- &
  ---
\\
$V={\thetaToV}_{\unit\comma\time}(\tau,x)$ &
  --- &
  --- &
  \multicolumn{2}{c|}{$\tau^2$}  &
  --- &
  --- &
  $\tau^2$
\\
\hline
\rule{0mm}{4.5mm}effort (core mins, $U=100$) & 
  0.9 & 
  2.1 & 
  4.2 & 
  0.3  & 
  0.3 &
  0.2 &
  0.0 
\\
effort  (core mins,  $U=80$) & 
  1.4 & 
  3.2 & 
  12.1 & 
  0.7 & 
  0.6 &
  0.3 &
  0.1 
\\
effort  (core mins, $U=60$) & 
  2.2 & 
  6.0 & 
  32.0 & 
  1.9 & 
  1.3 &
  0.6 &
  0.2 
\\
effort  (core mins,  $U=30$) & 
  4.0 & 
  11.6 & 
  87.1 & 
  6.1 & 
  3.2 &
  1.3 &
  0.4   
\\
effort  (core mins,  $U=10$) & 
  7.6 & 
  26.1 & 
  311.6 & 
  47.1 & 
  9.1 &
  3.2 &
  1.1 
\\
\hline

\end{tabular}
\caption{Algorithmic settings for the correlated Brownian motion numerical example. 
Computational effort is measured in core minutes for running one filter, corresponding to a point on Figure~{\MainFigureBmScaling} in the main text. 
The time taken for computing a single point using the parallel {\UBF}, {\ABF} and {\ABFIR} implementations is the effort divided by the number of cores, here $40$.
The time taken for computing a single point using the single-core GIRF, PF, BPF and EnKF implementations is equal to the effort in core minutes.
}\label{tab:bm}
\end{center}
\end{table}

\clearpage



\section{\secTitleSpace The measles example}


\begin{table}
\renewcommand{\arraystretch}{1.2}
  \centering
  \begin{tabular}{|crcl|}
    \hline
    parameter & value & unit & description  \\  \hline
    $\meanBeta$ & 1560.6 & year$^{-1}$& mean contact rate\\
    $\mu_{\bullet D}^{-1}$ & 50.0 &year& mean duration in the population \\
    $\mu_{EI}^{-1}$ & 7.0 &day& latent period  \\
    $\mu_{IR}^{-1}$ & 7.0 &day& infectious period \\
    $\sigma_{SE}$ & 0.150 &year$^{1/2}$& process noise\\
    $\amplitude$ & 0.500 &--- & amplitude of seasonality  \\
    $\alpha$ & 1 &---& mixing exponent \\
    $\tau$ & 4 & year & delay from birth to entry into susceptibles\\    
    $\rho$ & 0.5 &---& reporting probability \\
    $\psi$ & 0.15 &---& reporting overdispersion \\
    $G$ & 400 &---& gravitation constant \\
    $S_{\unit}(0), \unit \in 1:\Unit$ & 0.032 &---& initial susceptible fraction \\
    $E_{\unit}(0), \unit \in 1:\Unit$ & 0.00005 &---& initial exposed fraction \\
    $I_{\unit}(0), \unit \in 1:\Unit$ & 0.00004 &---& initial infectious fraction \\
\hline
  \end{tabular}
  \caption{\label{tab:measles_parameters} Parameters for the spatiotemporal measles transmission model}
\end{table}


Table~\ref{tab:measles_parameters} gives the model parameter values and Table~\ref{tab:mscale} gives the algorithmic settings used for the filters.
The times in Table~\ref{tab:mscale} give the total time required by each algorithm to calculate all its results for Figure~{\MainFigureMeaslesScaling} in the main text, using 36 cores.
The expected forecast function $\mu(x,s,t)$ needed for {\ABFIR} and GIRF was computed using a numerical solution to the deterministic skeleton of the stochastic model, i.e, a system of ODEs with derivative matching the infinitesimal mean function of the stochastic dynamic model.
In the specifications of $h_\unit\big(x)$, $\VtoTheta_\unit(V,x,\theta)$ and $\thetaToV_\unit(\theta,x)$, the latent process value $x$ contains a variable $C$ giving the cumulative removed infections in the current observation interval.

In Table~\ref{tab:mscale}, we see that the effort allocated to {\UBF}, {\ABF} and PF scales linearly with $\Unit$, since the number of bootstrap replications and particles is fixed in this experiment. 
GIRF computational effort scales fairly linearly in $\Unit$, since its effort is dominated by the guide simulations (which are linear in $\Unit$) rather than by the intermediate timestep calculations (which are quadratic in $\Unit$ since we carry out $\Unit$ intermediate calculations each of size linear in $\Unit$).
The effort allocated to {\ABFIR} scales with $\Unit^2$, since {\ABFIR} is more parsimonious with guide simulations (all particles in one bootstrap replication share the same guide simulations) and so the intermediate timestep calculations dominate the effort.
To obtain stable variance in the log likelihood estimate, the number of particles and bootstrap replications would have to grow with $\Unit$.
However, given a constraint on total computational resources, the number of particles and bootstrap replications would have to shrink as $\Unit$ increases.
The limit studied in this experiment is a balance between the two: the assumption is that one is prepared to invest a growing amount of computational effort as the data grow, but this should not grow too fast.
{\ABFIR} was permitted the greatest computational effort, but the following two considerations balance this:
\begin{enumerate}
\item Parallelization. {\UBF}, {\ABF} and {\ABFIR} are trivially parallelizable. 
The value of parallelization depends, among other things, on how many replications are being computed simultaneously and on how many cores are available.
Nevertheless, it is helpful that the core minute effort requirement for {\ABF} and {\ABFIR} can be divided by the number of available cores to give the computational time. 
Parallelizations of GIRF and PF can be constructed \citep{park20} but these involve non-trivial interaction between processors leading to additional algorithmic complexity and computational overhead. 
\item Memory. 
The intermediate timestep calculations in {\ABFIR} and GIRF do not add to the memory requirement, and the memory demands of {\UBF}, {\ABF} and {\ABFIR} are distributed across the parallel computations. 
A basic PF implementation for a large model can become constrained by its memory requirement (linear in the number of particles) before it can match the processor effort employed by the other algorithms.
\end{enumerate}

\begin{table}
\begin{center}
\begin{tabular}{|l|c|c|c|c|c|c|c|}
  \hline & 
  {\UBF}  & 
  {\ABF} & 
  {\ABFIR}  & 
  GIRF &
  EnKF &
  PF &
  BPF 
\\ 
  \hline 
  particles, $\Np$    \rule{0mm}{4.5mm} & 
  1 &
  500 & 
  200 & 
  2000  & 
  10000 &
  100000 &
  20000 
\\
  replicates, $\Rep$ & 
  20000 & 
  500 & 
  200 &
  --- & 
  --- &
  --- &
  ---
\\
  guide simulations, $\Nguide$ &
  --- & 
  --- &
  --- & 
  40 & 
  --- &
  --- &
  ---
\\
  lookahead lag, $L$ &
  --- &
  --- &
  --- & 1 & 
  --- &
  --- &
  ---
\\ 
  intermediate steps, $S$ & 
  --- & 
  --- &
  $\Unit/2$ & 
  $\Unit$ & 
  --- &
  --- &
  ---
\\
  \cline{2-4}
\hspace{-3mm} \begin{tabular}{l}
neighborhood, $B_{\unit\comma\time}$ \\
or block size
\end{tabular}
&
  \multicolumn{3}{c|}{
    \rule[-3mm]{0mm}{8mm}$\big\{(\unit,\time-1),(\unit,\time-2)\big\}$
  } &
  --- &
  --- &
  --- &
  2 
\\
  \cline{2-6}
  
  measurement mean, $h_{\unit\comma\time}\big(x)$ &
  --- &
  --- &
  \multicolumn{3}{c|}{\rule[-3mm]{0mm}{8mm}$\rho C$} &
  --- &
  --- 
\\  
  $V=\thetaToV_{\unit\comma\time}(\psi,\rho,x)$ &
  --- &
  --- &
  \multicolumn{3}{c|}{\rule[-3mm]{0mm}{8mm}$\rho(1-\rho)C + \rho^2C^2\psi^2$} &
  --- &
  ---
\\
  \cline{2-6}

  forecast mean, $\myvec{\mu}(\myvec{x},s,t)$ &
  --- &
  --- &
  \multicolumn{2}{c|}{\rule{0mm}{5mm}ODE model} &
  --- &
  --- &
  ---
\\
  $\psi=\VtoTheta_{\unit\comma\time}(V,x)$ &
  --- &
  --- &
  \multicolumn{2}{c|}{$\frac{\sqrt{V - \rho(1-\rho)C}}{\rho C}$} &
  --- &
  --- &
  ---
\\
\hline
  effort (core mins, $U=2$)   \rule{0mm}{4.5mm}  & 
  28.4 & 
  11.5 & 
  6.4 & 
  2.6  & 
  0.3  & 
  3.2 &
  0.8 
\\
  effort (core mins, $U=4$)  & 
  35.7 & 
  19.0 & 
  19.0 & 
  5.5  & 
  0.6  & 
  5.9 &
  1.5 
\\
  effort (core mins, $U=8$)  & 
  52.5 & 
  35.0 & 
  60.1 & 
  12.7  &
  1.2  &   
  11.5 &
  2.9 
\\
  effort (core mins, $U=16$)   & 
  87.9 & 
  66.7 & 
  217.8 & 
  36.3  & 
  2.4  & 
  22.5 &
  5.8 
\\
  effort (core mins, $U=32$)   & 
  155.2 & 
  133.3 & 
  1032.1 & 
  133.7  & 
  4.7  & 
  45.3 &
  11.6  
\\
\hline

\end{tabular}
\caption{Algorithmic settings for the measles example calculations in Figures~{\MainFigureMeaslesScaling} and~{\MainFigureMeaslesSlice}. 
Computational effort is measured in core minutes for running one filter, corresponding to a point on Figure~\MainFigureMeaslesScaling.
The time taken for computing a single point using the parallel {\UBF}, {\ABF} and {\ABFIR} implementations is the effort divided by the number of cores, here $36$.
The time taken for computing a single point using the single core GIRF, EnKF, PF and BPF implementations is equal to the effort in core minutes.
}\label{tab:mscale}
\end{center}
\end{table}

\clearpage

\clearpage

\section{\secTitleSpace Varying the neighborhood for measles}


\begin{knitrout}
\definecolor{shadecolor}{rgb}{1, 1, 1}\color{fgcolor}\begin{figure}

{\centering \includegraphics[width=3in]{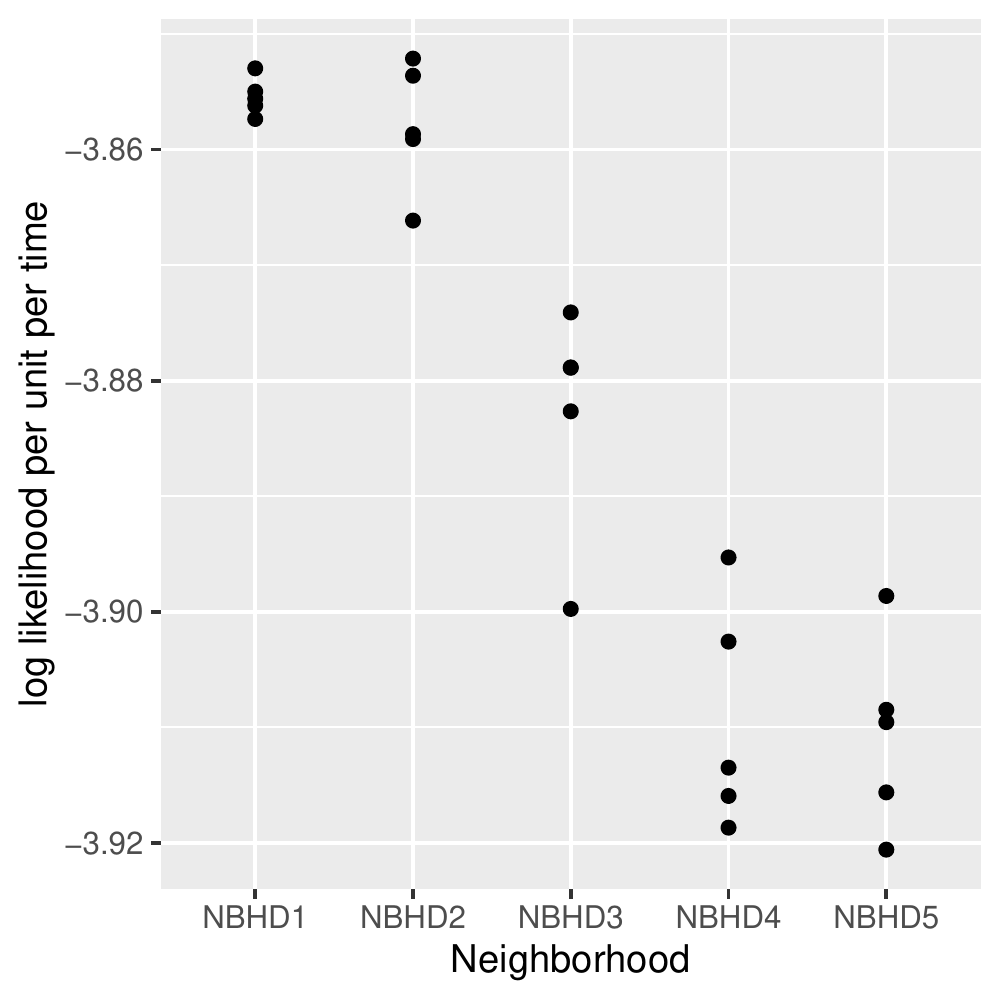} 

}

\caption[log likelihood estimates for simulated data from the measles model using {\ABF}, with varying neighborhoods]{log likelihood estimates for simulated data from the measles model using {\ABF}, with varying neighborhoods.}\label{fig:abfNbhd_loglik_plot}
\end{figure}

\end{knitrout}

We compared five different neighborhoods for the measles model:
\begin{center}
\begin{tabular}{lll}
\hline
NBHD1 & One co-located lag & $\{(\unit,\time-1)\}$ 
\\
NBHD2 & Two co-located lags & $\{(\unit,\time-1),(\unit,\time-2)\}$ 
\\
NBHD3 & Three co-located lags & $\{(\unit,\time-1),(\unit,\time-2),(\unit,\time-3)\}$ 
\\
NBHD4 & Two co-located lags and the previous city& $\{(\unit,\time-1),(\unit,\time-2),(\unit-1,\time)\}$ 
\\
NBHD5 & Two co-located lags and London& $\{(\unit,\time-1),(\unit,\time-2),(1,\time)\}$ 
\\
\hline
\end{tabular}
\end{center}
We filtered simulated data for $U=40$ and $N=130$, with $10$ replications.
We used {\ABF} with $200$ particles on each of $1000$ bootstrap replications.
The results are shown in Fig.~\ref{fig:abfNbhd_loglik_plot}.
Larger neighborhoods should increase the expected likelihood, but their increased Monte Carlo variability can decrease the expected log likelihood due to Jensen's inequality.
In this case, we see that a neighborhood of two co-located lags provides a reasonable bias-variance tradeoff.
The time taken for the above calculation was insensitive to the size of the neighborhood.
The total run time for each neighborhood in Fig.~\ref{fig:abfNbhd_loglik_plot} was 
223.0 mins for NBHD1,
225.5 mins for NBHD2,
225.4 mins for NBHD3,
218.6 mins for NBHD4,
234.8 mins for NBHD5.

\clearpage


\clearpage

\section{\secTitleSpace A Lorenz-96 example}
\label{sec:lorenz}

Our primary motivation for {\ABF} and {\ABFIR} is application to population dynamics arising in ecological and epidemiological models.
Geophysical models provide an alternative situation involving spatiotemporal data analysis.
We compare methods on the Lorenz-96 model, a nonlinear chaotic system providing a toy model for global atmospheric circulation \citep{lorenz96,vankekem18}.
We consider a stochastic Lorenz-96 model with added Gaussian process noise \citep{park20} defined
as the solution to the following system of stochastic differential equations,
\begin{equation}
  dX_{\unit}(t) =
\big \{
\big(X_{\unit+1}(t) - X_{\unit-2}(t)\big)
\cdot X_{\unit-1}(t) - X_{\unit}(t) + F
\big\} dt
+ \sigma_p dB_{\unit}(t), \qquad \unit \in \seq{1}{\Unit}.
  \label{eqn:stoLorenz}\end{equation}
We define $X_{0} = X_{\Unit}$, $X_{-1} = X_{\Unit-1}$, and $X_{\Unit+1} = X_{1}$ so that the $\Unit$ spatial locations are placed on a circle.
The terms $\{B_{\unit}(t), \unit \in \seq{1}{\Unit} \}$ denote $\Unit$ independent standard Brownian motions.
$F$ is a forcing constant, and we use the value $F{\,=\,}8$ which was demonstrated by \citet{lorenz96} to induce chaotic behavior.
The process noise parameter is set to $\sigma_p=1$.
The system is started with initial state $X_{\unit}(0)$ drawn as an independent normal random variable with mean $5$ and standard deviation $2$ for $\unit\in \seq{1}{\Unit}$.
This initialization leads to short transient behavior.
Observations are independently made for each dimension at 
$t_\time = \time$ 
for $\time \in \seq{1}{\Time}$ with Gaussian measurement noise of mean zero and standard deviation $\tau=1$,
\begin{equation}
Y_{\unit, \time} = X_{\unit}(t_{\time}) + \eta_{\unit, \time}\, \hspace{20mm} \eta_{\unit, \time} \sim N(0, \tau^2).
\end{equation}
We used an Euler-Maruyama method for numerical approximation of the sample paths of $\{\myvec{X}(t)\}$, with timestep of $0.005$. 
A simulation from this model is shown in Fig.~\ref{fig:lz_image_plot}.

The ensemble Kalman filter (EnKF) is a widely used filtering method in weather forecasting for high dimensional systems \citep{evensen96}.  
EnKF involves a local Gaussian approximation which is problematic in highly nonlinear systems \citep{ades15}.
Methods that make local Gaussian assumptions like EnKF are necessary to scale up to the dimensions of the problems in weather forecasting. 
Figure~\ref{fig:lz_loglik_plot} shows that for a small number of units, the basic particle filter (PF) and GIRF out-perform EnKF.
Then, as the number of spatial units increases, the performance of PF rapidly deteriorates whereas GIRF continues to perform well up to a moderate number of units.
{\UBF}, {\ABF}, and particularly {\ABFIR}, scales well despite under-performing EnKF on this example.
The additive Gaussian observation and process noise in the Lorenz-96 model is well suited to the approximations involved in EnKF.
By contrast, it is less clear how to apply EnKF to discrete population non-Gaussian models such as the measles example, and how effective the resulting approximations might be.

\begin{knitrout}
\definecolor{shadecolor}{rgb}{1, 1, 1}\color{fgcolor}\begin{figure}

{\centering \includegraphics[width=4.5in]{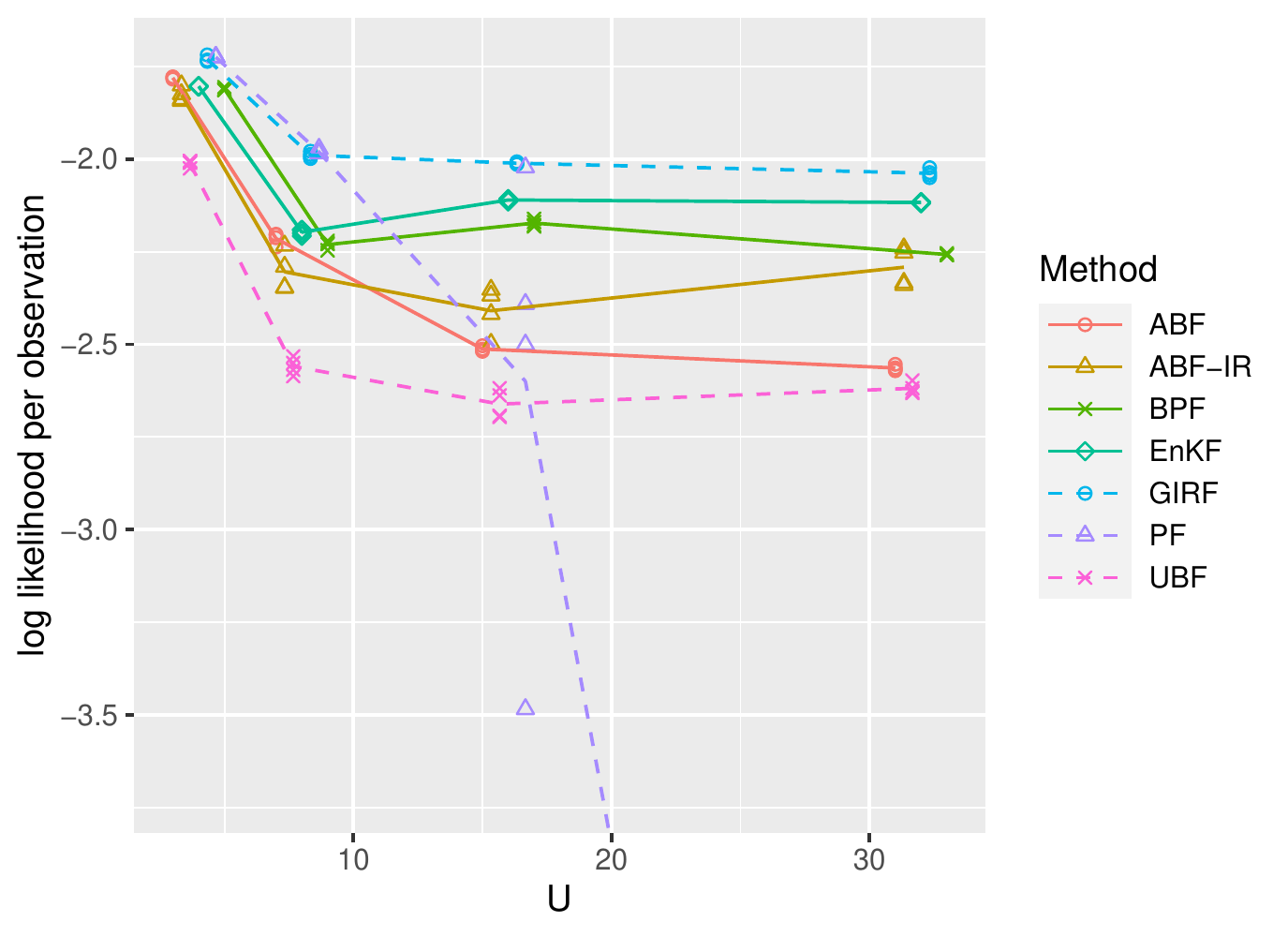} 

}

\caption[Log likelihood estimates for a Lorenz-96 model of various dimensions]{Log likelihood estimates for a Lorenz-96 model of various dimensions. {\UBF}, {\ABF} and {\ABFIR} are compared with a guided intermediate resampling filter (GIRF), a standard particle filter (PF), a block particle filter (BPF) and an ensemble Kalman filter (EnKF).}\label{fig:lz_loglik_plot}
\end{figure}

\end{knitrout}

\begin{knitrout}
\definecolor{shadecolor}{rgb}{1, 1, 1}\color{fgcolor}\begin{figure}

\includegraphics[width=6.5in]{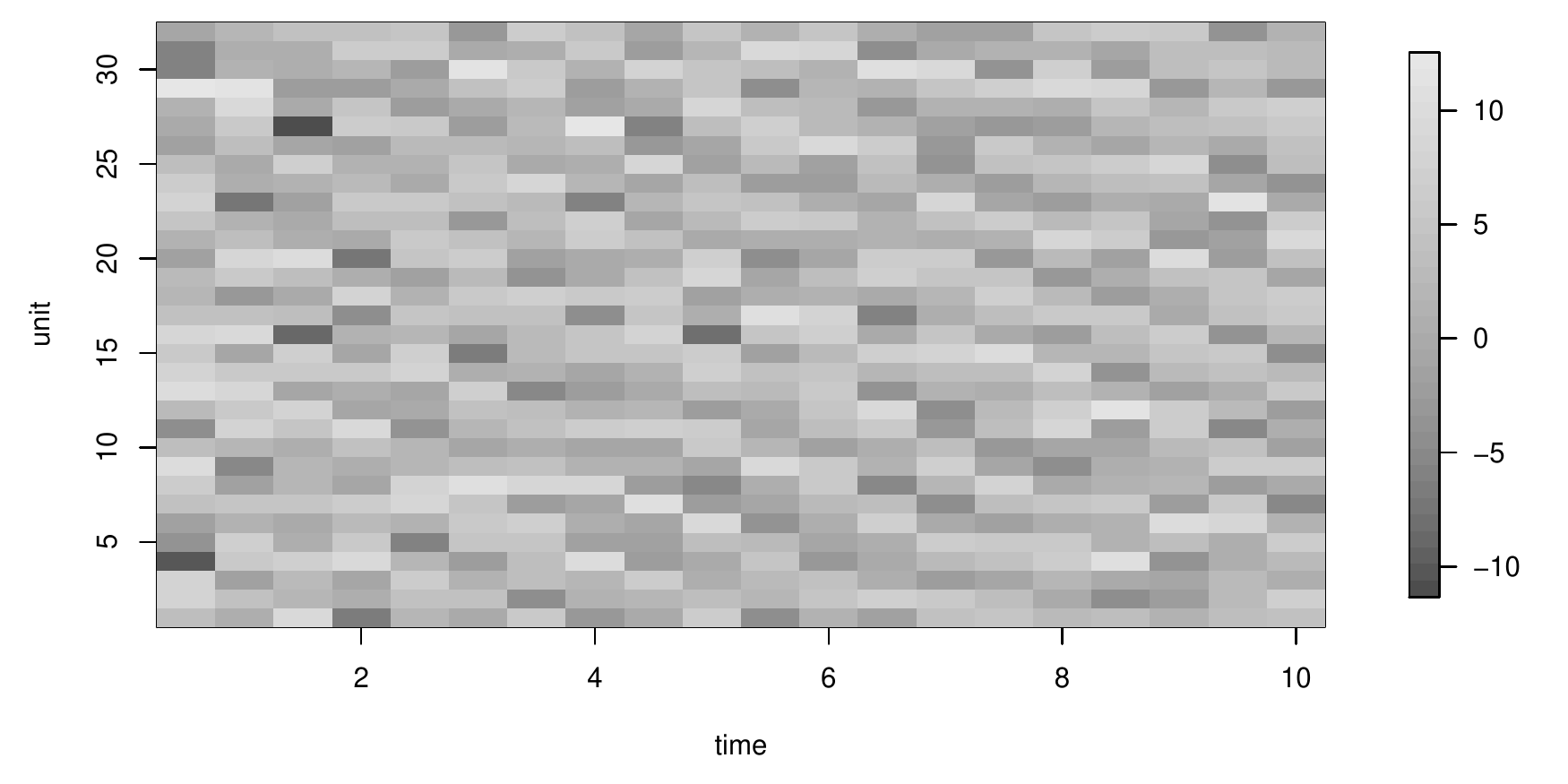} \hfill{}

\caption[Lorenz '96 simulation used in the main text]{Lorenz '96 simulation used in the main text}\label{fig:lz_image_plot}
\end{figure}

\end{knitrout}

\begin{table}
\begin{center}
\begin{tabular}{|l|c|c|c|c|c|c|c|}
\hline &
  {\UBF} &
  {\ABF} &
  {\ABFIR} &
  GIRF &
  PF &
  EnKF &
  BPF
\\ 
\hline 
\rule{0mm}{4.5mm}particles, $\Np$ &
  1 &
  400 &
  200 &
  1000  &
  100000 &
  10000 &
  10000 
\\
bootstrap replicates, $\Rep$ &
  40000 &
  400 &
  200 &
  --- &
  --- &
  --- &
  ---
\\
guide simulations, $\Nguide$ &
  --- &
  --- &
  --- &
  50 &
  --- &
  --- &
  ---
\\
lookahead lag, $L$ &
  --- &
  --- &
  --- &
  2 &
  --- &
  --- &
  ---
\\ 
intermediate steps, $S$ &
  --- &
  --- &
  $\Unit/2$ &
  $\Unit$ &
  --- &
  --- &
  ---
\\
\cline{2-4}

\hspace{-2mm}\begin{tabular}{l}
neighborhood, $B_{\unit\comma\time}$ \\
or block size
\end{tabular}
&
  \multicolumn{3}{c|}{
    $ \rule[-4mm]{0mm}{11mm}
    \begin{array}{c} 
    $\big\{(\unit,\time-1),(\unit,\time-2),$ \\
    $(\unit-1,\time),(\unit-2,\time)\big\}$
    \end{array}$
  } &
  --- &
  --- &
  --- &
  4
\\
\cline{2-5}
forecast mean, $\myvec{\mu}(\myvec{x},s,t)$ &
  --- &
  --- &
  \multicolumn{2}{c|}{ODE model} &
  --- &
  --- &
  ---
\\
measurement mean, $h_{\unit\comma\time}(x)$ &
  --- &
  --- &
  \multicolumn{2}{c|}{$x$} &
  --- &
  $x$ &
  ---
\\
\cline{4-5}
$\tau={\VtoTheta}_{\unit\comma\time}(V,x)$ &
  --- &
  --- &
  \multicolumn{2}{c|}{\rule{0mm}{5mm}$\sqrt{V}$}  &
  --- &
  --- &
  ---
\\
$V={\thetaToV}_{\unit\comma\time}(\tau,x)$ &
  --- &
  --- &
  \multicolumn{2}{c|}{$\tau^2$} &
  --- &
  $\tau^2$ &
  ---
\\
\hline
\rule{0mm}{4.5mm}effort (core mins, $U=4$) & 
  34.5 & 
  5.2 & 
  5.1 & 
  2.0  & 
  2.0 & 
  0.2 &
  0.2
\\
effort  (core mins,  $U=6$) & 
  40.9 & 
  7.3 & 
  8.2 & 
  3.0 &
  2.9 & 
  0.3 &
  0.3  
\\
effort  (core mins,  $U=10$) & 
  54.5 & 
  11.2 & 
  16.3 & 
  5.0 & 
  4.8 & 
  0.5 &
  0.5
\\
effort  (core mins,  $U=16$) & 
  75.5 & 
  16.8 & 
  33.7 & 
  8.3 & 
  7.6 & 
  0.8 &
  0.9
\\
effort  (core mins,  $U=50$) & 
  202.3 & 
  48.6 & 
  174.8 & 
  34.4 & 
  23.7 & 
  2.5 &
  2.7  
\\
\hline

\end{tabular}
\caption{Algorithmic settings for the Lorenz-96 numerical example. 
Computational effort is measured in core minutes for running one filter, corresponding to a point on Figure~ Figure~\ref{fig:lz_loglik_plot}. 
The time taken for computing a single point using the parallel {\UBF}, {\ABF} and {\ABFIR} implementations is the effort divided by the number of cores, here $36$.
The time taken for computing a single point using the single-core GIRF, PF, EnKF and BPF implementations is equal to the effort in core minutes.
}\label{tab:lz}
\end{center}
\end{table}


\clearpage


\section{\secTitleSpace A memory-efficient representation of {\ABF}}

The {\ABF} pseudocode in the main text emphasizes the logical structure of the mathematical quantities computed, rather than describing a specific implementation. 
Arguably, an algorithm should specify not only what is to be computed, but also details of what variables should be created and saved to carry out these computations efficiently, and how computations and storage are shared across multiple locations when the algorithm is parallelized.
We use the term algorithm to denote a higher level description of the quantities to be calculated and we will say that alternative pseudocodes arriving at the same quantities are representations of the algorithm. 
Here, we present an alternative representations of {\ABF} which we call {\ABF}$_2$, and we refer to the representation in the main text as {\ABF}$_1$.
The most concrete form of an algorithm is the actual computer code implementing the algorithm in a programming language.
The implementation of {\ABF} in \texttt{spatPomp}, used for numerical results in this paper, uses an embarrassingly parallel approach based on the representation {\ABF}$_1$. 
This strategy facilitates robust and simple parallelization, and is appropriate when memory constraints are not limiting. 
To develop {\ABF}$_2$, we set up notation similar to that used for the mathematical theory.
Let 
\begin{equation}
\gamma^{}_{\unit,\time,\rep,k}=\frac{1}{\Np}\sum_{\np=1}^{\Np}\hspace{2mm}\prod_{\altUnit:(\altUnit,\time-k)\in B_{\unit,\time}} w^M_{\altUnit,\time-k,\rep,\np}
\end{equation}
Also, let 
\begin{equation}
\gamma^{+}_{\unit,\time,\rep,0}=\frac{1}{\Np}\sum_{\np=1}^{\Np}\prod_{(\altUnit,\time)\in B^{+}_{\unit,\time}} w^M_{\altUnit,\time,\rep,\np}
\end{equation}
Now set $K$ to be the largest value of $k$ for which $B^{[n-k]}_{\unit,\time}$ is nontrivial for some $(\unit,\time)$, i.e., $K$ is the largest temporal lag in any neighborhood. 
With this notation, we can write \myeqref{eq:gamma:def} as
\begin{eqnarray*}
\gamma^{MC,\rep}_{B_{\unit,\time}} &=& \prod_{k=0}^{K} \gamma^{}_{\unit,\time,\rep,k}
\\
\gamma^{MC,\rep}_{B^{+}_{\unit,\time}} &=& \gamma^{+}_{\unit,\time,\rep,0} \, \prod_{k=1}^{K} \gamma^{}_{\unit,\time,\rep,k}
\end{eqnarray*}
This motivates the following representation of {\ABF}.

\begin{center}
\noindent\begin{tabular}{l}
\hline
{\bf 
{{\ABF}$_2$}. Adapted bagged filter, representation $2$.}\inputSpace\\
\hline
\firstLineSpace
Initialize adapted simulation: $\myvec{X}^{\IF}_{0,\rep} \sim f_{\myvec{X}_0}(\myvec{x}_0)$
\\
For $\time\ \mathrm{in}\ \seq{1}{\Time}$
\\
\asp  Proposals:
    $\myvec{X}_{\time,\rep,\np}^{\IP} \sim 
    f_{\myvec{X}_{\time}|X_{1:\Unit,\time-1}} 
    \big( \myvec{x}_{\time}\given \myvec{X}^{\IF}_{\time-1,\rep}\big)$
\\
\asp Measurement weights:
  $w^M_{\unit,\time,\rep,\np} = 
    f_{Y_{\unit,\time}|X_{\unit,\time}} 
    \big (\data{y}_{\unit,\time}\given X^{\IP}_{\unit,\time,\rep,\np}\big)$
\\
\asp  Adapted resampling weights:
  $w^{\IF}_{\time,\rep,\np} = 
    \prod_{\unit=1}^{\Unit} w^M_{\unit,\time,\rep,\np}$
\\
\asp
      Resampling:
        $\prob\big[\resampleIndex({\rep})=a \big] = w^{\IF}_{\time,\rep,a}
  \Big( 
  \sum_{\altNp=1}^{\Np} w^{\IF}_{\time,\rep,\altNp}
  \Big)^{-1}$
\\
\asp 
$\myvec{X}^{\IF}_{\time,\rep} = \myvec{X}^{\IP}_{\time,\rep,r(\rep)}$ 
\\
\asp Compute $\gamma^{}_{\unit,\time+k,\rep,k}$  for $k\ \mathrm{in}\ \seq{0}{K}$,
\\
\asp Compute $\gamma^{+}_{\unit,\time,\rep,0}$
\\
End for
\\
$\displaystyle \MC{\loglik}_{\unit,\time}= 
\log\Bigg(
\frac{
\sum_{\rep=1}^\Rep \gamma^{+}_{\unit,\time,\rep,0}\prod_{k=1}^K
\gamma^{}_{\unit,\time,\rep,k}
}{
\sum_{\rep=1}^\Rep \prod_{k=0}^K
\gamma^{}_{\unit,\time,\rep,k}
}
\Bigg)
$
\vspace{1mm}
\\
\hline
\end{tabular}
\end{center}

It is clearer from the {\ABF}$_2$ representation than from the {\ABF}$_1$ representation that we do not have to save every individual particle and its weight $w^M_{\unit,\time,\rep,\np}$ after the quantities $\myvec{X}^A_{\time,\rep}$, $\gamma^{}_{\unit,\time+k,\rep,k}$ and $\gamma^{+}_{\unit,\time,\rep,0}$ have been computed.
{\ABF}$_2$ has an embarrassingly parallel implementation, since the replicates do not need to interact until  they are combined to compute $\displaystyle \MC{\loglik}_{\unit,\time}$.
An embarrassingly parallel implementation therefore requires $O\big(\Unit(K\Time+\Np)\big)$ memory for each replicate, and $O\big(\Unit\Time \Rep K\big)$ memory when the results from each replicate are collected together.

By using additional communication, for example when the implementation is designed for a single core or multiple cores with shared memory, it is possible to further reduce the memory requirement.
At time $\time$, we need to save only $\myvec{X}^A_{\time,\rep}$ and $\gamma^{}_{\unit,\time-K:\time+K,\rep,k}$ to compute $\displaystyle \MC{\loglik}_{\unit,\time}$ and all subsequent quantities.
Computing  $\myvec{X}^A_{\time,\rep}$ requires $O(\Unit\Np)$ storage.
Therefore, {\ABF} can be implemented with memory requirement $O\big(\Unit (K\Rep+\Np) \big)$, independent of $\Time$.


\section{\secTitleSpace Bagged filters for functions of the latent states}

The theory and methodology in the main article focused on filtering for likelihood estimation.
Here, we describe extensions to other filtering problems.
Let $\{\filtFunc_{\filts}:\myvec{X}^{\Unit}\to\R, \mbox{ $\filts$ in $\seq{1}{\Filts}$}\}$ be a collection of functions where $\filtFunc_{\filts}$ depends only on a subset of units $\filtFuncSet_{\filts}\subset \seq{1}{\Unit}$.
We suppose that there exist neighborhoods
\begin{equation}
B^\prime_{\filts,\time}\subset A^\prime_{\time}=(\seq{1}{\Unit})\times(\seq{0}{\time})
\end{equation}
such that $\filtFunc_{\filts}(\myvec{X}_{\time})$ is approximately independent of $\{Y_{\unit,\time}:(\unit,\time) \in B^{\prime c}_{\filts,\time}\}$ given  $\{Y_{\unit,\time}: (\unit,\time) \in B^\prime_{\filts,\time}\}$, where $B^{\prime c}_{\filts,\time}$ is a complement in $A^\prime_{\time}$.
Unlike the sets $A_{\unit,\time}$ and $B_{\unit,\time}$ defined for likelihood estimation, the sets $A^\prime_{\time}$ and $B^\prime_{\filts,\time}$ can include any locations at time $\time$.
We consider Monte Carlo estimation of
\begin{equation}
\label{eq:filt1}
\overline{\filtFunc}_{k,\time}= \E\big[\filtFunc_{\filts}(\myvec{X}_\time) \big| \myvec{Y}_{1:\time}\big], \quad \filts \mbox{ in } \seq{1}{\Filts}.
\end{equation}
For example, if we set $\filtFunc_{\unit}(\myvec{x}_\time)=x_{\unit,\time}$ and $\Filts=\Unit$, the collection of quantities $\{\overline{\filtFunc}_{\unit,\time}\}$ estimated in \eqref{eq:filt1} corresponds to a vector of filter means.
Setting $\filtFunc_{\filts}(\myvec{x}_\time)=x_{\unit_{\filts},\time}\, x_{\altUnit_{\filts},\time}$ enables calculation of a collection of filter covariances between units $\unit_k$ and $\altUnit_k$ for $k$ in $\seq{1}{\Filts}$.

We consider bagged filtering approaches to estimation of $\{\overline{\filtFunc}_{\filts,\time}, \filts \mbox{ in } \seq{1}{\Filts}\}$.
Although each $\overline{\filtFunc}_{\filts,\time}$ is required to have only local dependence, some global quantities such as filter means and their variances across all units, can be expressed in terms of collections of such quantities.
For bagged filtering to operate successfully, the neighborhoods $\{B^\prime_{\filts,\time}\}$ should not be large.
Since  $B^\prime_{\filts,\time}$ will typically be larger than $\{\filtFuncSet_\filts\}$, this rules out estimation of filtered quantities that cannot be adequately represented by a collection of localized filtering calculations.
We now present variants of the pseudocode in the main text, targeted at estimation of $\{\overline{\filtFunc}_{k,\time},k \mbox{ in } 1{\mycolon}K\}$.
We do not prove theorems about these algorithms, but we conjecture from their similarity to the algorithms in the main text that comparable theoretical results should exist.
Table~\ref{tab:filter:inputs} lists the inputs, outputs and ranges of the implicit loops for the following algorithms.

\begin{table}[htbp]
\begin{center}
\noindent\begin{tabular}{l}
\hline
\inputSpace {\bf Latent state estimation via bagged filters.}\\
\hline
\inputSpace {\bf input:}
\\
collection of functions, $\filtFunc_{\filts}$\\
neighborhoods, $B^\prime_{\filts,\time}$\\
simulator for $f_{\myvec{X}_0}(\myvec{x}_0)$ and $f_{\myvec{X}_{\time}|\myvec{X}_{\time-1}}(\myvec{x}_{\time}\given \myvec{x}_{\time-1})$\\
    evaluator for $f_{{Y}_{\unit\comma\time}|{X}_{\unit\comma\time}}({y}_{\unit\comma\time}\given {x}_{\unit\comma\time})$\\
    number of replicates, ${\Rep}$\\
    data, $\data{\myvec{y}}_{1:\Time}$\\
{\ABF} and {\ABFIR}:    particles per replicate,  $\Np$\\
{\ABFIR}: number of intermediate timesteps, $\Ninter$ \\
{\ABFIR}: measurement variance parameterizations, ${\VtoTheta}_{\unit\comma\time}$ and ${\thetaToV}_{\unit\comma\time}$\\
{\ABFIR}: approximate process and observation mean functions, $\myvec{\mu}$ and $h_{\unit\comma\time}$\\
\inputSpace {\bf output:}\\
  filter estimate, $\MC{\overline{\filtFunc}}_{\filts,\time}$\\
\inputSpace {\bf implicit loops:}\\
$\unit \mbox{ in } \seq{1}{\Unit}$, 
$\; \time \mbox{ in } \seq{1}{\Time}$, 
$\; \rep \mbox{ in } \seq{1}{\Rep}$, 
$\; \np \mbox{ in } \seq{1}{\Np}$,
$\; \filts \mbox{ in } \seq{1}{\Filts}$
\lastLineSpace \\
\hline
\end{tabular}
\end{center}
\caption{\label{tab:filter:inputs}
Notation for bagged filter for latent state estimation: inputs, outputs and implicit loops.}
\end{table}

\clearpage

\begin{center}
\noindent\begin{tabular}{l}
\hline
\inputSpace {\bf {\UBF}. Unadapted bagged filter for latent state estimation.}\\
\hline
\firstLineSpace 
Simulate $\myvec{X}^{\tif}_{0:\Time,\rep}\sim f_{\myvec{X}_{0:\Time}}(\myvec{x}_{0:\Time})$\\ 
Measurement weights,
$w^M_{\unit,\time,\rep}=f_{Y_{\unit,\time}|X_{\unit,\time}}(\data{y}_{\unit,\time}\given X^{\tif}_{\unit,\time,\rep})$
\\
Filtering weights, 
$w^F_{\filts,\time,\rep}=\prod_{(\tilde \unit,\tilde n)\in B^\prime_{\filts,\time}}
w^M_{\tilde\unit,\tilde n,\rep}$\\
$\displaystyle
\MC{\overline{\filtFunc}}_{\filts,\time}= 
\frac{
  \sum_{\rep=1}^\Rep \filtFunc_\filts(\myvec{X}^{\tif}_{\time,\rep}) \,
  w^F_{\filts,\time,\rep}
}{
  \sum_{\rep=1}^\Rep 
  w^F_{\filts,\time,\rep}
}
$
\lastLineSpace
\\
\hline
\end{tabular}
\end{center}

\begin{center}
\noindent\begin{tabular}{l}
\hline
{\bf 
{\ABF}. Adapted bagged filter for latent state estimation.}\inputSpace\\
\hline
\firstLineSpace
Initialize adapted simulation: $\myvec{X}^{\IF}_{0,\rep} \sim f_{\myvec{X}_0}(\myvec{x}_0)$
\\
For $\time\ \mathrm{in}\ \seq{1}{\Time}$
\\
\asp  Proposals:
    $\myvec{X}_{\time,\rep,\np}^{\IP} \sim 
    f_{\myvec{X}_{\time}|X_{1:\Unit,\time-1}} 
    \big( \myvec{x}_{\time}\given \myvec{X}^{\IF}_{\time-1,\rep}\big)$
\\
\asp Measurement weights:
  $w^M_{\unit,\time,\rep,\np} = 
    f_{Y_{\unit,\time}|X_{\unit\comma\time}} 
    \big (\data{y}_{\unit\comma\time}\given X^{\IP}_{\unit\comma\time,\rep,\np}\big)$
\\
\asp  Adapted resampling weights:
  $w^{\IF}_{\time,\rep,\np} = 
    \prod_{\unit=1}^{\Unit} w^M_{\unit,\time,\rep,\np}$
\\
\asp
      Resampling:
        $\prob\big[\resampleIndex({\rep})=a \big] = w^{\IF}_{\time,\rep,a}
  \Big( 
  \sum_{q=1}^{\Np} w^{\IF}_{\time,\rep,q}
  \Big)^{-1}$
\\
\asp 
$\myvec{X}^{\IF}_{\time,\rep} = \myvec{X}^{\IP}_{\time,\rep,r(\rep)}$ 
\\
\asp 
  $w^{F}_{\filts,\time,\rep,\np}= \displaystyle
  \prod_{\altTime=1}^{\time-1}
  \Big[
    \frac{1}{\Np}\sum_{q=1}^{\Np}
    \hspace{1mm}
       \prod_{\altUnit:(\altUnit,\altTime)\in B^\prime_{\filts,\time}} 
    \hspace{-1mm}
        w^M_{\altUnit,\altTime,\rep,q}
  \Big] \prod_{\altUnit:(\altUnit,\time)\in B^\prime_{\filts,\time}} 
    \hspace{-1mm}
        w^M_{\altUnit,\time,\rep,\np}$
\\
End for
\\
$\displaystyle \MC{\overline{\filtFunc}}_{\filts,\time}=
\frac{
\sum_{\rep=1}^\Rep \sum_{\np=1}^{\Np} \filtFunc_\filts(\myvec{X}^{\IP}_{\time,\rep,\np}) \,
w^F_{\filts,\time,\rep,\np}
}{
\sum_{\rep=1}^\Rep \sum_{\np=1}^{\Np} w^F_{\filts,\time,\rep,\np}
}
$
\vspace{1mm}
\\
\hline
\end{tabular}
\end{center}

\begin{center}
\noindent\begin{tabular}{l}
\hline
{\bf {\ABFIR}. Adapted bagged filter with intermediate resampling}
\\
{\bf \hspace{2cm} for latent state estimation.} 
\firstLineSpace \\  
\vspace{0.4mm} \\
\hline
\firstLineSpace
Initialize adapted simulation: $\myvec{X}^{\IF}_{0,\rep} \sim f_{\myvec{X}_0}(\myvec{x}_0)$
\\
For $\time\ \mathrm{in}\ \seq{1}{\Time}$
\\
\asp Guide simulations:
    $\myvec{X}_{\time,\rep,\npgir}^{G} \sim 
    f_{\myvec{X}_{\time}|\myvec{X}_{\time-1}} 
    \big( \myvec{x}_{\time}\given \myvec{X}^{\IF}_{\time-1,\rep} \big)$
\\
\asp Guide sample variance: $V_{\unit,\time,\rep}=
      \var \big\{
        h_{\unit\comma\time}\big( {X}_{\unit,\time,\rep,\npgir}^{G}\big), \npgir \mbox{ in } \seq{1}{\Npgir}
      \big\}$ 
\\
\asp $\guideFunc^{\resample}_{\time,0,\rep,\np}=1 \; \; $ and
$\; \myvec{X}_{\time,0,\rep,\np}^{\GR}=\myvec{X}^{\IF}_{\time-1,\rep}$
\\
\asp For $\ninter  \,\, \mathrm{in} \,\, \seq{1}{\Ninter}$
\\
\asp\asp Intermediate proposals:
        ${\myvec{X}}_{\time,\ninter,\rep,\np}^{\GP}
          \sim {f}_{{\myvec{X}}_{\time,\ninter}|{\myvec{X}}_{\time,\ninter-1}}
          \big(\mydot|{\myvec{X}}_{\time,\ninter-1,\rep,\np}^{\GR}\big)$ 
\\
\asp\asp 
        $\myvec{\mu}^{\GP}_{\time,\ninter,\rep,\np} 
           = \myvec{\mu}\big( \myvec{X}^{\GP}_{\time,\ninter,\rep,\np},t_{\time,\ninter},t_{\time} \big)$
\\
\asp\asp      
        $V^{\mathrm{meas}}_{\unit,\time,\ninter,\rep,\np}
           = \thetaToV_{\unit}(\theta,\mu^{\GP}_{\unit,\time,\ninter,\rep,\np})$
\\
\asp\asp  
        $V^{\mathrm{proc}}_{\unit,\time,\ninter,\rep}
          = V_{\unit,\time,\rep} \,
          \big(t_{\time}-t_{\time,\ninter}\big) \Big/
          \big(t_{\time}-t_{\time,0}\big)$ 
\\
\asp\asp
        $\theta_{\unit,\time,\ninter,\rep,\np}= 
          \VtoTheta_{\unit}\big(
            V^{\mathrm{meas}}_{\unit,\time,\ninter,\rep,\np} + V^{\mathrm{proc}}_{\unit,\time,\ninter,\rep}, 
            \, \mu^{\GP}_{\unit,\time,\ninter,\rep,\np}
          \big)$
\\
\asp\asp  
        $
\guideFunc_{\time,\ninter,\rep,\np}=
          \prod_{\unit=1}^{\Unit}
          f_{Y_{\unit,\time}|X_{\unit,\time}}
          \big(
            \data{y}_{\unit,\time}\given \mu^{\GP}_{\unit,\time,\ninter,\rep,\np} \giventh \theta_{\unit,\time,\ninter,\rep,\np} 
          \big)$
\\
\asp\asp Guide weights:
$w^G_{\time,\ninter,\rep,\np}= \guideFunc^{}_{\time,\ninter,\rep,\np}
         \big/ \guideFunc^{\resample}_{\time,\ninter-1,\rep,\np}$
\\
\asp\asp
      Resampling:
        $\prob\big[\resampleIndex({\rep,\np})=a \big] = w^G_{\time,\ninter,\rep,a}
\Big( \sum_{q=1}^{\Np}w^G_{\time,\ninter,\rep,q}\Big)^{-1}$
\\
\asp\asp
        $\myvec{X}_{\time,\ninter,\rep,\np}^{\GR}=\myvec{X}_{\time,\ninter,\rep,\resampleIndex({\rep,\np})}^{\GP}\; \; $ and
        $\; \guideFunc^{\resample}_{\time,\ninter,\rep,\np}= \guideFunc^{}_{\time,\ninter,\rep,\resampleIndex({\rep,\np})}\,$
\\
\asp
End For
\\
\asp
  Set $\myvec{X}^{\IF}_{\time,\rep}=\myvec{X}^{\GR}_{\time,\Ninter,\rep,1}$ 
\\ 
\asp Measurement weights:
  $w^M_{\unit,\time,\rep,\npgir} = 
    f_{Y_{\unit,\time}|X_{\unit,\time}} 
    \big (\data{y}_{\unit,\time}\given X^{G}_{\unit,\time,\rep,\npgir} \big)$
\\
\asp 
  $w^{F}_{\filts,\time,\rep,\npgir}= \displaystyle
  \prod_{\altTime=1}^{\time-1}
  \Big[
    \frac{1}{\Npgir}\sum_{q=1}^{\Npgir}
    \hspace{1mm}
       \prod_{\altUnit:(\altUnit,\altTime)\in B^\prime_{\filts,\time}} 
    \hspace{-1mm}
        w^M_{\altUnit,\altTime,\rep,q}
  \Big] \prod_{\altUnit:(\altUnit,\time)\in B^\prime_{\filts,\time}} 
    \hspace{-1mm}
        w^M_{\altUnit,\time,\rep,\npgir}$
\\
End for
\\
$\displaystyle \MC{\overline{\filtFunc}}_{\filts,\time}=
\frac{
\sum_{\rep=1}^\Rep \sum_{\np=1}^{\Np} \filtFunc_\filts(\myvec{X}^{\IP}_{\time,\rep,\np}) \,
w^F_{\filts,\time,\rep,\np}
}{
\sum_{\rep=1}^\Rep \sum_{\np=1}^{\Np} w^F_{\filts,\time,\rep,\np}
}
$
\vspace{1mm}
\\
\hline
\end{tabular}
\end{center}


\section{\secTitleSpace An iterated bagged filter for parameter estimation}

This paper focuses on evaluating the likelihood function for SpatPOMP models via filtering.
Although the likelihood function is fundamental for inference, evaluation alone is not sufficient.
Likelihood maximization enables calculation of the maximum likelihood estimate, profile likelihood confidence intervals, likelihood ratio tests and likelihood-based model selection critera.
Iterated filtering methodology \citep{ionides06-pnas,ionides11,ionides15} provides an approach to extending filtering algorithms to likelihood maximization algorithms. 
Here, we demonstrate one such extension in the context of bagged filters.
This demonstration provides a proof of concept to motivate future work.

Iterated filtering approaches apply filtering to a modified version of the model where parameters are perturbed at each time point.
The filtering procedure directs the perturbed parameters toward values consistent with the data.
At the end of each filtering operation, a parameter updating rule is applied and a new filtering iteration is started with reduced perturbation variance.
Under suitable conditions, iterative procedures of this type converge to a neighborhood of a maximum likelihood parameter despite the presence of Monte Carlo filtering error.
We implemented an iterated bagged filter procedure, described by the pseudocode below.
The pseudocode is presented for an iterated adapted bagged filter (IABF) but the iterated unadapted filter (IUBF) corresponds to the case $\Np=1$ and the iterated bagged filter with intermediate resampling (IABF-IR) follows by adding the intermediate resampling procedure used by ABF-IR.
The filtering step of IABF uses ABF to estimate the likelihood at the $K$ perturbed parameter sets.
The selection step does not resample parameters based on these estimated likelihood.
Rather, it selects the parameters with the top $p$ quantile of likelihoods and copies them appropriately to get $K$ new parameters for the next filtering step.
This quantile-based resampling allows us to maintain the diversity of the $K$ parameter sets and avoid \textit{parameter degeneracy}, whereby very few parameters are resampled, leading to an inefficient search of parameter space.

For simplicity, in this description, we assume that parameters are transformed so that their values are unconstrained.
Our software implementation, provided the R package \texttt{spatPomp} \citep{asfaw21arxiv}, provides facilities for carrying out such transformations.
The Gaussian distribution used for perturbations, and the geometric perturbation variance reduction factor,  $\alpha$, are convenient specifications but are not required in theory \citep{ionides15}.
As another simplification, the pseudocode for IABF represents the logical structure of the algorithm without attending to issues of memory management and parallelization.
For implementation issues, we refer to \texttt{spatPomp} \citep{asfaw21arxiv}.

In Figure~5 of the main text, we use this IABF implementation to construct a profile likelihood for the measles model. We use $\Np=1$, $\Rep=30000$, $K=250$, $p=0.8$, $\alpha=0.5$, $M=15$ and $\Sigma$ set to be a diagonal matrix with perturbation variance for each non-initial value parameter set to 0.02. For this exercise, we fix the initial value parameters at their true values.

The resulting Monte Carlo parameter estimates have Monte Carlo uncertainty in both likelihood evaluation and maximization.
Therefore, we used a Monte Carlo adjusted profile likelihood \citep{ionides17,ning21} that accounts for this uncertainty.
Fig.~\MainFigureMeaslesProfile gives empirical demonstration of this procedure on the measles model.

\begin{table}
\caption{Iterated bagged filter inputs, outputs and implicit loops, extending ABF table in main text.}
\begin{center}
\noindent\begin{tabular}{l}
\hline
\inputSpace {\bf input: same as ABF table in main text plus}
\\
\asp Number of maximization iterations, $M$
\\
\asp Number of parameter vectors, $K$
\\
\asp Perturbation variance, $\Sigma$
\\
\asp Variance reduction factor after 50 iterations, $\alpha$
\\
\asp Starting parameters, $\theta_{1:K}^{(0)}$
\\
\asp Resampling proportion, $p$
\\
\inputSpace {\bf output:}\\
\asp Parameter estimates approaching the maximum likelihood estimate, $\theta^{(M)}_{1:K}$
\\  
\inputSpace {\bf implicit loop:}\\
\asp $\; k \mbox{ in } \seq{1}{K}$, $\; \rep \mbox{ in } \seq{1}{\Rep}$
\lastLineSpace \\
\hline
\end{tabular}
\end{center}
\end{table}

\begin{center}
\noindent\begin{tabular}{l}
\hline
{\bf 
{IABF}. Iterated adapted bagged filter. }\inputSpace
\\
\hline
\firstLineSpace
For $m\ \mathrm{in}\ \seq{1}{M}$
\\
\asp $\theta_{0,1:K}^{F} = \theta_{1:K}^{(m-1)}$
\\
\asp Initialize adapted simulation: $\myvec{X}^{F}_{0,\rep,k} \sim f_{\myvec{X}_0}\big(\myvec{x}_0 \giventh \theta_{0,k}^{F}\big)$
\\
\asp For $\time\ \mathrm{in}\ \seq{1}{\Time}$
\\
\asp\asp $\theta_{\time,k}^{P}\sim\normal\big[\theta_{\time-1,k}^{F}, \alpha^{2m/50} \Sigma\big]$
\\
\asp\asp 
    $\myvec{X}_{\time,\rep,\np,k}^{\IP} \sim 
    f_{\myvec{X}_{\time}|\myvec{X}_{\time-1}} 
    \big( \myvec{x}_{\time}\given \myvec{X}^{F}_{\time-1,\rep,k} \giventh \theta_{\time,k}^{P}\big)$
\\
\asp\asp Measurement weights:
  $w^M_{\unit,\time,\rep,\np,k} = 
    f_{Y_{\unit,\time}|X_{\unit\comma\time}} 
    \big (\data{y}_{\unit\comma\time}\given X^{\IP}_{\unit\comma\time,\rep,\np,k}
      \giventh \theta_{\time,k}^{P} \big)$
\\
\asp\asp  Adapted resampling weights:
  $w^{\IF}_{\time,\rep,\np,k} = 
    \prod_{\unit=1}^{\Unit} w^M_{\unit,\time,\rep,\np,k}$
\\
\asp\asp
      State resampling:
        $\prob\big[\resampleIndex({\rep},k)=a \big] = w^{\IF}_{\time,\rep,a,k}
  \Big( 
  \sum_{\xi=1}^{\Np} w^{\IF}_{\time,\rep,\xi,k}
  \Big)^{-1}$
\\
\asp\asp
$\myvec{X}^{\IF}_{\time,\rep,k} = \myvec{X}^{\IP}_{\time,\rep,r(\rep,k),k}$ 
\\
\asp\asp 
  $w^{\LCP}_{\unit,\time,\rep,\np,k}= \displaystyle
  \prod_{\altTime=1}^{\time-1}
  \Big[
    \frac{1}{\Np}\sum_{\xi=1}^{\Np}
    \hspace{1mm}
       \prod_{\altUnit:(\altUnit,\altTime)\in B_{\unit,\time}} 
    \hspace{-1mm}
        w^M_{\altUnit,\altTime,\rep,\xi,k}
  \Big] \prod_{\altUnit:(\altUnit,\time)\in B_{\unit,\time}} 
    \hspace{-1mm}
        w^M_{\altUnit,\time,\rep,\np,k}$
\\
\asp\asp
$\displaystyle \MC{\loglik}_{\time,k}= 
\sum_{\unit=1}^\Unit\log\Bigg(
\frac{
\sum_{\rep=1}^\Rep \sum_{\np=1}^{\Np} w^M_{\unit,\time,\rep,\np,k}w^P_{\unit,\time,\rep,\np,k}
}{
\sum_{\rep=1}^\Rep \sum_{\np=1}^{\Np} w^P_{\unit,\time,\rep,\np,k}
}
\Bigg)
$
\\
\asp\asp Select the highest $p K$ likelihoods:
  find $s$ with \\~~~~~
  $
   \{s(1),\dots,s(\lceil p K\rceil)\}=
   \big\{k: \sum_{\tilde{k}=1}^{K}{\mathbf{1}} 
   \{\MC{\loglik}_{\time,\tilde{k}} > \MC{\loglik}_{\time,k} \} < (1-p) K
   \big\}$
\\
\asp\asp
Make $1/p$ copies of successful parameters,
$\theta_{\time,k}^{F}=\theta_{\time,s(\lceil p k\rceil)}^{P}$
\\
\asp\asp
$\myvec{X}^{F}_{\time,\rep,k} = \myvec{X}^{A}_{\time,\rep,s(\lceil p k\rceil)}$ 
\\
\asp End for
\\
\asp
$\theta_{k}^{(m)}=\theta_{\Time,k}^{F}$
\\
End for
\vspace{1mm}
\\
\hline
\end{tabular}
\end{center}


\section{\secTitleSpace Replicates versus particles for the measles model}

\begin{knitrout}
\definecolor{shadecolor}{rgb}{1, 1, 1}\color{fgcolor}\begin{figure}

{\centering \includegraphics[width=3.5in]{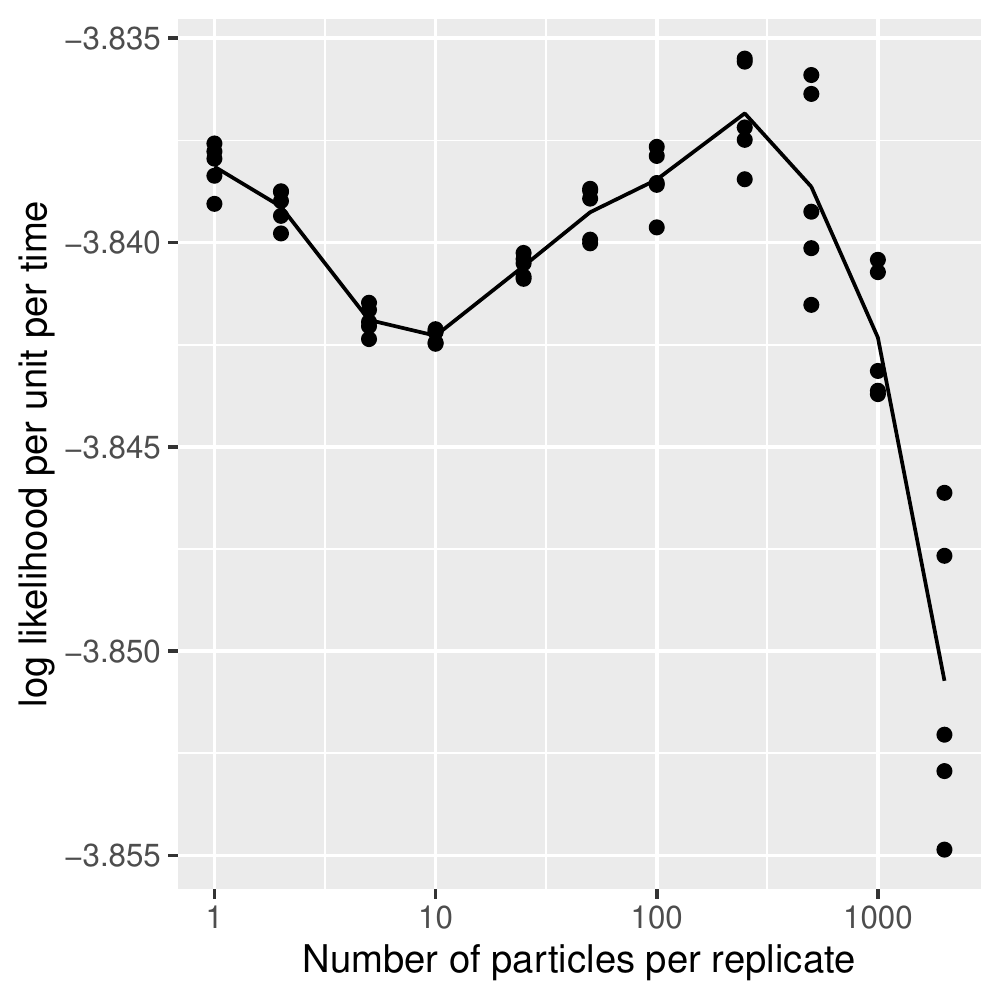} 

}

\caption[log likelihood estimates for simulated data from the measles model using {\ABF}, with varying algorithmic parameters]{log likelihood estimates for simulated data from the measles model using {\ABF}, with varying algorithmic parameters.}\label{fig:ij_loglik_plot}
\end{figure}

\end{knitrout}

It is necessary in practice to decide whether computational resources are best directed toward a large number of replicates, $\Rep$, or a large number of particles per replicate, $\Np$.
Computational effort for ABF and ABF-IR is approximately proportional to $\Np\Rep$, and UBF corresponds to ABF with $\Np=1$.
Suitable algorithmic parameters may depend on the model under consideration, and here we consider resource allocation for the measles model studied in the main text, using simulated data with $U=40$ and $N=104$.
From Figure~\MainFigureMeaslesSlice, we know that this implementation of the model is well suited to ABF and UBF.
ABF-IR performs less well, and the weak performance of GIRF suggests that the weakness may be due to an inadequate guide function.
Figure~\ref{fig:ij_loglik_plot} investigates the tradeoff between UBF and ABF by plotting evaluated log likelihood against $\Rep$ with $\Np$ chosen to give approximately a constant computational effort.
Algorithmic settings and run times are reported in Table~\ref{tab:ij}.
For our implementation, choosing $\Np$ very low and $\Rep$ correspondingly high led to greater computation time, perhaps because the code was written to parallelize nicely when $\Np$ is relatively large.

We interpret the bimodal curve as follows.
When ABF is carried out with an inadequate number of particles for each bootstrap replicate, the algorithm cannot make a good representation of a draw from the adapted distribution.
In that case, there is an advantage to using UBF, which does not attempt to carry out adapted simulation.
For very small numbers of particles per replicate, the ABF algorithm behaves like a not-quite-properly-weighted version of UBF.
Large numbers of particles per replicate presumably lead to improved Monte Carlo representation of draws from the adapted distribution, but computational cost constraints prevent combining this with a large number of replicates.
We see a mode around 500 particles per replicate where ABF out-performs UBF.
On this problem, UBF is relatively successful, presumably because the measles dynamics in each city are strongly attracted toward relatively few stable cycles (annual epidemics, or peaks in odd years, or peaks in even years) and a tractable number of simulations can represent all these scenarios.
The Lorenz model of Sec.~\ref{sec:lorenz} provides an alternative situation, where adaptation has more advantages.
Also, the plug-and-play guide function appears to operate successfully in this case, as evidenced by relatively strong performance from ABF-IR and GIRF.

\begin{table}
\begin{center}
\begin{tabular}{ccc}
\hline
  $\Np$ & $\Island$ & time\\
\hline
2000 & 500  & 70.7 \\
1000 & 1000  & 71.3 \\
500 & 2000  & 72.4 \\
250 & 4000  & 75.0 \\
100 & 10000  & 80.9 \\
50 & 20000  & 93.8 \\
25 & 40000  & 114.9 \\
10 & 40000  & 71.4 \\
5 & 40000  & 56.1 \\
2 & 40000  & 47.7 \\
1 & 40000  & 43.0 \\
\hline
\end{tabular}
\end{center}
\caption{Bootstrap replications, $\Rep$, particles per replicate, $\Np$, and computational time (total minutes for the five points) for the results presented in Figure~\ref{fig:ij_loglik_plot}}\label{tab:ij}
\end{table}

\clearpage

\bibliography{../bib-iif}